\documentclass[twocolumn]{aastex631}
\usepackage{amsmath}
\usepackage{gensymb}
\usepackage{xcolor}
\usepackage{xcolor}
\begin{document}
\title{Generation of fast magnetoacoustic waves in the corona  by impulsive bursty reconnection}

\author{Sripan Mondal}
\affiliation{Department of Physics, Indian Institute of Technology (BHU), Varanasi-221005, India.}
\author{A.K.~Srivastava}
\affiliation{Department of Physics, Indian Institute of Technology (BHU), Varanasi-221005, India. Email:- asrivastava.app@iitbhu.ac.in}
\author{David I. Pontin}
\affiliation{School of Information and Physical Sciences, University of Newcastle, Australia.}
\author{Eric R. Priest}
\affiliation{Mathematics Institute, St Andrews University, KY16 9SS, St Andrews, UK.}
\author{R.~Kwon}
\affiliation{Korea Astronomy and Space Science Institute, Daejeon 34055, Republic of Korea.}
\author{Ding~Yuan}
\affiliation{Shenzhen Key Laboratory of Numerical Prediction for Space Storm, Institute of Space Science and Applied Technology, Harbin Institute of Technology, Shenzhen, Guangdong, People's Republic of China. Email:- yuanding@hit.edu.cn}




\begin{abstract}
Fast-mode magnetohydrodynamic (MHD) waves in the solar corona are often known to be produced by solar flares and eruptive prominences. We here simulate the effect of the interaction of an external perturbation on a magnetic null in the solar corona which results in the formation of a current sheet (CS). Once the CS undergoes a sufficient extension in its length and squeezing of its width, it may go unstable to the formation of multiple impulsive plasmoids. Eventually, the plasmoids merge with one another to form larger plasmoids and/or are expelled from the sheet. The formation, motion and coalescence of plasmoids with each other and with magnetic Y-points at the outer periphery of the extended CS are found to generate wave-like perturbations. An analysis of the resultant quasi-periodic variations of pressure, density, velocity and magnetic field at certain locations in the model corona indicate that these waves are predominantly fast-mode magnetoacoustic waves. For typical coronal parameters, the resultant propagating waves carry an energy flux of $\mathrm{10^{5}~\mathrm{erg~cm^{-2}~s^{-1}}}$ to a large distance of at least 60 Mm away from the current sheet. In general, we suggest that both waves and reconnection play a role in heating the solar atmosphere and driving the solar wind and may interact with one another in a manner that we refer to as a $"$Symbiosis of WAves and Reconnection (SWAR)$"$.
\end{abstract}


\section{Introduction} 
Magnetohydrodynamic (MHD) waves are ubiquitous in three distinct modes in the solar corona, namely, slow, fast and Alfv\'{e}n modes. Under the influence of normal collisional dissipative properties (e.g., viscosity and resisitivity), transverse waves such as Alfv\'{e}n waves are difficult to dissipate in the inner corona \citep{Holl07}. Alfv\'{e}n modes can be dissipated during nonlinear mode conversion to compressive modes which further results in the formation of shocks and associated heating \citep{1998ApJ...493..474O,2010ApJ...712..494A}. Also, the effective dissipation of Alfv\'{e}n modes may contribute to heating the corona when it is subjected to non-ideal effects such as phase mixing \citep{1983A&A...117..220H} and resonant absorption \citep{1987ApJ...317..514D}. Likewise damping and dissipation of {\it in-situ} generated slow and fast mode waves are capable of heating  the corona and chromosphere if they have significant energy fluxes \citep{1992SoPh..140....7E,1994ApJ...435..482P,2001MNRAS.326..675P}. Thus, dissipation of all  three modes is often studied as a mechanism for coronal and chromospheric heating, but it is a topic of continual debate and refinement (see the review articles by \citet{2020SSRv..216..140V}, \citet{2021JGRA..12629097S} and references therein). Moreover, the study of these MHD waves is important for acquiring physical insights about solar eruptions, acceleration of the solar wind, and the physical properties of the solar atmosphere by coronal seismology \citep{2005LRSP....2....3N}. The presence and propagation of large-scale coronal disturbances (often fast-mode MHD waves) have been observed in optically thin coronal EUV and SXR emission in addition to Thompson-scattered white-light radiation. The dawn of high-resolution and high-temporal cadence observational facilities such as SOHO/EIT \citep{1995SoPh..162..291D}, STEREO/EUVI \citep{2004SPIE.5171..111W} and SDO/AIA \citep{2012SoPh..275...17L} has produced on-disk and off-limb observations of large-scale propagation of fast-mode waves which carry energy far away from the source regions and eventually dissipate to cause heating (see review articles \citet{2011SSRv..158..365G,2012SoPh..281..187P,2014SoPh..289.3233L,2015LRSP...12....3W} and so on). Indeed, recently \citet{2024ApJ...960...51P} have suggested that reconnection from small-scale flux cancellation may both generate  the solar wind and heat the corona. 

Even though large-scale fast MHD waves are omnipresent in the solar corona, the mechanisms behind their generation  are still debated. Solar flares, CMEs, and filament eruptions are possible sources in the neighbourhood of active regions (ARs) even though the signatures of  magnetic reconnection are difficult to detect \citep[e.g.,][]{2011ApJ...736L..13L,2012ApJ...753...52L,2013SoPh..288..585S,2013A&A...554A.144Y,2014A&A...569A..12N,2017ApJ...844..149K,2018ApJ...858L...1Z,2023ApJ...957..110L}. Nevertheless, quasi-periodic pulsations in flare emission indicate that  a common mechanism may be generating fast MHD waves and the time variability of flare energy release. This similarity points towards the possible role of impulsive reconnection. For example, \citet{2018ApJ...868L..33L} reported the generation of quasi-periodic propagating large-scale disturbances from a magnetic reconnection site between coronal loops viewed with SDO/AIA.

To study a possible connection between quasi-periodic reconnection and the generation of large-scale fast MHD waves, several numerical studies have been carried out so far. \citet{2015ApJ...800..111Y} studied the excitation of fast-mode waves by the interaction of  plasmoids and the ambient magnetic field in the outflow region.  \citet{2016ApJ...823..150T} demonstrated the excitation of quasi-periodic fast wave trains by above-the-loop-top oscillations driven by reconnection outﬂow in an elongated straight current sheet. Similarly, \citet{2017ApJ...847...98J} reported the generation of fast MHD waves via merging of plasmoids in a vertical gravitationally stratified current sheet. Even though \citet{2022A&A...666A..28S} did not reported generation of waves due to plasmoid coalescence, they showed that thermal and tearing instabilities can reinforce each other to  significantly increase instability growth rate and produce plasmoid-trapped condensations. Therefore, it is obvious that presence of non-adiabatic effects such as radiative cooling plays important role in formation and evolution of plasmoids in coronal current sheets which may further have important implication in wave generation. All of these studies have been conducted for a non-viscous corona either in the presence or  absence of anisotropic thermal conduction and radiative cooling. Viscosity can have two important roles here. One is on the relative movement of plasmoids and their coalescence and another is on the dissipation of the resulting waves. Therefore, it is interesting to simulate wave generation from such plasmoid coalescence  and to analyze the impacts on coronal heating.  

It has long been suggested that waves may drive reconnection \cite[e.g.,][]{1983JPlPh..30..109S,1991ApJ...371L..41C} and that time-dependendent reconnection may drive waves \cite[e.g.,][]{2007PhPl...14l2905L}. 
Both waves and reconnection may play a role in heating the chromosphere and corona, and may interact with one another, which we refer to here as a $"$Symbiosis of WAves and Reconnection (SWAR)$"$. The work of this paper is one example of such an interaction, and we aim to describe others in future.

We present a recipe for generating fast MHD waves in the corona via coalescence between plasmoids in a dynamic current sheet (CS) that is formed in response to interaction of velocity perturbations with a magnetic null initiated by a far away source in an anisotropically thermally conductive, resistive and viscous solar corona. Eventually the waves, generated by plasmoid coalescence, propagate as multiple arc-shaped bright fronts  across the ambient coronal magnetic field. As a result of their large-scale propagation, they carry energy far from the current sheet to  distant regions. In Sect.~2, the physics-based numerical model used for this study is described. In Sect.~3, the resulting dynamics are discussed in detail. Lastly, in Sect.~4, we summarize the findings, compare them with existing studies and discuss the importance of this numerical experiment in the light of a causal connection between MHD waves and magnetic reconnection (namely, SWAR). Some complementary details of the modelling and estimation supporting the main scientific findings are outlined in various appendices.

\begin{figure*}
\hspace{-0.8 cm}
\includegraphics[scale=0.8]{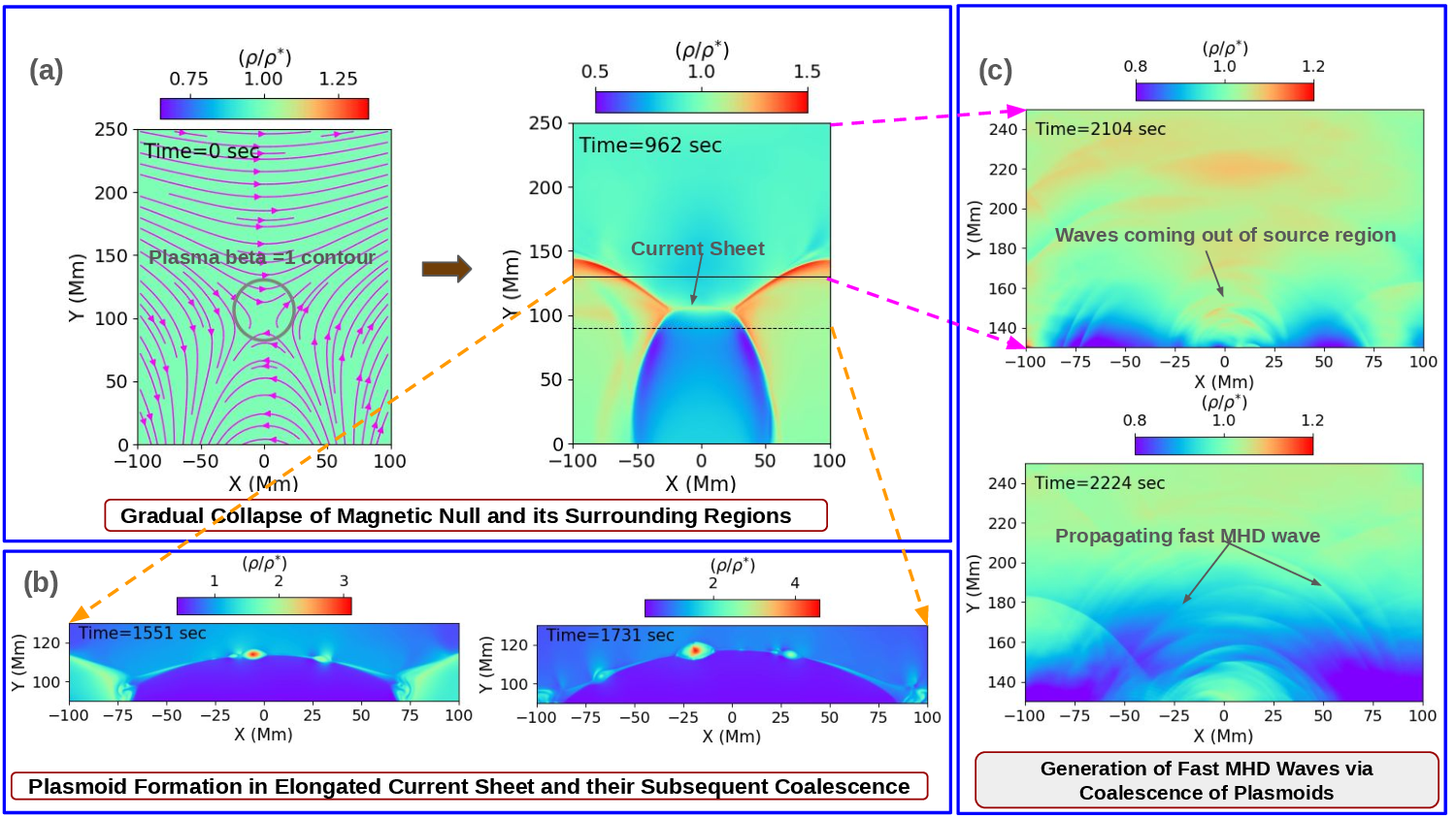}
\caption{Panel (a) shows the initial magnetic field lines with a plasma of uniform density shown in green together with the formation of a horizontal current sheet (CS) at time t=962 s. Panel (b) exhibits multiple plasmoid formation and their coalescence in the thinned and elongated CS due to nonlinear resistive instabilities. Panel (c) demonstrates the large-scale propagation of fast MHD waves emitted by the coalescing plasmoids in the CS. An animation covering the propagation of the initial velocity perturbation, its distortion and the formation of the current sheet from the start of the simulation to 1070 s is available in the online version. The real-time duration of the animation is 9 s.}
\label{label 1}
\end{figure*}

\begin{figure*}

\mbox{
\hspace{-0.9 cm}
\includegraphics[height=7.5 cm,trim={0 0 0 0},clip]{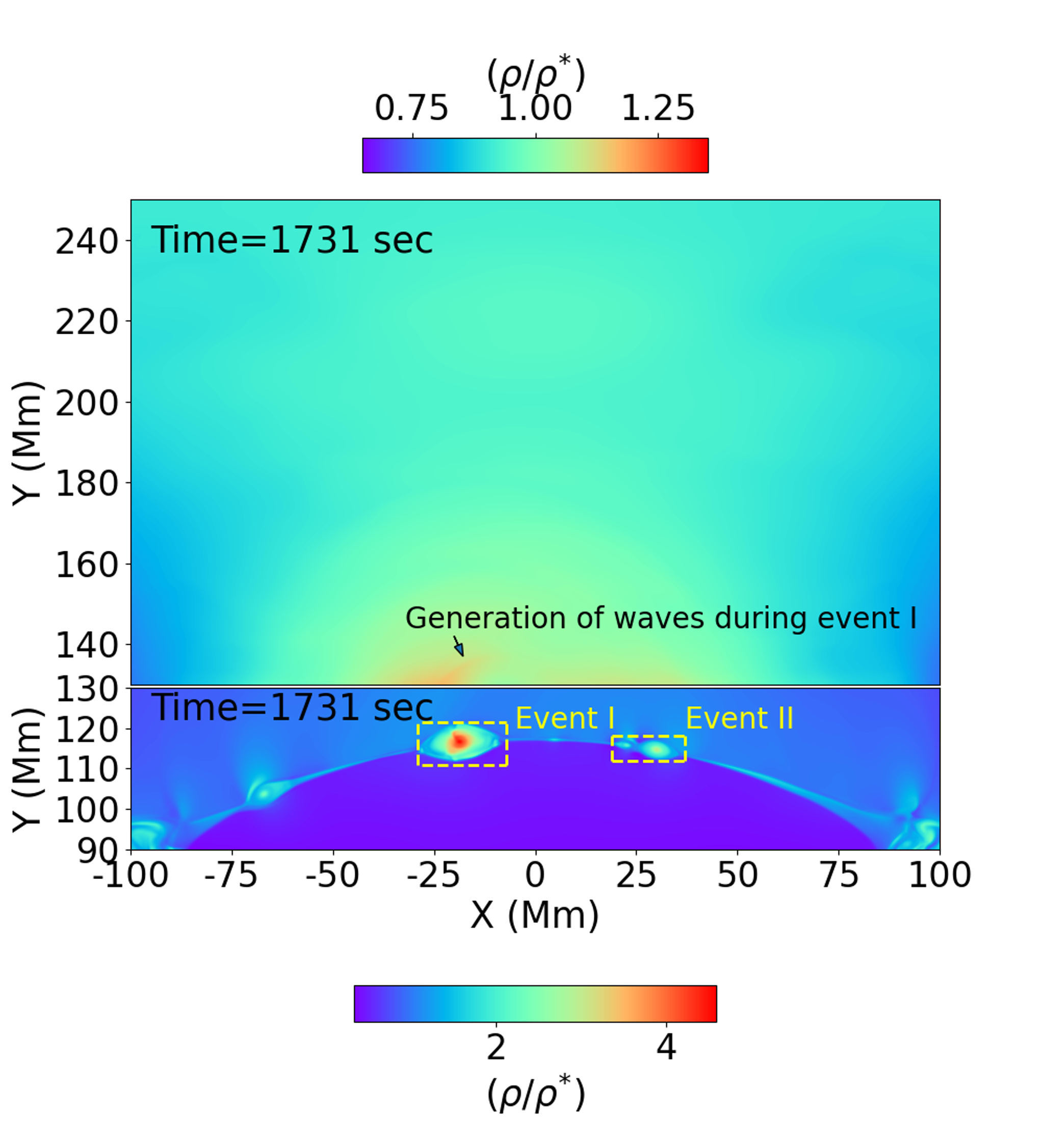}
\includegraphics[height=7.5 cm,trim={0 0 0 0},clip]{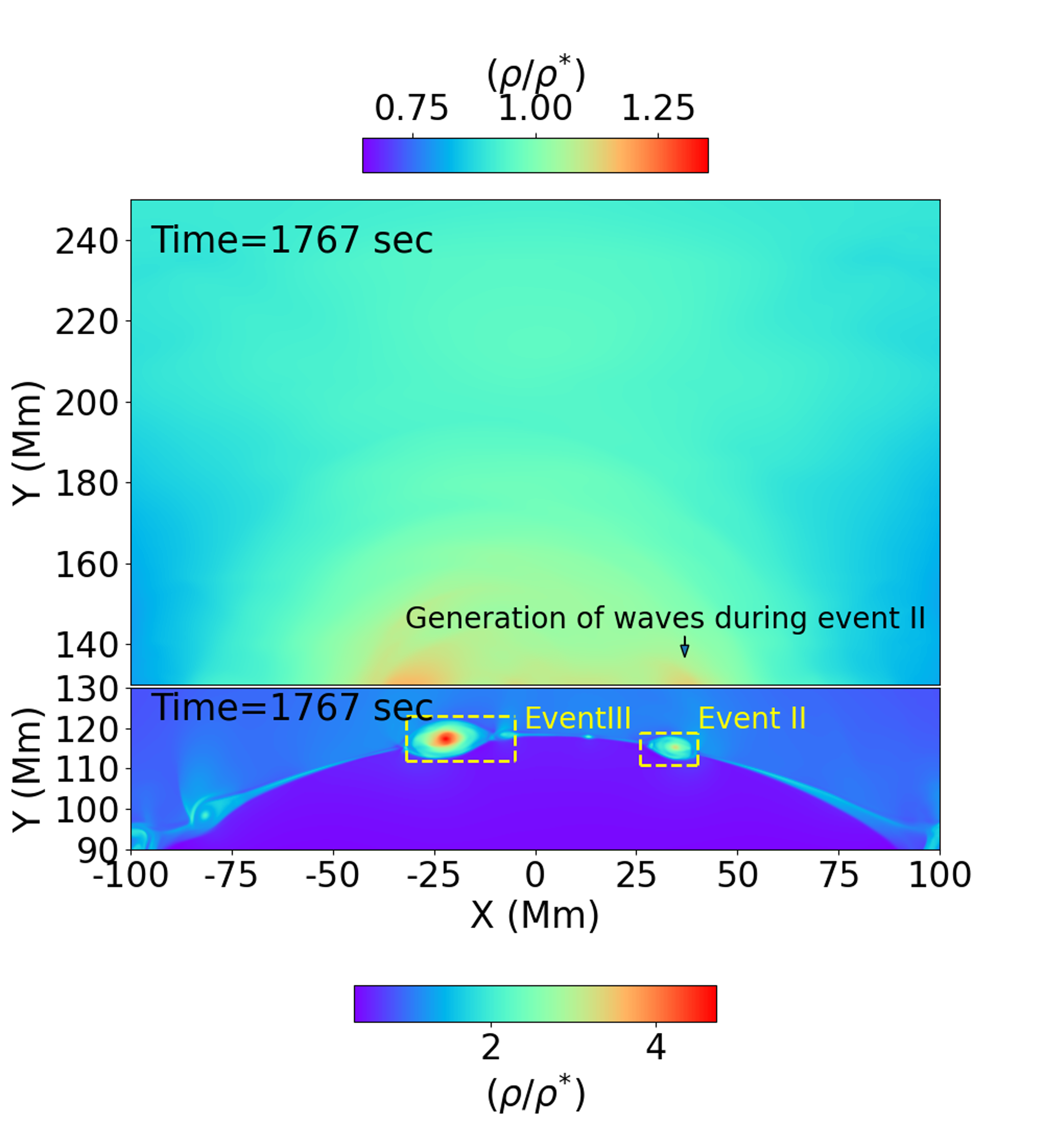}
\includegraphics[height=7.5 cm,trim={0 0 0 0},clip ]{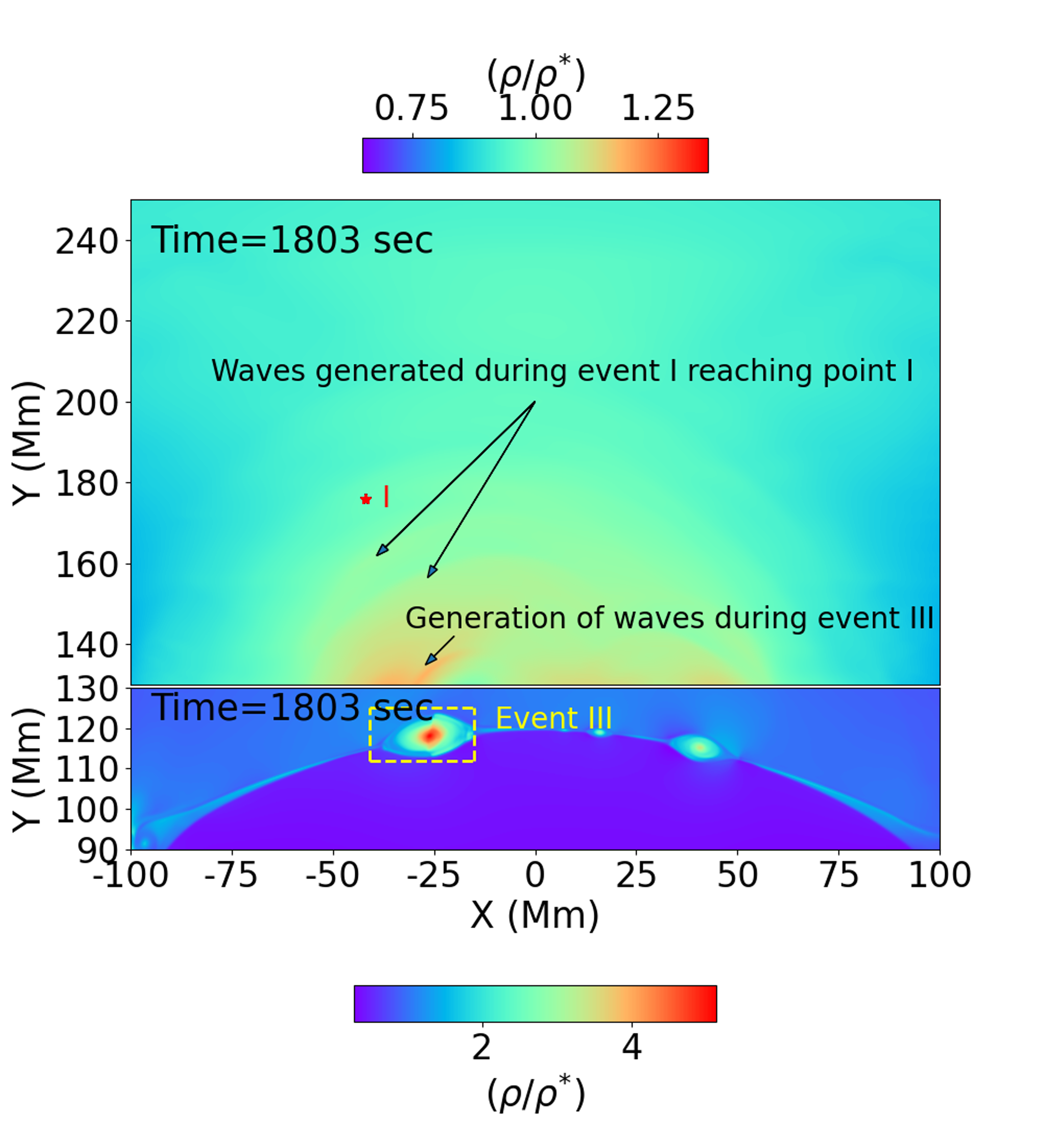}
}

\mbox{
\hspace{-0.9 cm}
\includegraphics[height=7.5 cm,trim={0 0 0 0},clip]{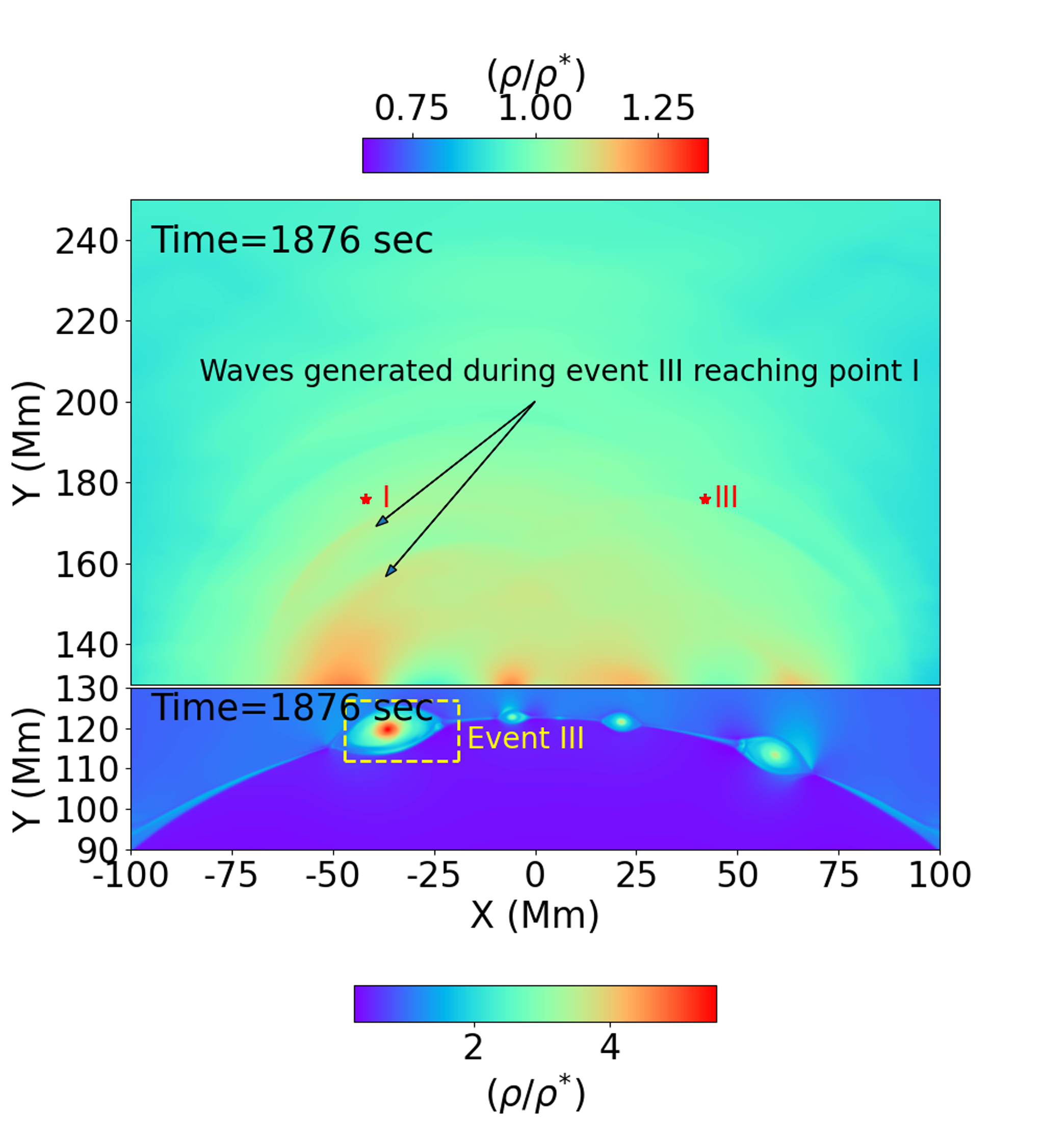}
\includegraphics[height=7.5 cm,trim={0 0 0 0},clip]{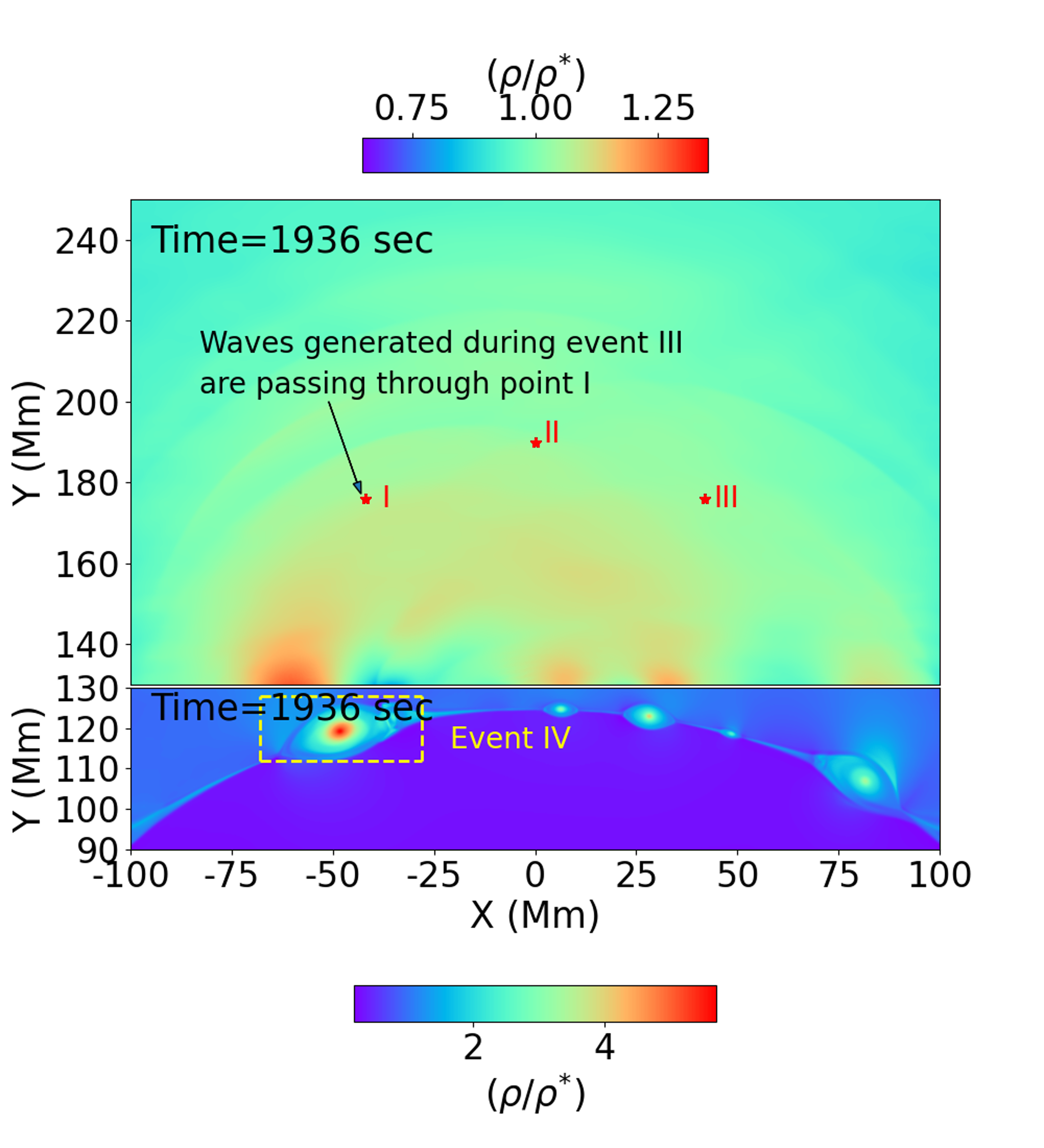}
\includegraphics[height=7.5 cm,trim={0 0 0 0},clip]{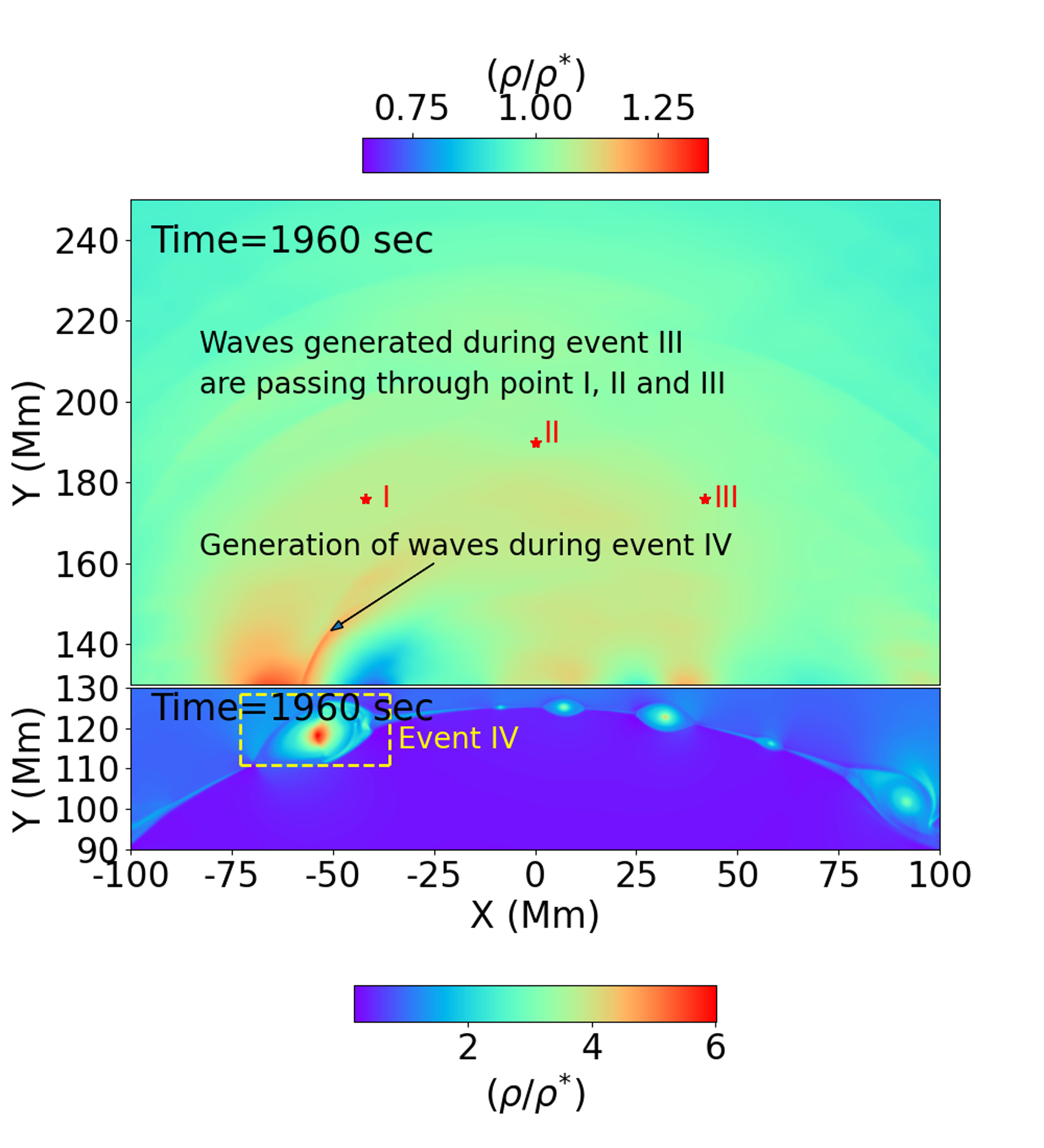}
}

\mbox{
\hspace{-0.9 cm}
\includegraphics[height=7.5 cm,trim={0 0 0 0},clip]{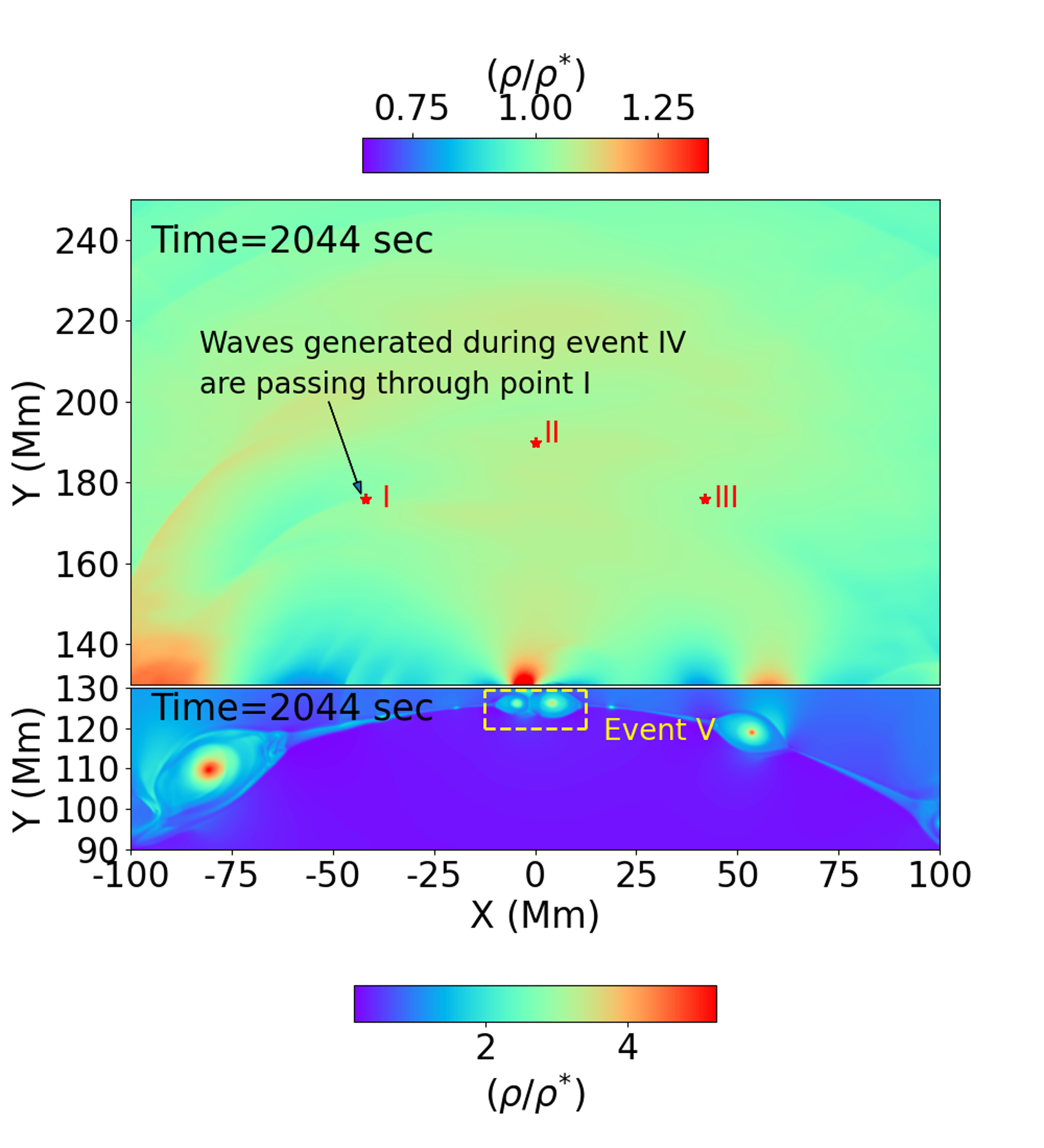}
\includegraphics[height=7.5 cm,trim={0 0 0 0},clip]{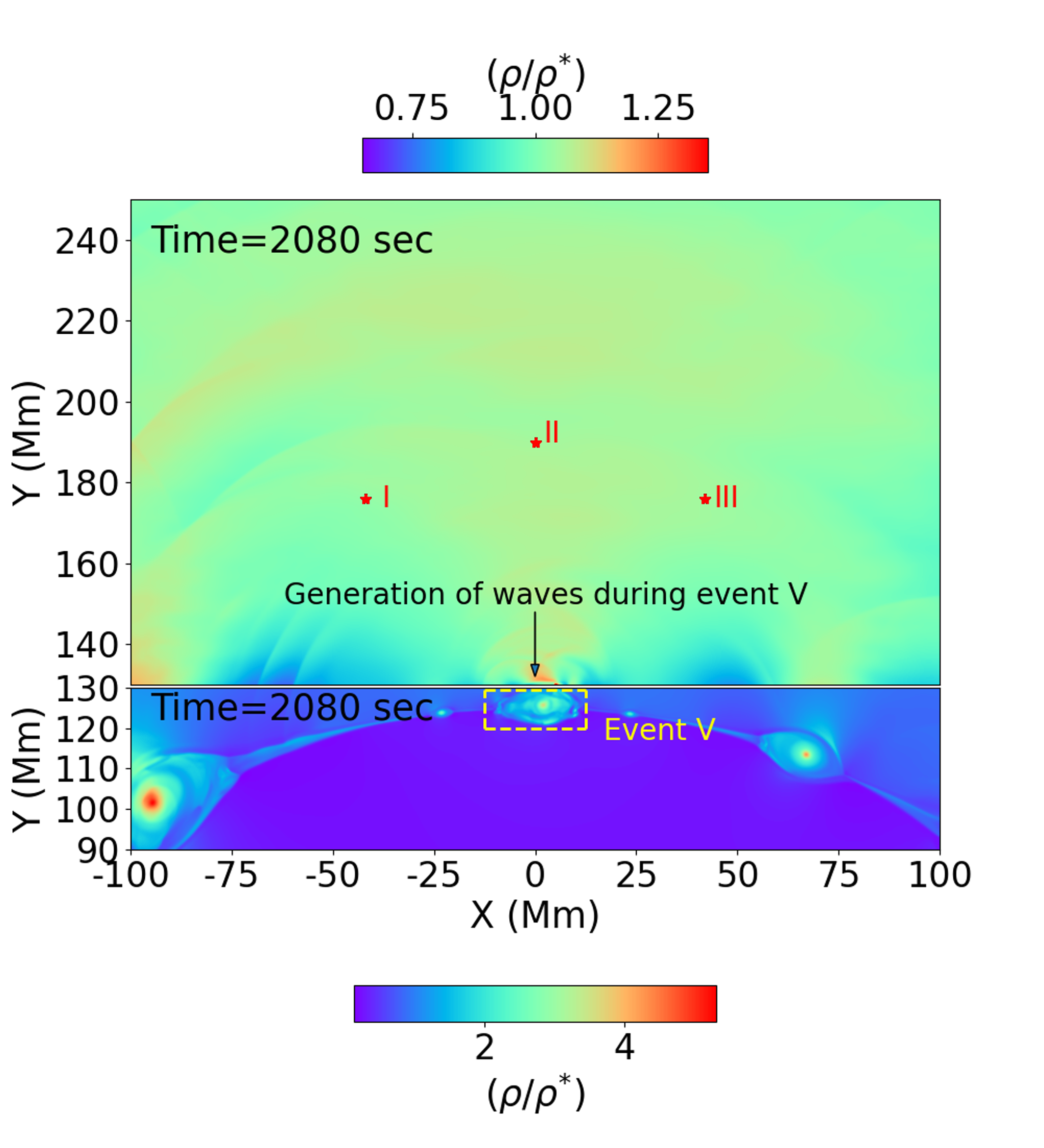}
\includegraphics[height=7.5 cm,trim={0 0 0 0},clip ]{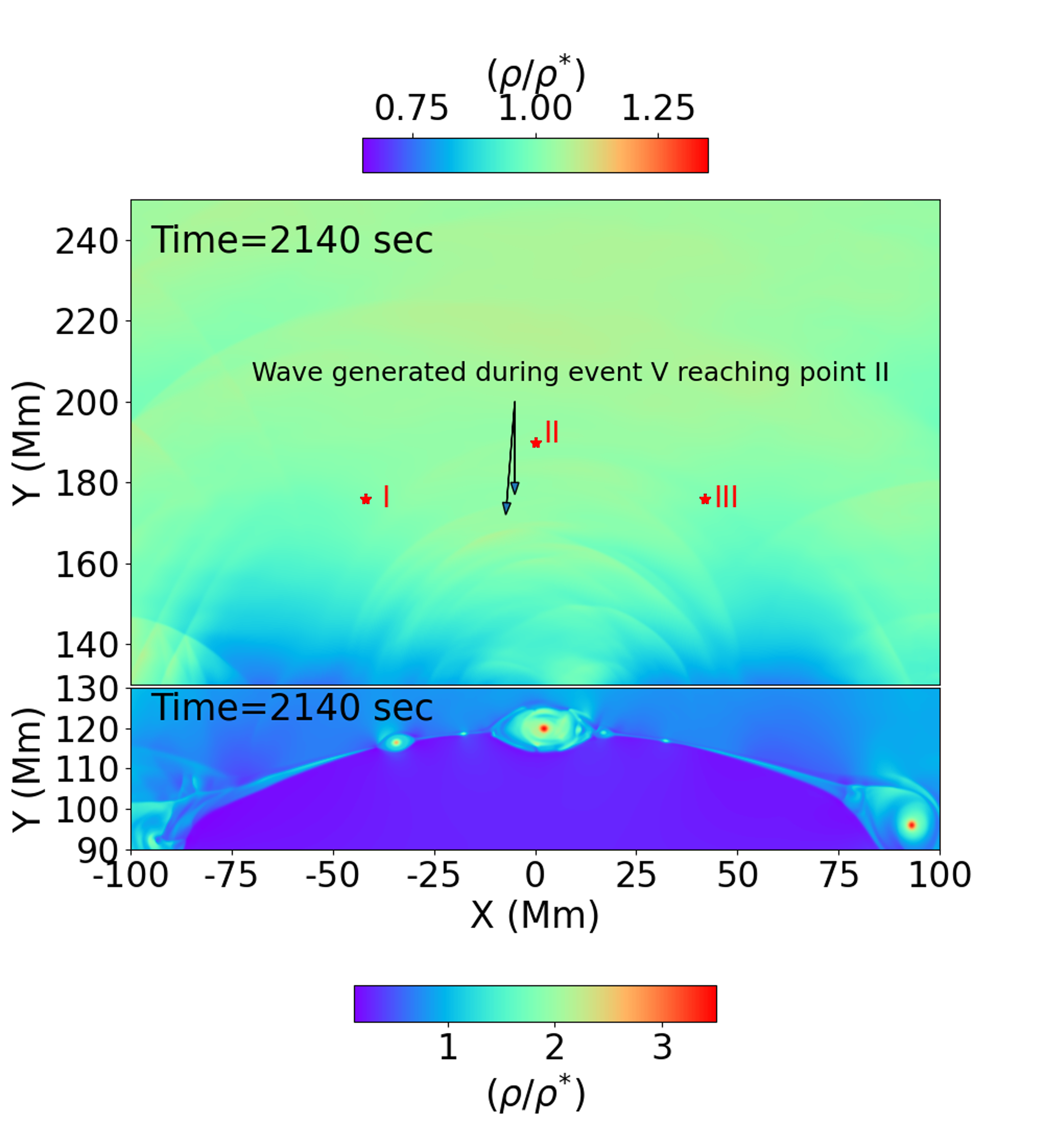}
}

\caption{Temporal evolution of coalescing plasmoids in the curved CS and the generation and propagation of fast-mode MHD waves from each coalescence event. The field of view  of the plasmoids in the CS and the large-scale diffuse corona are exhibited with different color scales in order to aid visualization of both the regions. This figure clearly demonstrates a causal connection between plasmoid coalescence  and the generation of fast-mode MHD waves. The labels I, II and III denote three locations for measuring properties of the wavefronts discussed in detail later. An animation covering the entire dynamics, including plasmoid coalescence and the generation and propagation of waves from 1082 s to the end of the simulation is available in the online version. The real-time duration of the animation is 9 s.}
\label{label 2}
\end{figure*}

\section{Model for Corona around a Dynamic Current Sheet} 
We assume homogeneous plasma properties  in the solar corona, namely, a uniform number density of $10^{9}~\mathrm{cm^{-3}}$ and temperature of 1 MK throughout the entire domain of $x~=~[-100, 100]$~Mm and $y$~=~[0, 250]~Mm being considered. We also assume a uniform  diffusivity ($\eta$) in Equations (6) and (7)) of $2.4\times 10^{8}~ \mathrm{m^{2}s^{-1}}$ and an initial magnetic field that is current-free and consists of a closed magnetic arcade that has emerged into an overlying horizontal magnetic field ($B_0\bar{\bf x}$), namely, \citep{2014masu.book.....P}:

\begin{equation}
B_{y}(x,y)+iB_{x}(x,y) = iB_{0}+iD/[(x-a_{1})+i(y-a_{2})]^{2}
\end{equation}
\begin{equation}
B_{z}(x,y)=0.
\end{equation}
The strength ($B_{0}$) of the overlying horizontal field  is taken to be 10 Gauss, while the values of other parameters used in the above expression are: $a_{1}= 0~\mathrm{Mm}$, $a_{2}= -60~\mathrm{Mm}$ and the strength of the magnetic dipole is $D= 2.6\times 10^{5}~\mathrm{Gauss~Mm^{2}}$. As a result, the magnetic field possesses a null point at $ x = 0~\mathrm{Mm}$, $y = 102~\mathrm{Mm}$. Thus, we have a  magnetohydrostatic equilibrium with a uniform plasma in a current-free magnetic field. A test run confirms that numerically this remains in equilibrium with no dynamic behaviour. 

Following the same procedure as \citet{2024ApJ...963..139M},  collapse of the null and subsequent reconnection is driven by imposing initially a localised Gaussian velocity pulse in the form:
\begin{equation}
   V_{x} = V_{0}~\mathrm{exp} \left(- \frac{(x-x_{0})^{2}}{w_{x}^{2}}-\frac{(y-y_{0})^{2}}{w_{y}^{2}}\right),
\end{equation}
 where $x_{0}= -40~\mathrm{Mm}$, $y_{0}= 100~\mathrm{Mm}$, $w_{x}= 10~\mathrm{Mm}$, $w_{y}= 2~\mathrm{Mm}$. $V_{0}$ is taken to be $850~\mathrm{km~s^{-1}}$. The leading edge undergoes steepening to form a fast mode shock when the perturbation is moving towards the magnetic null (See Figure~\ref{label 7} in Appendix and associated animation). Also, the amplitude of the velocity decreases during its passage in model corona. The amplitude of the shock wave-like perturbation during its interaction with the magnetic null is found to be approximately $60~\mathrm{km~s^{-1}}$ (See Figure~\ref{label 7} in Appendix and associated animation). As such, our choice of initial perturbation  mimics the interaction of a fast-mode shock wave with the magnetic null.
 
To simulate the plasma dynamics in the corona around a current sheet (CS) undergoing impulsive bursty reconnection, we solve the following non-ideal magnetohydrodynamic (MHD) equations in the presence of thermal conduction and viscosity \citep{2014masu.book.....P,2017ApJ...841..106Z,2019ApJ...870L..21G,2020ApJ...891...62L,2023A&A...678A.132S,2024ApJ...973...21L}:
 \begin{equation} 
\frac{\partial \rho}{\partial t} + \vec{\nabla} \cdot ( \rho \vec{V} ) = 0,
\end{equation}
\begin{equation}
  \frac{\partial}{\partial t}(\rho \vec{V}) + \vec{\nabla} \cdot \left [ \rho \vec{V}\vec{V}  + p_{tot}\vec{I} - \frac{\vec{B}\vec{B}}{4\pi} \right ] = \mu \vec{\nabla} \cdot [2S-\frac{2}{3}(\vec{\nabla} \cdot \vec{V})\vec{I}] ,
\end{equation}

\begin{equation}
\begin{split}
\frac{\partial e}{\partial t} +  \vec{\nabla} \cdot \left( e\vec{V} + p_{tot}\vec{V} -\frac{\vec{B}\vec{B}}{4\pi} \cdot \vec{V}\right)  = \eta \vec{J^{2}}-
   \vec{B} \cdot \vec{\nabla} \times (\eta \vec{J})\\
   +\vec{\nabla}_{\parallel} \cdot (\kappa_{\parallel} \vec{\nabla}_{\parallel} T)+\mu [2S^{2}-\frac{2}{3}(\vec{\nabla} \cdot \vec{V})^{2}],
\end{split}   
\end{equation}

\begin{equation}
  \frac{\partial \vec{B}}{\partial t} + \vec{\nabla} \cdot \left(\vec{V}\vec{B} - \vec{B}\vec{V}\right)+ \vec{\nabla} \times (\eta \vec{J}) = 0,
\end{equation}

\quad \textrm{where} \quad
\begin{equation}
p_{tot} = p + \frac{B^2}{8\pi}, ~~e = \frac{p}{\gamma-1} + \frac{1}{2}\rho V^{2} + \frac{B^2}{8\pi}
\end{equation}
\quad \textrm{and} \quad
\begin{equation}
  \vec{J} = \frac{\vec{\nabla} \times \vec{B}}{4\pi}, ~~\vec{\nabla} \cdot \vec{B} =0.
\end{equation} 
The thermal conduction acts only parallel to the magnetic field with $\kappa_{\parallel}=10^{-6}~T^{5/2}~\mathrm{erg~cm^{-1}~s^{-1}~K^{-1}}$ being the component of the thermal conduction tensor along the field. The dynamic viscosity coefficient ($\mu$) is taken to be $0.027~\mathrm{g~cm^{-1}~s^{-1}}$. The strain rate tensor ($S_{ij}$) is defined as $\frac{1}{2}(\partial V_{i}/\partial x_{j}+\partial V_{j}/\partial x_{i})$. Since the MHD equations are coupled partial differential equations, we numerically solve them using open source MPI-AMRVAC 3.0\footnote{https://amrvac.org/} \citep{2023A&A...673A..66K}, which simulates the evolution of all the dimensionless physical variables numerically. Therefore, all of the physical quantities in Eqs.~(1-9) are further subsequently normalized with respect to their typical values in the numerical code, namely, $L^{*}= 10^{9}~\mathrm{cm}$, $\rho^{*}= 2.34\times10^{-15}~\mathrm{g~cm^{-3}}$, $V^{*}= 1.16\times10^{7}~\mathrm{cm~s^{-1}}$, $T^{*}= 10^{6}~\mathrm{K}$, $P^{*}=0.3175~\mathrm{dyne~cm^{-2}}$, and $B^{*}= 2~\mathrm{Gauss}$.

MPI-AMRVAC possesses the facility to impose adaptive mesh refinement (AMR) on a uniform initial grid structure depending upon--[i] whether the difference in chosen physical variables such as density, magnetic field, velocity and pressure from one time step to the next is higher than a user-defined threshold or not in each grid and [ii] the user-defined spatial locations within the entire simulation domain. This facility ensures that the resolution becomes high at only the locations which are dynamic or in which we are interested. Since our simulation involves impulsive bursty magnetic reconnection due to coalescence of plasmoids in an elongated CS, we use five AMR levels depending on the above criterion [i], which results in an effective number of grid points of 4096 × 5120, with the smallest grid size being 48 km in both directions. Temporal integration is carried out via a $``$two-step$"$ method, and a $``$Harten-Lax-van Leer (HLL)$"$ Riemann solver \citep{1983JCoPh..49..357H} is used to estimate the flux at cell interfaces. Continuous boundary conditions are set at all the boundaries to ensure zero gradients for all the variables across them. A second-order symmetric total variation diminishing (TVD) limiter $``$vanleer$"$ \citep{1979JCoPh..32..101V} is employed to suppress spurious numerical oscillations while solving the MHD equations. Divergence cleaning of the magnetic field is carried out using Powell's eight-wave method \citep{1999JCoPh.154..284P}. Absence of any unusual forces, currents at the boundaries, and non-negative values of pressure, temperature, density throughout the simulation, suggest that the  dynamics are  physical and free from large numerical errors.

\section{Results}
In this section, we describe the three stages of the simulation and analyse in detail the waves that are generated by impulsive bursty reconnection. Such highly time-dependent reconnection occurs when a CS goes unstable to a secondary resistive instability such as the tearing mode. The term ``impulsive bursty" was coined by \citet{1986MitAG..65...41P} based on its discovery with relevance to the Sun by \citet{1982SoPh...81..303F,1983SoPh...84..169F,1987RvGeo..25.1583F} as well as \citet{1982PhLA...87..357B}. More recently, with the advent of more powerful codes there was a rediscovery of its importance by \citet{2007PhPl...14j0703L}, \citet{2009PhPl...16k2102B} and others in the laboratory and magnetospheric community when it was called ``plasmoid instability" instead.

\subsection{Formation and Reconnection in Coronal Current Sheet}
A coronal magnetic configuration having a magnetic null point near the center of the domain is perturbed by an initial Gaussian velocity pulse in the $x$-direction. This anisotropic perturbation possesses a width that is 5 times larger in the x-direction than  in the y-direction as described in Eq. 3. This mimics the effect of an EUV wave propagating from an eruption site elsewhere in the distant corona \citep{2024ApJ...963..139M}. It results in the collapse of the null to form an initially straight CS extending in the $x$-direction. The unbalanced forces stretch the CS, while the magnetic pressure in the initially closed loop causes a bending of the field lines to create a curvature in the CS. Later, the field lines of the closed loop and the oppositely-directed overlying field lines reconnect. This process can be viewed in the online animation associated with panel (a) of Figure~\ref{label 1}. The CS is recognizable as such from simulation time 660~s, after the waves generated by the initial velocity perturbation have escaped. Up to 1070~s, the CS undergoes gradual thinning and bending in the $y$-direction with a simultaneous stretching in the $x$-direction.

\subsection{Fragmentation of the Current Sheet and Coalescence of the Plasmoids}
Once the CS has thinned and stretched sufficiently, tearing mode instability  sets in \citep[e.g.,][]{1987RvGeo..25.1583F}, so that the CS fragments with the formation of successive magnetic X-points and O-points along it. We estimated the Lundquist number at 1070 s just before the visibility of plasmoids since it is an important parameter related to the onset of tearing mode. We find the system size, i.e., CS length and the average Alfv\'en speed inside the CS to be 51 Mm and $150~\mathrm{km~s^{-1}}$ respectively. The magnetic diffusivity is uniform and gives a Lundquist number of $3.2 \times 10^{4}$, which is higher than the lower threshold of $3 \times 10^{4}$ \citep{2009PhPl...16k2102B}. Moreover, the aspect ratio is estimated to be 113 at that instance, when fragmentation of the CS has  already started. The magnetic O-points evolve as dense plasma blobs or plasmoids due to the accumulation of plasma. They are basically regions with parallel currents which attract each other. But the plasma outflows and magnetic tension force associated with reconnection at the magnetic X-point between two O-points tend to oppose this process. Eventually, the attraction between the parallel current-carrying plasma blobs dominates and they coalesce to form a larger plasmoid \citep{1977PhFl...20...72F,1980PhRvL..44.1069B} (See Appendix A and Figure~\ref{label 8} for relevant discussions and a representative case). As this larger plasmoid moves outward along the CS, the part of the CS behind it becomes thinner and results in further fragmentation and plasmoid formation. Such a scenario is repeated multiple times.

In the simulation, plasmoid formation commences at around 1082~s. The formation and subsequent interaction of the plasmoids is illustrated in Figure~\ref{label 1}(b). Overall, the behaviour is characterised by formation and subsequent rapid coalescense of multiple plasmoids, as follows. At around 1082 s, two small plasmoids form and move away from each other. During its outward movement, the rightward moving plasmoid grows a little in its dimensions. Around 1298 s, a single plasmoid forms in the central area of the CS which exhibits slow growth with time. Around 1407 s, a small plasmoid coalesces with the plasmoid formed at 1298 s from right side. Around 1551 s, another small plasmoid coalesces with that plasmoid from left side (See left panel of Figure~\ref{label 1}(b)). At 1623 s, another small plasmoid merges with the bigger plasmoid. At 1695 s, this plasmoid undergoes another coalescence from the right side. These subsequent coalescences  grow the plasmoid which also moves left along the CS. This plasmoid eventually undergoes two more coalescences during its visibility in the field of view (FOV) at 1767 and 1924 s. At around 1731 s, two different plasmoids coalesce with each other to grow in size at around $x=25$~Mm (See right panel of Figure~\ref{label 1}(b)), the resultant plasmoid  eventually moves outward in the right direction without being subjected to any further coalescence. At 2044 s, two plasmoids merge with each other at the central region of the curved CS. The merged plasmoid further undergoes subsequent coalescences  during 2128-2176 and 2188-2236 s. The entire dynamics of multiple plasmoid behaviour, including their movement and growth following  coalescence are exhibited in FOV of y =[90 Mm, 130 Mm] in a few selective snapshots in Figure~\ref{label 2} (See the corresponding animation in the online version for more details). 

\begin{figure*}
\mbox{
\hspace{4.5 cm}
\includegraphics[scale=0.12]{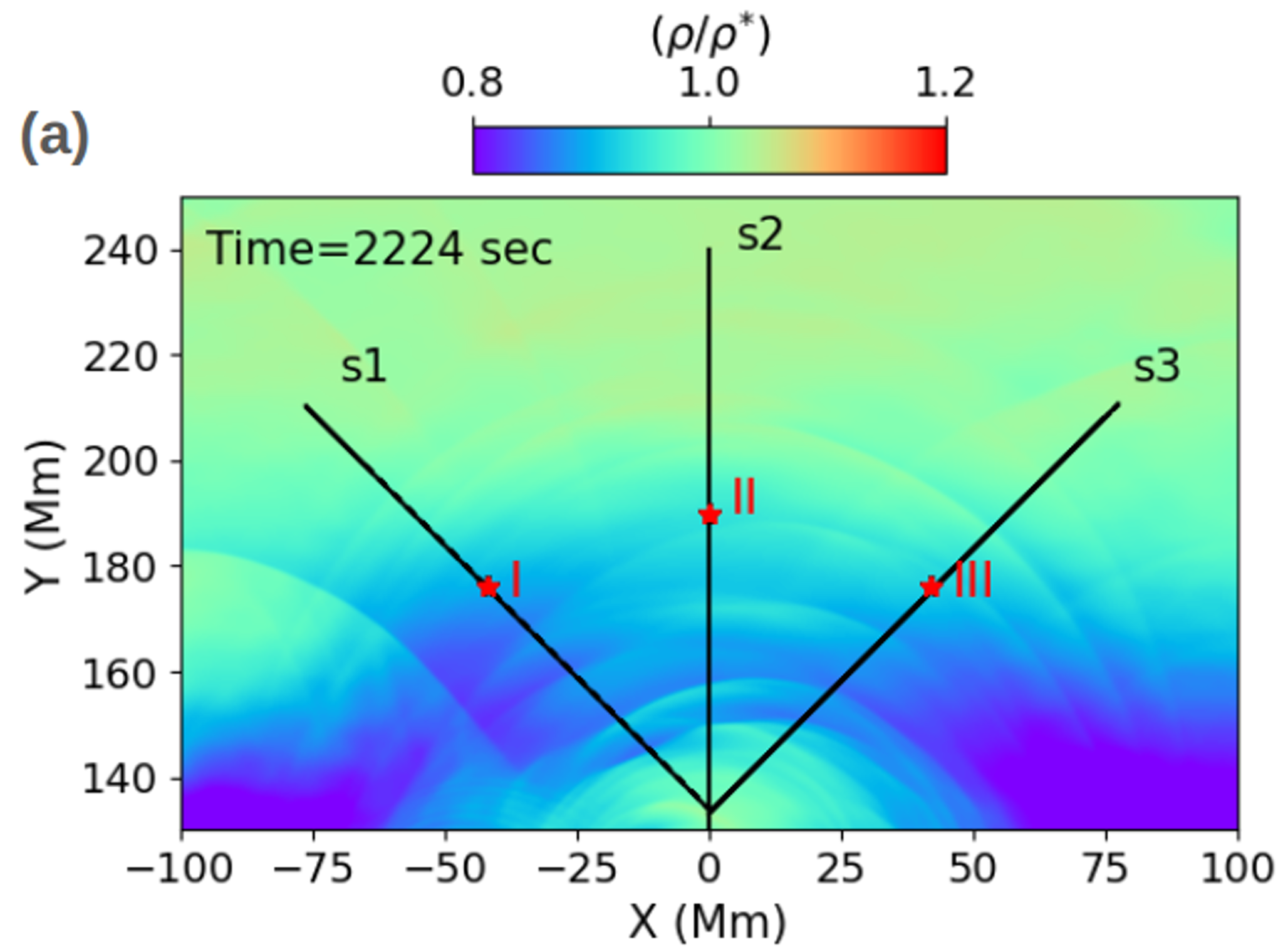}
}
\mbox{
\hspace{-2 cm}
\includegraphics[width=1.2\textwidth]{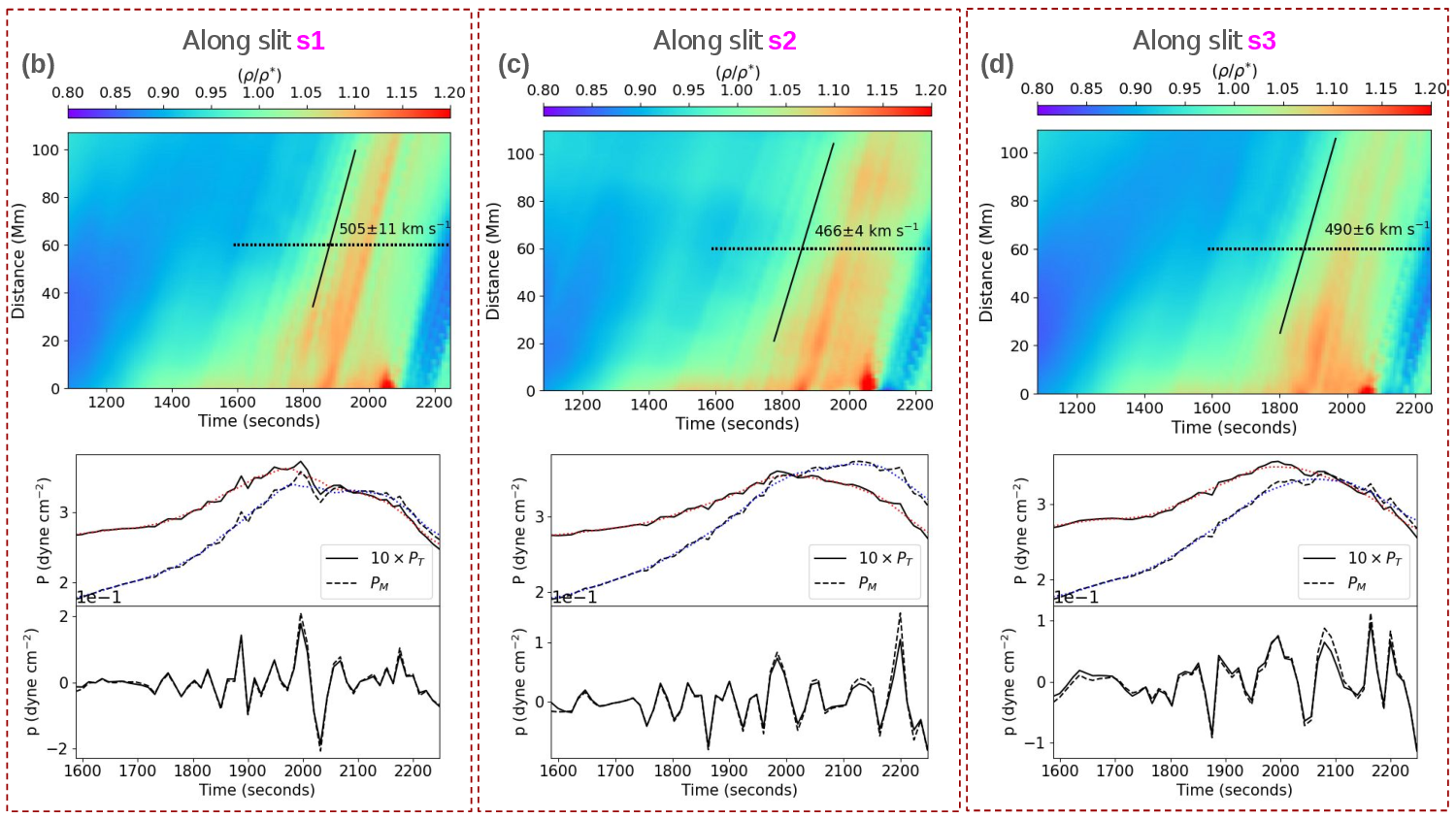}
}
\caption{Three slits `s1', `s2' and `s3' are taken in three different directions to capture the dynamics of fronts. The orientations of the slits are shown in panel (a). Distance-time diagrams in density show  slanted features which are signatures of outwardly propagating waves (see top panels of panel (b), (c) and (d)). The estimated propagation speeds are consistent with the characteristic fast mode speed shown in Figure~\ref{label 4}. The perturbations in thermal pressure ($P_{T}$) and magnetic pressure ($P_{M}$) exhibited in the bottom panels of panel (b), (c) and (d) are in phase with one another, which also confirms the fast-mode nature of the waves. The profiles are measured at a distance 60 Mm along slits `s1', 's2' and 's3' (corresponding to locations I, II and III) starting from 1587 s, as shown by horizontal dotted lines. The small-scale fluctuations in thermal and magnetic pressure are extracted by subtracting the long-term background trends shown as red and blue dotted curves.}
\label{label 3}
\end{figure*}

\subsection{Generation and Propagation of Fast Magnetoacoustic Waves in the Diffused Model Corona}
Arc-shaped, propagating velocity disturbances are observed both above and below the curved CS once plasmoid formation commences. These waves are mainly associated with rapid coalescences, but could also be produced by plasmoid motion and their merger with magnetic Y points at the ends of the extended CS (See top left panel of Figure~\ref{label 9} in Appendix and associated animation). The propagating disturbances are most clearly identifiable above the current sheet, therefore, we focus on analyzing this particular region-of-interest (ROI). The reason for this upward and downward differences in the visibility of the waves is discussed in detail in Appendix B. In most instances, the disturbances generated by the merging of multiple plasmoids  possess higher velocity amplitudes than other causes (i.e., moving plasmoids without merging).  Figure~\ref{label 2} and its associated animation clearly exhibits a cause-effect relationship between the merging of plasmoids in the CS and the generation of outward-propagating wave-fronts in the corona.

It is important to point out that no residuals of the initial velocity perturbations are present within the simulation domain in the time window for studying the generation and propagation of these wavefronts in the large-scale corona. We shall discuss  measurements of periodicities and wave energy fluxes at three locations I, II and III (See Figure~\ref{label 3}(a)) in subsections 3.3.3 and 3.3.4, respectively, but first we determine a rough spatio-temporal correlation between multiple coalescence events and the peaks in wave energy fluxes  which will further confirm the link between propagating waves and plasmoid coalescence. The peaks in wave energy fluxes can be largely attributed to coalescence events. However, those peaks will also have  contributions from the waves generated during the movement of plasmoids. 

\subsubsection{Possible Temporal Correlation between Plasmoid Coalescences and Wave Propagation}
To determine the relation during five coalescence events between the merging of plasmoids and the resulting fluctuations in physical variables such as density, velocity and magnetic field, we describe those events as follows.
\newline

[I] EVENT I (1695-1755 s): During this time, the coalesced plasmoid (shown within the left yellow-dashed box in the top-left panel of Figure~\ref{label 2}) moves from [$x,y$]=[$-12$ Mm, 118 Mm] to [$x,y$]=[$-22$ Mm, 122 Mm] (see animation related to Figure~\ref{label 2}). So, the distance to closest location of measurement, i.e., location I will decrease from 63 Mm to 57 Mm. The distances to location II and III will also vary with time. Now, for a rough average propagation speed of  $500 ~\mathrm{km~s^{-1}}$ (see subsection 3.3.2), the generated waves during this event will reach the location I in a time window of 1819-1869~s as denoted via pale pink shaded region bounded via red dashed lines in Figure~\ref{label 6}(a). During this time, there are two peaks evident in wave energy flux at 1827 and 1851~s at location I. Similarly, pale pink shaded regions in Figure~\ref{label 6}(c), (e) correspond to rough arrival times of waves generated during this event to location II and III namely, 1839-1898~s and 1853-1922~s, respectively (if they will arrive there). These time windows are calculated using the distances of location II and III from the instantaneous positions of the coalesced plasmoid associated with this event. Also, for simplicity, it has been assumed that the wavefronts will propagate with the same speed in all directions. Nevertheless, since we considered the time window of the coalescence event from the start of the coalescence to just before the next coalescence, this assumption will give us the possible arrival times of waves. Careful observation suggests that the peak at around 1863 s at location II is caused by the wavefront associated with this event (See Figure~\ref{label 6}(c)).
\newline

[II] EVENT II (1731-1779 s): During this time, another two small plasmoids (shown within the right yellow-dashed box in the top-left and middle panels of Figure~\ref{label 2}) merge,  while the combined structure moves from [$x,y$]=[32 Mm, 115 Mm] to [$x,y$]=[37 Mm, 118 Mm]. So the distance to location III decreases from 61 Mm to 57 Mm. Similarly, the distances to locations I and II also change with time. So, as before, we find the arrival time window for the waves generated during this event at location III to be 1855-1895~s, as depicted by a pink shaded region bounded by pink dashed lines in Figure~\ref{label 6}(e). During this window, a peak is found around 1876 s in wave energy flux at location III. The wavefront generated during event I also reaches location III around the same time. (See the overlap of different shaded regions). So, it may contribute to the wave energy flux at that point at that time. Two small peaks at location II and one peak at location I fall within the estimated arrival time window for the waves generated by this event at those locations as shown by pink shaded regions in Figure~\ref{label 6}(a), (c). However, since they  overlap with the lime-green shaded regions  associated with event III mentioned below, we cannot exactly determine the relative contributions of the multiple events responsible for those peaks.
\newline

[III] EVENT III (1767-1912 s): In this period, the plasmoid (shown within the left yellow dashed box in snapshots at 1767, 1803 and 1876 s in Figure~\ref{label 2}) that is undergoing coalescence  moves from [$x,y$]=[-22 Mm, 122 Mm] to [$x,y$]=[-37 Mm, 128 Mm], which means the distance to location I changes from 57 Mm to 47 Mm. Similarly, the distances to other two locations are also calculated at start and end of this coalescence event. The estimated time-windows for arrival of waves generated during this event at locations I, II and III are 1882-2008~s, 1910-2056~s and 1934-2097~s, respectively, if they  propagate with the same speed in all directions (See limegreen shaded regions of Figure~\ref{label 6}(a), (c), (e)). During the respective time windows, as discussed above, multiple peaks have been observed at location I (1888~s, 1924~s, 1948~s and 1996~s), location II (1924~s, 1984~s, 2020~s and 2056~s) and location III (1948~s, 1996~s, 2044~s and 2080~s) in wave energy density (See limegreen shaded regions in panel (a), (c) and (e)  of Figure~\ref{label 6}). 
\newline

[IV] EVENT IV (1924-1996 s): The leftward moving plasmoid undergoes another coalescence in this time period (as shown within the yellow box in snapshots at 1936 and 1960 s in Figure~\ref{label 2}) during its movement from [$x,y$]=[-42 Mm, 128 Mm] to [$x,y$]=[-75 Mm, 120 Mm]. As a result, the distance to the location I changes from 47 Mm to 54 Mm. Like the previous cases, the distances to locations II and III will also vary due to the leftward movement of the coalesced plasmoid. Considering the varying distances, waves generated during this coalescence are estimated to reach  locations I, II and III between 2020-2126~s, 2074-2200~s and 2118-2248~s, respectively, as denoted by pale purple-shaded regions in Figure~\ref{label 6}(a), (c) and (e). The peaks at 2032~s, 2068~s and 2092~s at location I, and at 2128 s at location II, lie within the respective estimated wave arrival windows (See Figure~\ref{label 6}(a), (c)). 
\newline

[V] EVENT V (2044-2092 s): During this period, two plasmoids coalesce with each other around [$x,y$]=[0 Mm, 130 Mm] (as depicted within the rectangular box at 2044 s and 2080 s in Figure~\ref{label 2}) and getting internally restructured there without any significant movement left or right. So, if the waves  propagate isotropically as  circular arcs with the same propagation speeds in all directions, they will reach locations I, II and III within 2168-2216~s, 2164-2212~s and 2168-2216~s, respectively, as depicted via cyan-shaded regions in Figure~\ref{label 6}(a), (c) and (e). We find that there are peaks at 2176~s and 2200~s at location I, at 2164~s and 2188~s at locations II and III. Since, the peak at 2164~s is outside the estimated time window for location III, we carefully trace the wavefronts in the density map and find that a wavefront is indeed reaching location III around 2164~s. The discrepancy in the time-window is likely to be caused by an underestimation of approximate propagation speed. Also, the presence of dominant peaks at different times at different locations might be due to anisotropic propagation of different wavefronts in the time-evolving background.

Thus, we conjecture that each of these coalescence events  generate wavefronts which perturb the physical conditions at distances far away in the solar corona from those sources by delivering wave energy there. However, there are possibilities that waves generated due to motion of the individual plasmoids may also reach the detection points I, II and III. Therefore, the estimated wave energy fluxes at those points may not be intrinsically associated with waves generated due to coalescence of plasmoids only, but can additionally possess some contributions from the wave-like perturbations generated from individually moving plasmoids. However, since the amplitudes of waves generated due to coalescence are much higher and dominant at most instances than the contribution of the waves generated via other sources, the temporal variation of wave energy fluxes at three different locations are used to govern spatio-temporal correlation between coalescence events and resultant perturbations in physical variables there.

\begin{figure*}
\centerline{\hspace*{0.013\textwidth}
         \includegraphics[height=7.2 cm,trim={0 0 0 0},clip]{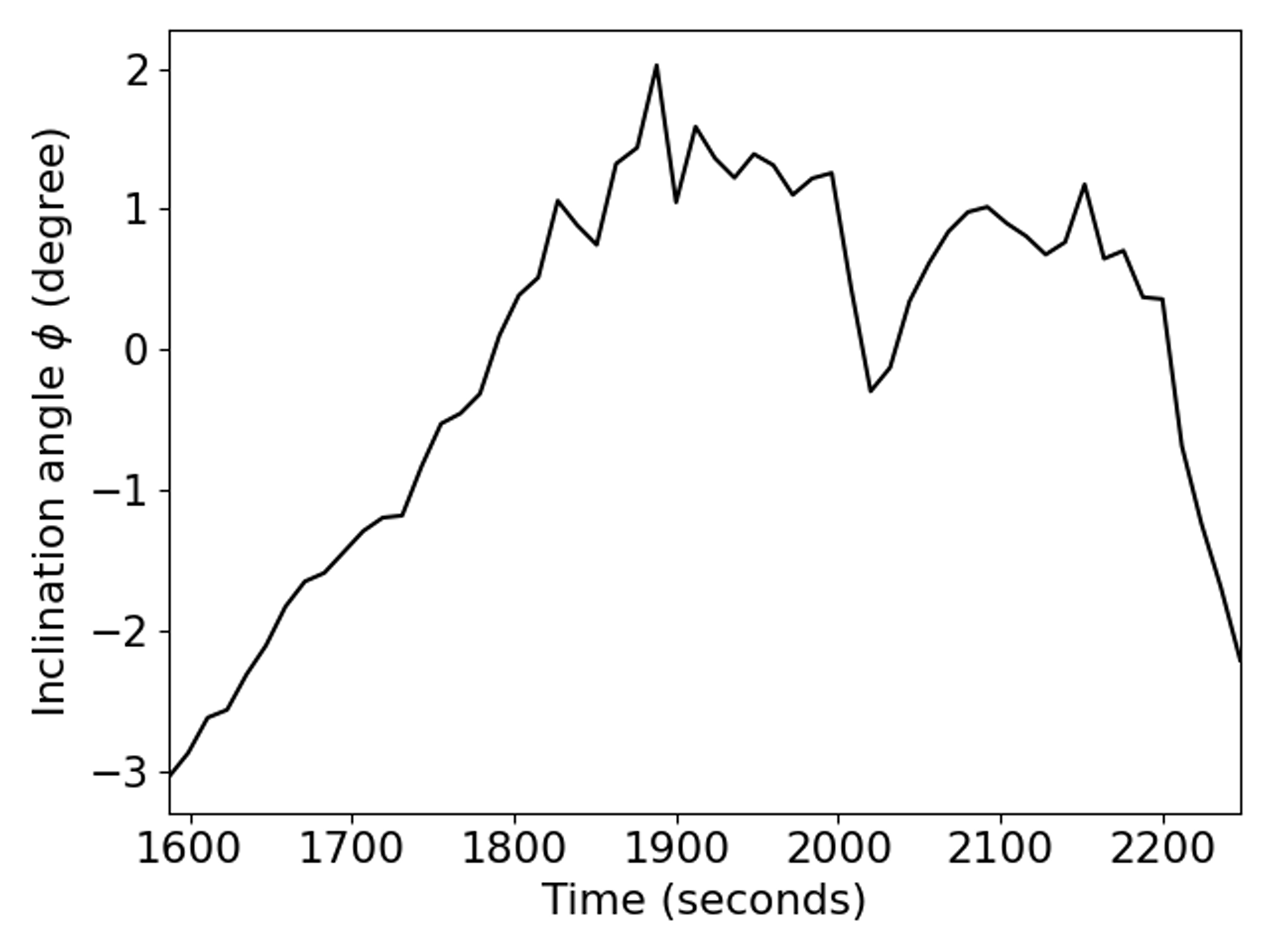}
         \hspace*{-0.001\textwidth}
         \includegraphics[height=7.2 cm,trim={0 0 0 0},clip]{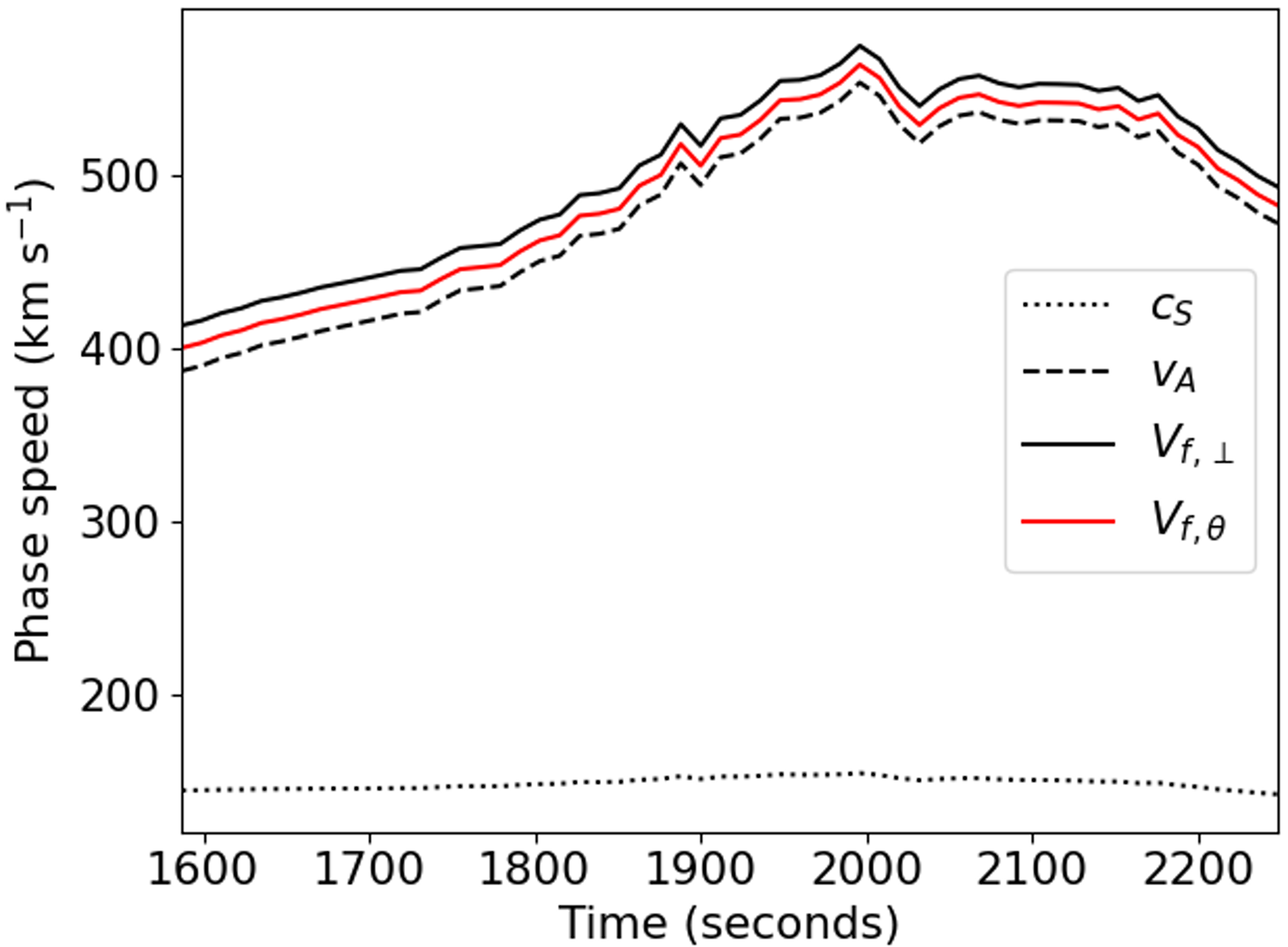}
       }
\vspace{-0.43\textwidth}
\centerline{ \large \bf      
\hspace{-0.03 \textwidth}  \color{black}{(a)}
\hspace{0.17 \textwidth}  \color{red}{Location I}
\hspace{0.16\textwidth}  \color{black}{(b)}
\hspace{0.18 \textwidth}  \color{red}{Location I}
   \hfill}
\vspace{0.43\textwidth}    
     
\centerline{\hspace*{0.013\textwidth}
         \includegraphics[height=7.2 cm,trim={0 0 0 0},clip]{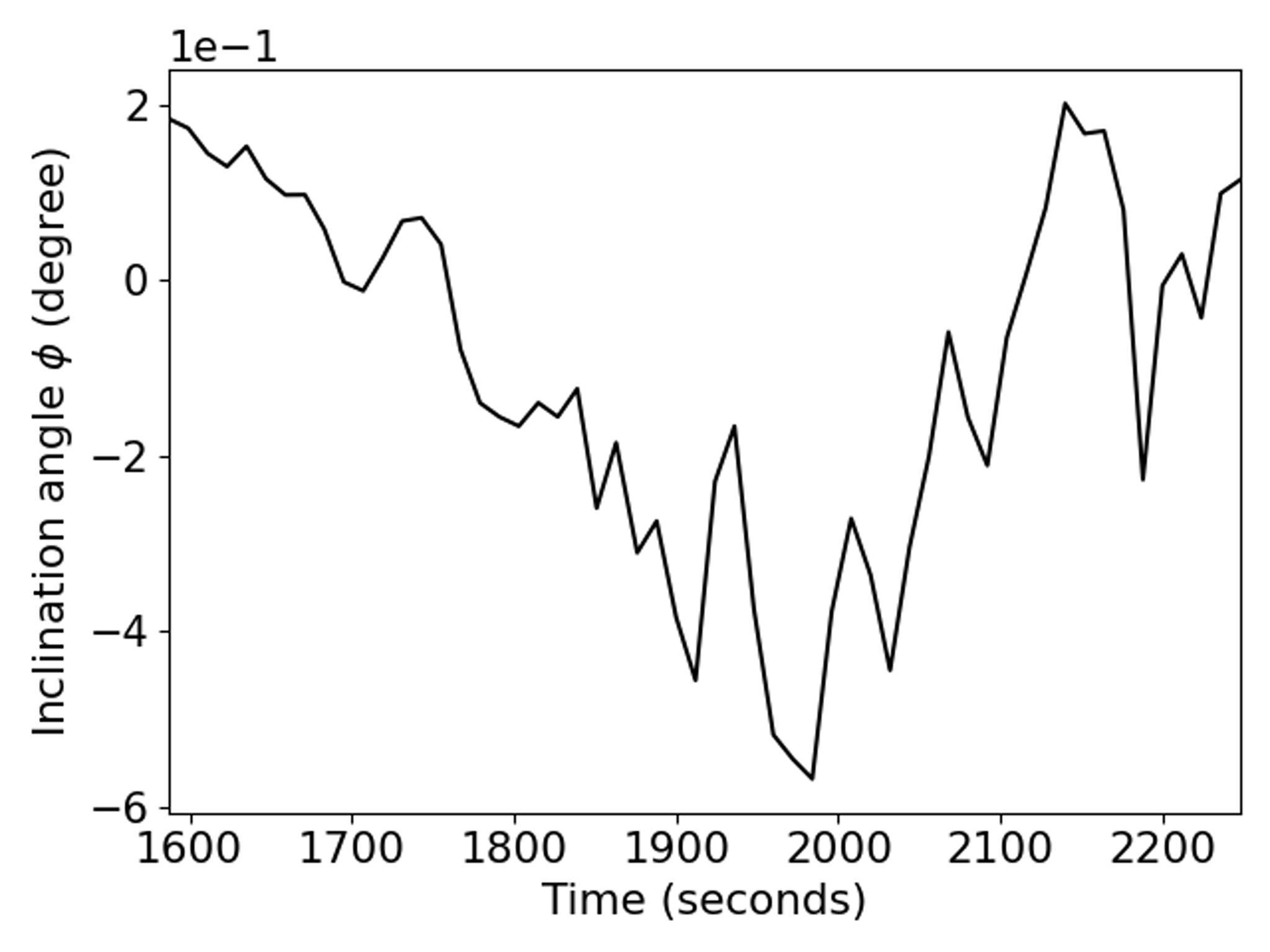}
         \hspace*{-0.001\textwidth}
         \includegraphics[height=7.2 cm,trim={0 0 0 0},clip]{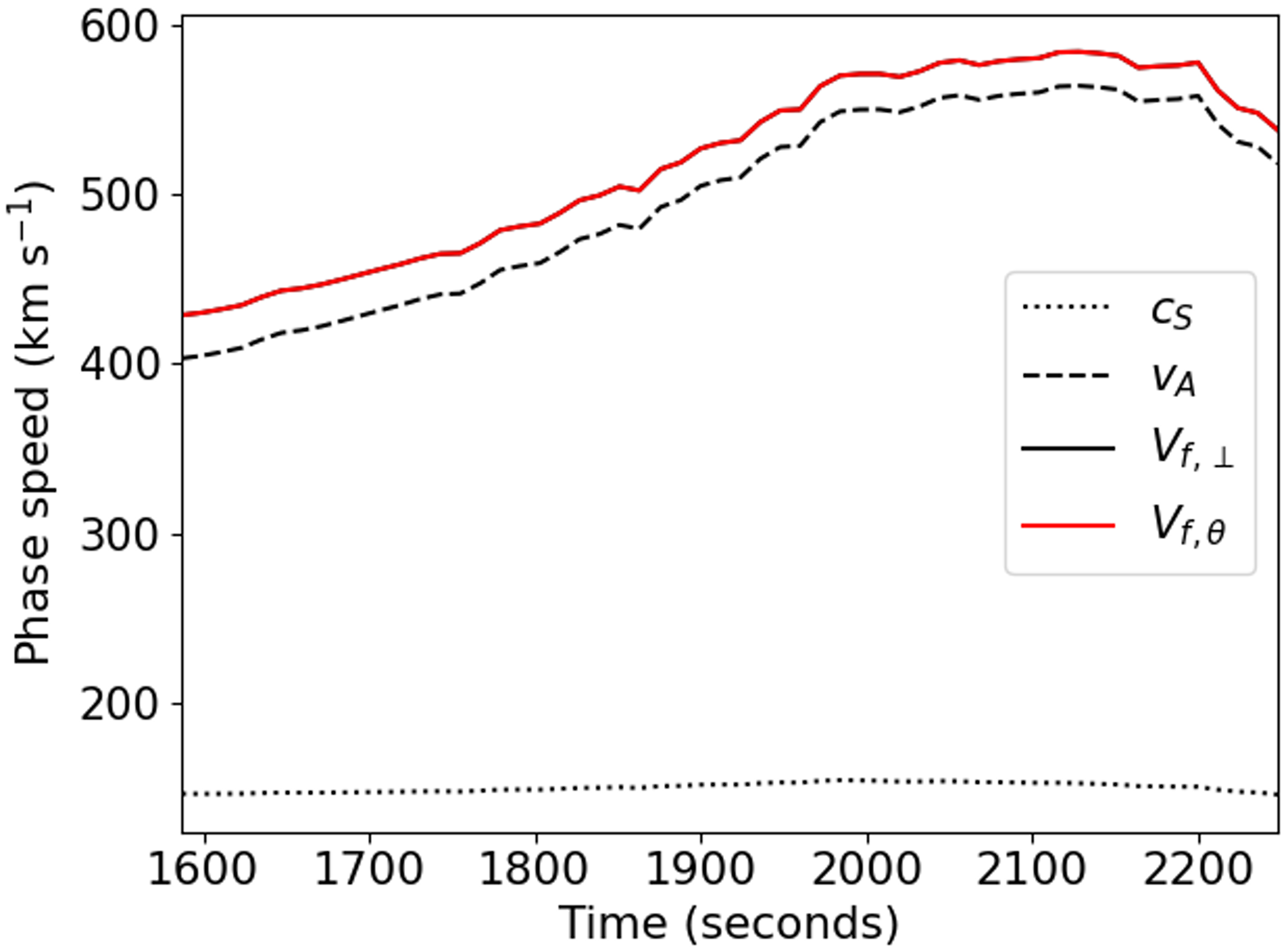}
        }
\vspace{-0.42\textwidth}
\centerline{ \large \bf      
\hspace{-0.03 \textwidth}  \color{black}{(c)}
\hspace{0.17 \textwidth}  \color{red}{Location II}
\hspace{0.15\textwidth}  \color{black}{(d)}
\hspace{0.18 \textwidth}  \color{red}{Location II}

   \hfill}
\vspace{0.42\textwidth}

\centerline{\hspace*{0.013\textwidth}
         \includegraphics[height=7.2 cm,trim={0 0 0 0},clip]{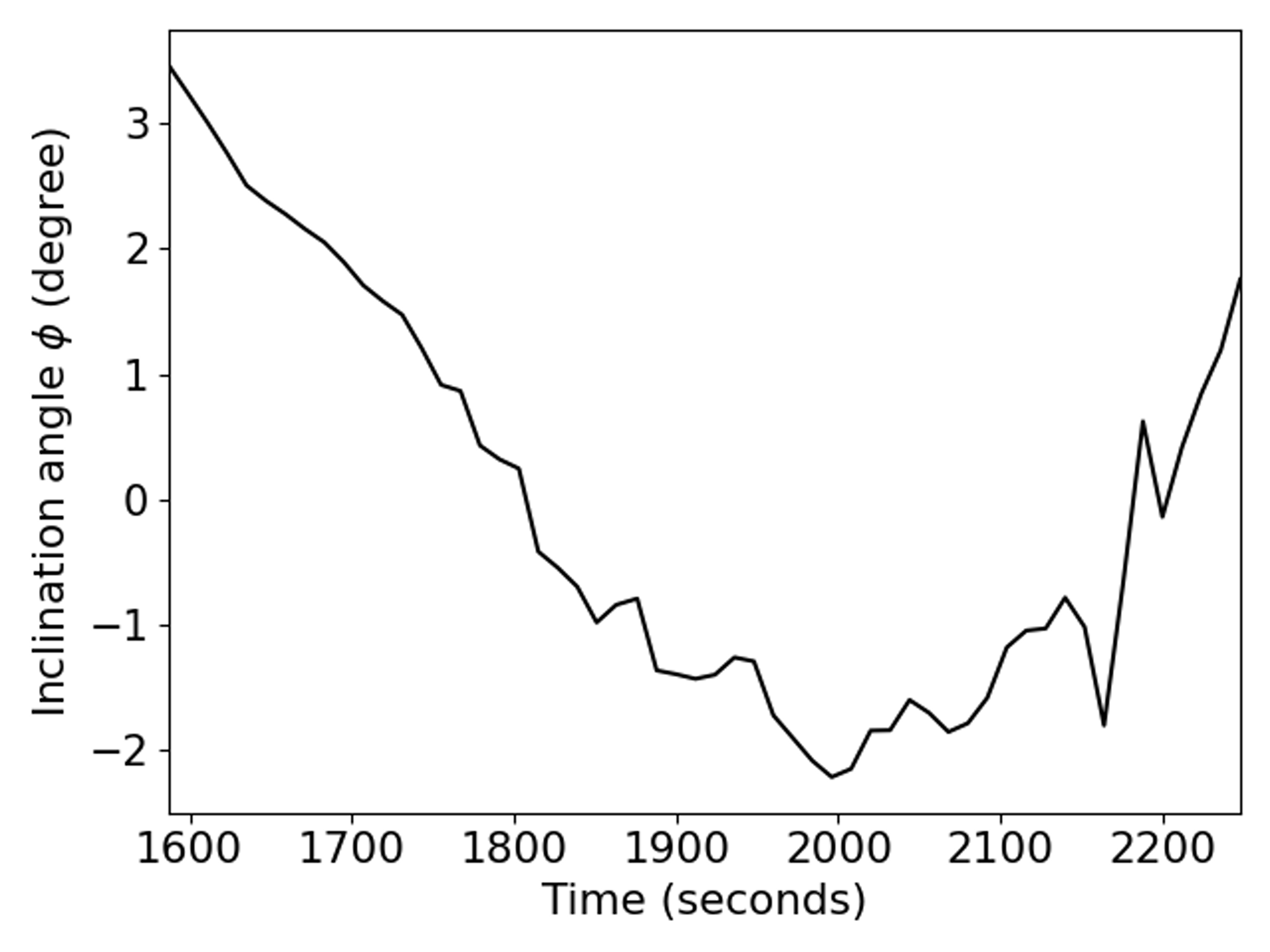}
         \hspace*{-0.001\textwidth}
         \includegraphics[height=7.2 cm,trim={0 0 0 0},clip]{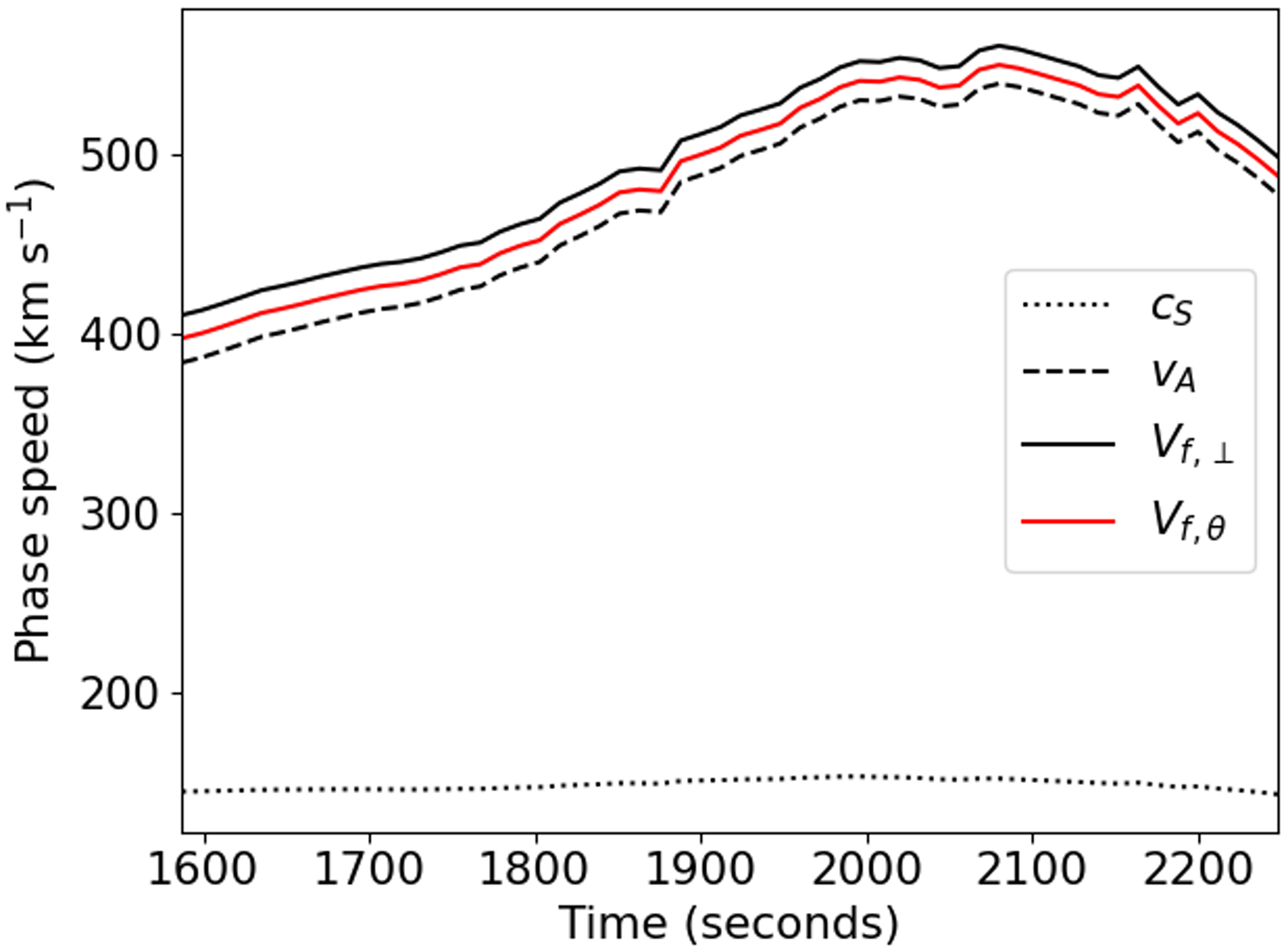}
        }
\vspace{-0.42\textwidth}
\centerline{ \large \bf      
\hspace{-0.03 \textwidth}  \color{black}{(e)}
\hspace{0.17 \textwidth}  \color{red}{Location III}
\hspace{0.16\textwidth}  \color{black}{(f)}
\hspace{0.17 \textwidth}  \color{red}{Location III}
   \hfill}
\vspace{0.39\textwidth}

\caption{Panels (a), (c) and (e) exhibit the temporal variation of the inclination of the magnetic field to the positive $x$-direction at locations I, II and III, respectively. Panels (b), (d) and (f) plot the local sound speed ($c_{S}$) (dotted curves), Alfv\'en speed ($v_{A}$) (dashed curves), the  fast-mode speed  ($V_{f,\perp}$) perpendicular to the magnetic field (solid black curves) and the  fast-mode speed  ($V_{f, \theta}$) along the slits at the locations  I, II and III. At location II on slit s2, the direction of  propagation is almost perpendicular to the  magnetic field, and so the  black solid curves is almost the same as the red solid curve in panel (d) ($V_{f, \theta} \approx V_{f,\perp}$).}
\label{label 4}
\end{figure*} 

\subsubsection{Dynamical and Fundamental Properties of the Propagating Wavefronts}
We determine the dynamic characteristics of the propagating wavefronts in three different directions along the  slits `s1', `s2' and `s3'  shown in Figure~\ref{label 3}(a). Top sub-panels of Figure~\ref{label 3}(b), (c), (d) present the distance-time diagrams in density along slit `s1', `s2' and `s3' respectively which clearly show the presence of successive slanted ridges, which are a manifestation of outwardly propagating perturbations due to waves along those directions. The estimated propagation speeds along `s1', `s2' and `s3' are $505 \pm 11~\mathrm{km~s^{-1}}$, $466 \pm 4~\mathrm{km~s^{-1}}$ and $490\pm 6~\mathrm{km~s^{-1}}$, respectively. Here, the uncertainties are basically one standard deviation of the measured slope via straight line fitting.  We calculate the thermal and magnetic pressure as $P_{T} = 2.3n_{H}k_{B}T$ and $P_{M} = B^2/(8\pi)$, respectively, where, $n_{H}$, $k_{B}$, T and B are the number density of hydrogen, Boltzmann constant, temperature and magnetic field magnitude, respectively. Actually, a fully ionised plasma is considered with 10:1 abundance of hydrogen and helium which corresponds to a mass density ratio of 10:4 with mass ratio being 1:4. Similarly, the number density ratio of ions and electrons is 11:12 due to charge neutrality. These particular ratios of number density and mass density results in a factor of 2.3 in equation of state used for estimation of thermal pressure \citep{2012ApJ...748L..26X,2017ApJ...841..106Z}. The spatio-temporal evolution of the magnetic field magnitude and temperature along the slits `s1', `s2' and `s3' is shown in Figure~\ref{label 12}(b), (c), (d) and Figure~\ref{label 13}(a), (b), (c), respectively, in the Appendix. We extract the temporal variation of fluctuations in thermal pressure, magnetic pressure and temperature due to wave propagation at 60 Mm distances along slit `s1', `s2' and `s3' from their respective starting points, as depicted by the same horizontal dashed lines on the distance-time diagrams of density in Figure~\ref{label 3}, magnetic field in Figure~\ref{label 12}, temperature in Figure~\ref{label 13}. Since the background is dynamic and time-evolving itself, we subtract the long-term trend to extract the small-scale fluctuations of these quantities, as described in Appendix C. We find that fluctuations in temperature and thermal pressure are in phase with each other at all three locations (See Figure~\ref{label 13}(d), (e), (f) in Appendix). The bottom sub-panels of Figure~\ref{label 3}(b), (c), (d) show in-phase relations between fluctuations in magnetic pressure and thermal pressure at those three locations denoted by I, II and III in Figure~\ref{label 3}(a). This in-phase relation is a basic characteristic of fast-mode wave propagation \citep{2017ApJ...847...98J}. In order to characterize the propagating wave as fast mode and to calculate the wave energy fluxes at locations I, II and III, we need a more precise estimate of the phase speed by including the angle of  propagation to the direction of the local magnetic field. Also, since the nearby current sheet and its associated wave generation is rather dynamic and time-dependent, the background corona  is varying with time during the propagation of the wavefront. Thus, we also need to calculate the variation of the local  sound speed and Alfv\'en speed in time. The fast-mode phase speed is given by \citep[e.g.,][]{1975RvGSP..13..263H}
\begin{equation}
 V_{f,\theta}^{2}=\frac{1}{2}[(c_{S}^{2}+v_{A}^{2})+[(c_{S}^{2}+v_{A}^{2})^{2}-4c_{S}^{2}v_{A}^{2}cos^{2}\theta_{kB_{0}}]^{\frac{1}{2}}],
\end{equation}
where $c_{S}$ and $v_{A}$ are the sound speed and Alfv\'en speed, respectively, and $\theta_{kB_{0}}$ is the angle between the propagation direction of the wave ($\vec{k}$) and the background magnetic field ($\vec{B_{0}}$).  The fast-mode speed varies  between $v_{A}$ for propagation  along the magnetic field ($\theta_{kB_{0}}= 0$),  as shown by the dashed black lines in Figure~\ref{label 4}(b), \ref{label 4}(d) and \ref{label 4}(f), and
\begin{equation}
V_{f,\perp} = (c_{S}^{2}+v_{A}^{2})^{\frac{1}{2}}
\end{equation}
for propagation perpendicular to the magnetic field ($\theta_{kB_{0}}= \pi/2$), as shown by  solid black curves in the same figures. For time-dependent multi-sourced wave generation from different spatial locations in our numerical experiment, it is difficult to follow propagation of each wavefront individually and estimate the directions of the propagation vectors accurately. Therefore, we consider wave propagation  along three different directions given by slits `s1', `s2' and `s3', and focus on the behaviour at points I, II and III, respectively, which lie at a distance of 60 Mm along those slits. So, the directions slits make angles 135\degree, 90\degree and 45\degree to the positive $x$-direction at locations I, II and III, respectively. We also estimate the angles subtended by the magnetic field vector ($\vec{B}$) to the positive $x$-direction, i.e., the inclination angle $\phi$   of the three locations at each time using
\begin{equation}
\phi = \arctan\left(\frac{B_{y}}{B_{x}}\right).    
\end{equation}
We find that the inclination angle $\phi$ varies from -3\degree to 2\degree at location I, -0.2\degree to -0.6\degree at location II, and 3\degree to -2\degree at location III (Figure~\ref{label 4}a, c, e).  From these we calculate $\theta_{kB_{0}}$  to estimate the time-varying fast mode speeds (red solid curves in Figure~\ref{label 4}b, d, f). Basically, we use 135\degree-$\phi_{I}$, 90\degree-$\phi_{II}$ and 45\degree-$\phi_{III}$ to estimate $\theta_{kB_{0}}$ at location I, II, and III, respectively, considering wave propagation along the slits `s1', `s2 and `s3'. These estimated speeds are only slightly smaller than the maximum speeds estimated using Equation (11)  and are greater than the instantaneous Alfv\'en speeds (Figure~\ref{label 4}b, d, f). It should also be noted that at location II, the wave is propagating almost perpendicular to the magnetic field, and so the estimated phase speeds $V_{f,\perp}$ and $V_{f, \theta}$ in Figure~\ref{label 4}(d) lie on top of each other (See panels (a1)-(a2) of Figure~\ref{label 12} in the Appendix for the orientation of magnetic field). 

Thus, we find that the observed average propagation speeds estimated via fitting tilted ridges in the distance-time diagrams in Figure~\ref{label 3} are consistent with the theoretically calculated fast-mode wave speeds, which vary from 400 to 564 $\mathrm{km~s^{-1}}$ at location I, 429 to 585 $\mathrm{km~s^{-1}}$ at location II, and 398 to 550 $\mathrm{km~s^{-1}}$ at location III, with corresponding temporal averages of 490, 520 and 486 $\mathrm{km~s^{-1}}$ at those locations in the modeled  corona (red solid curves in Figure~\ref{label 4}b, d, f). In particular, consider the times 1876 s, 1863 s, and 1876 s when the wavefronts fitted by tilted ridges in the middle panels of Figure~\ref{label 3} reach the locations I, II and III, respectively. At these particular times and locations, the theoretically calculated fast-mode wave speeds are 500, 502 and 480 $\mathrm{km~s^{-1}}$, which differ from the observed propagation speeds of waves by only 1\%, 7\% and 2\%, respectively. 

\begin{figure*}
\centerline{\hspace*{0.013\textwidth}
         \includegraphics[height=7.2 cm,trim={0 0 0 0},clip]{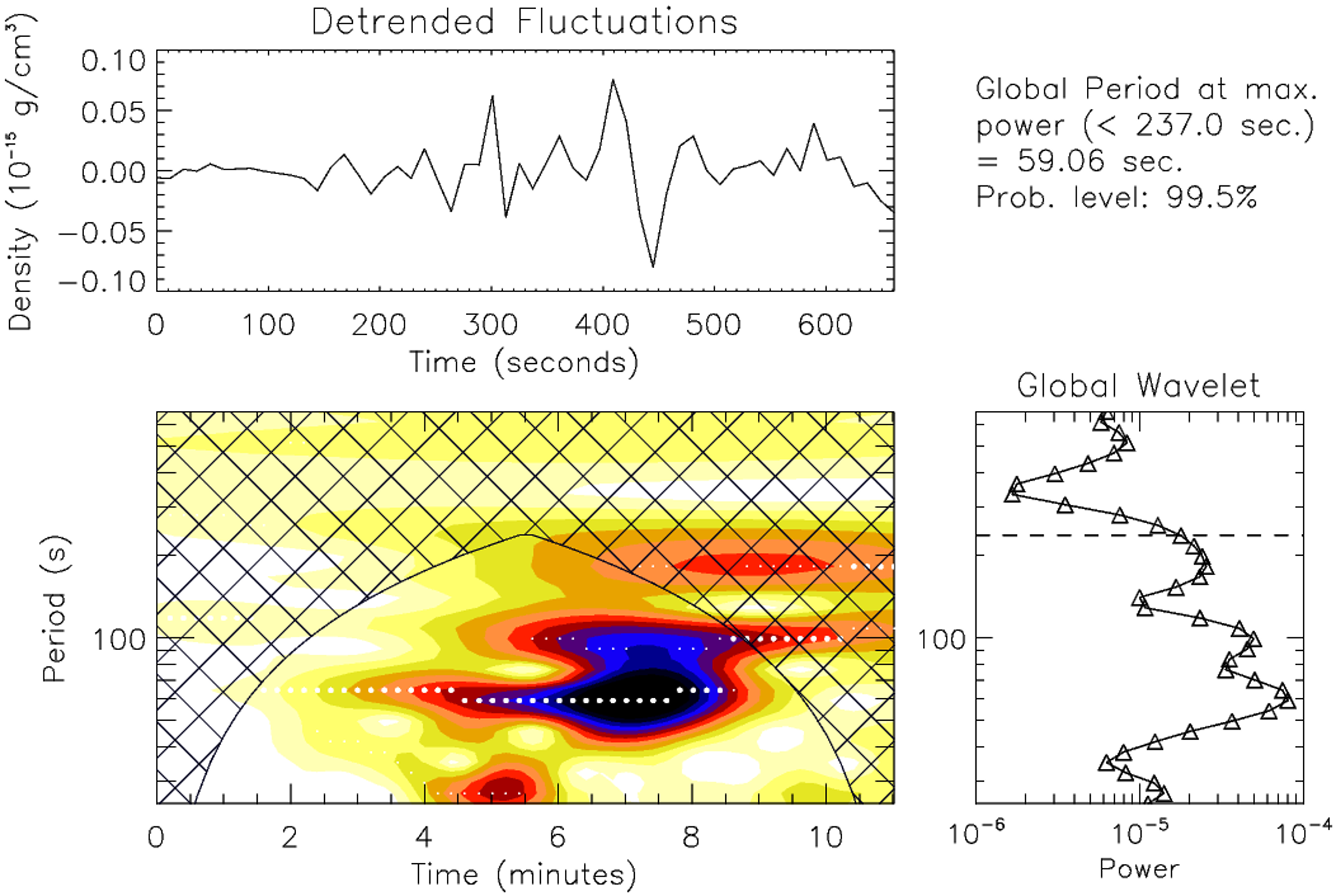}
         \hspace*{-0.01\textwidth}
         \includegraphics[height=7.2 cm,trim={0 0 0 0},clip]{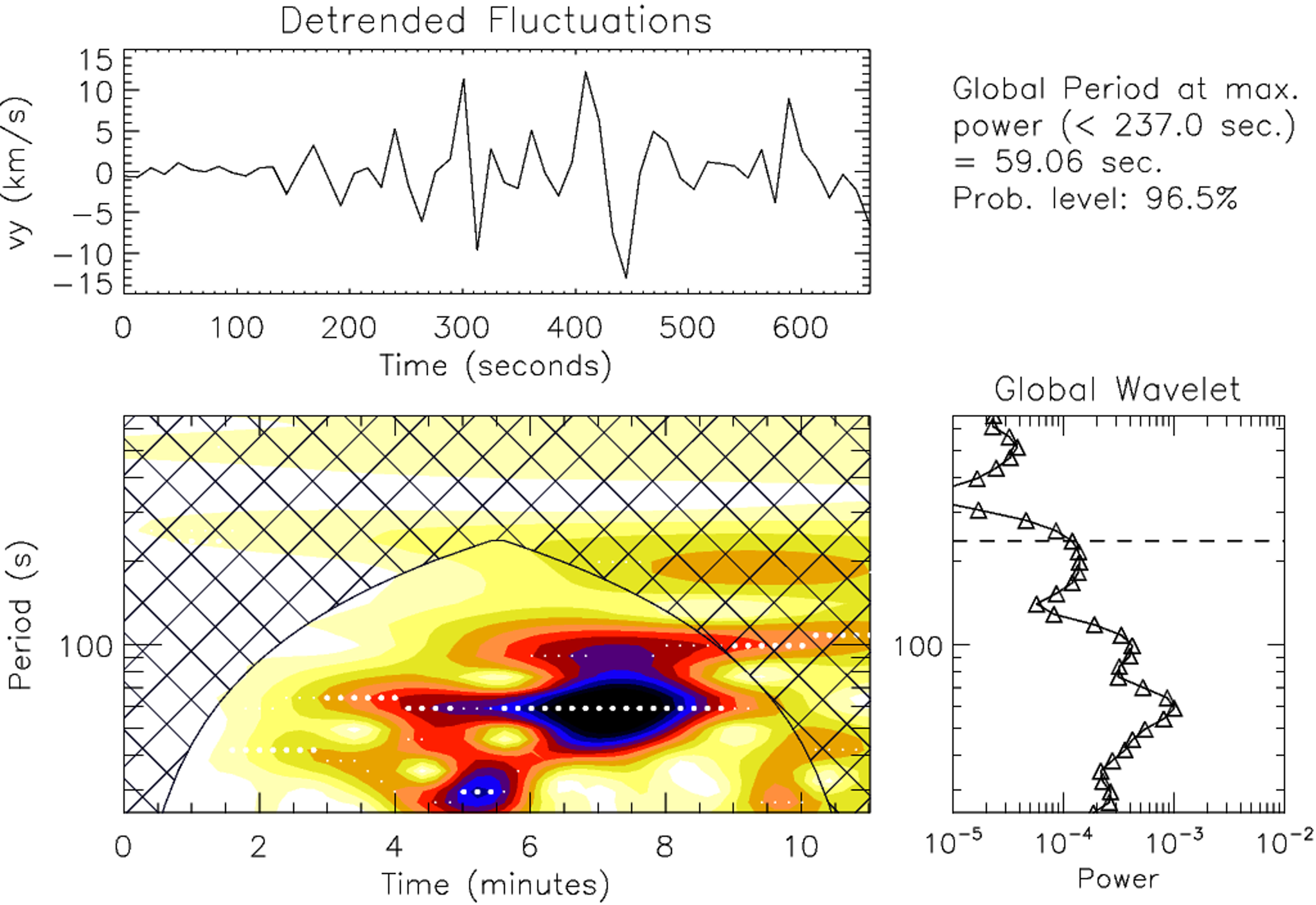}
       }
\vspace{-0.43\textwidth}
\centerline{ \large \bf      
\hspace{-0.1 \textwidth}  \color{black}{(a)}
\hspace{0.54\textwidth}  \color{black}{(b)}
   \hfill}
\vspace{0.42\textwidth}    
     
\centerline{\hspace*{0.013\textwidth}
         \includegraphics[height=7.2 cm,trim={0 0 0 0},clip]{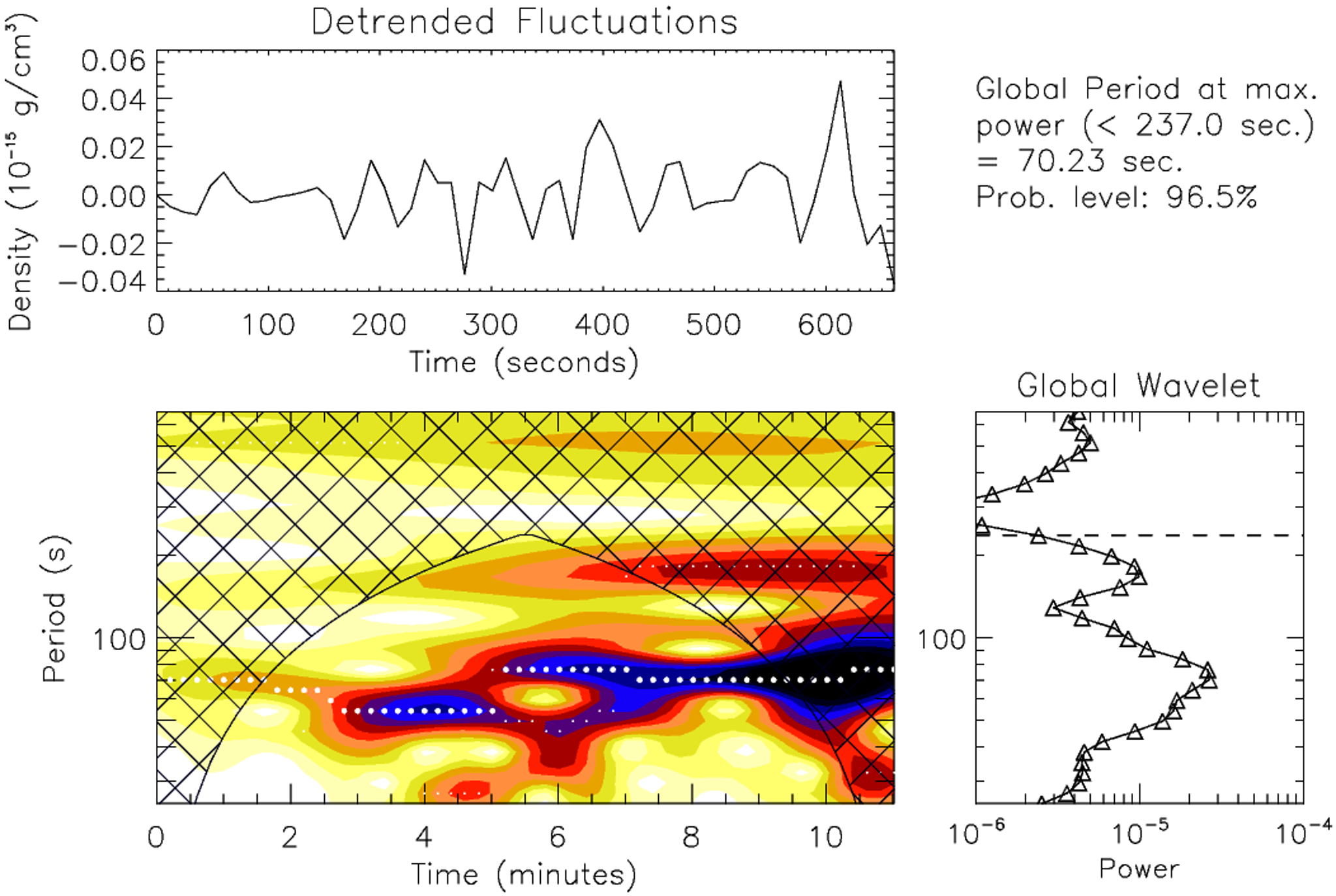}
         \hspace*{-0.01\textwidth}
         \includegraphics[height=7.2 cm,trim={0 0 0 0},clip]{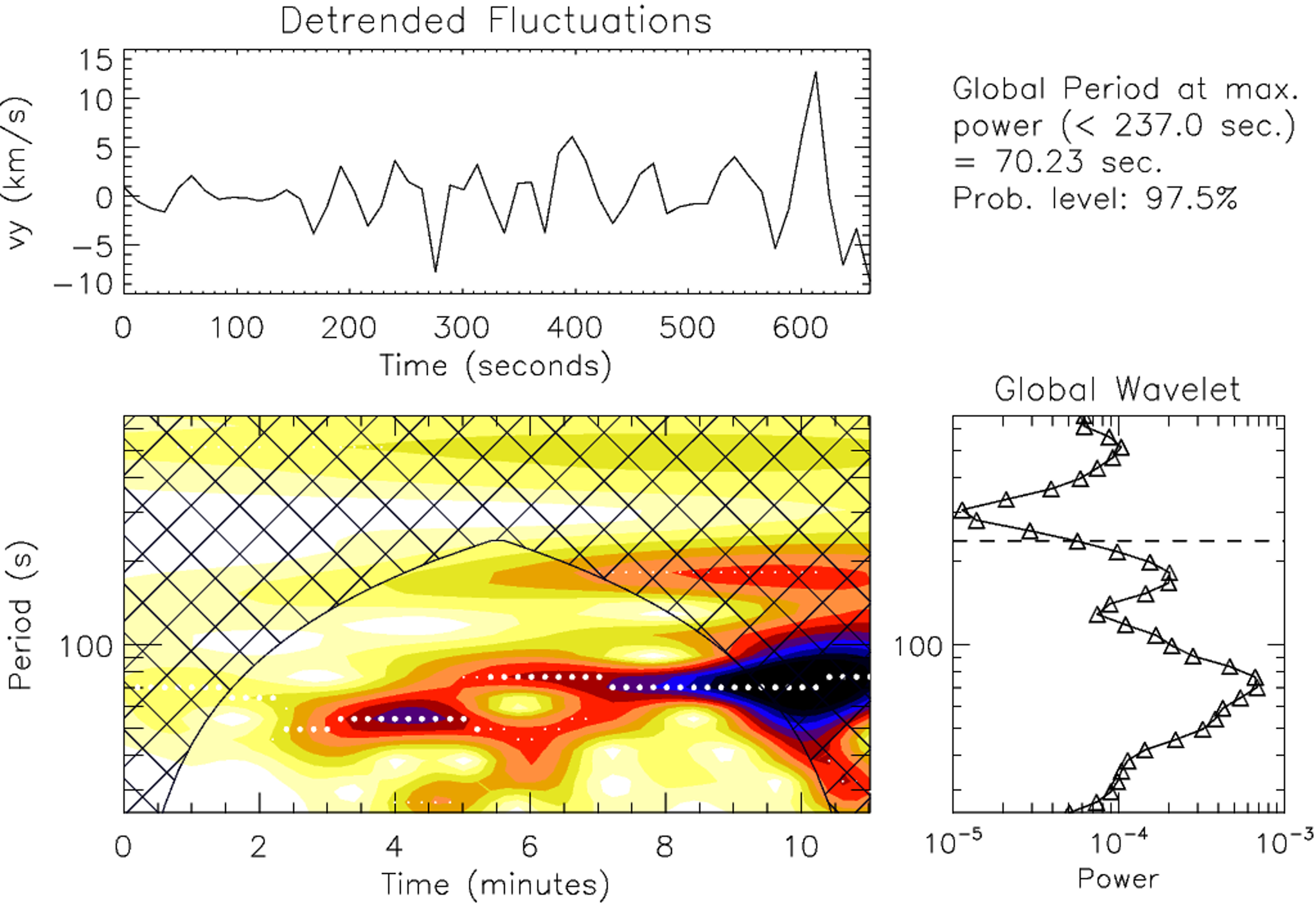}
        }
\vspace{-0.43\textwidth}
\centerline{ \large \bf      
\hspace{-0.1 \textwidth}  \color{black}{(c)}
\hspace{0.54\textwidth}  \color{black}{(d)}
   \hfill}
\vspace{0.42\textwidth}

\centerline{\hspace*{0.013\textwidth}
         \includegraphics[height=7.2 cm,trim={0 0 0 0},clip]{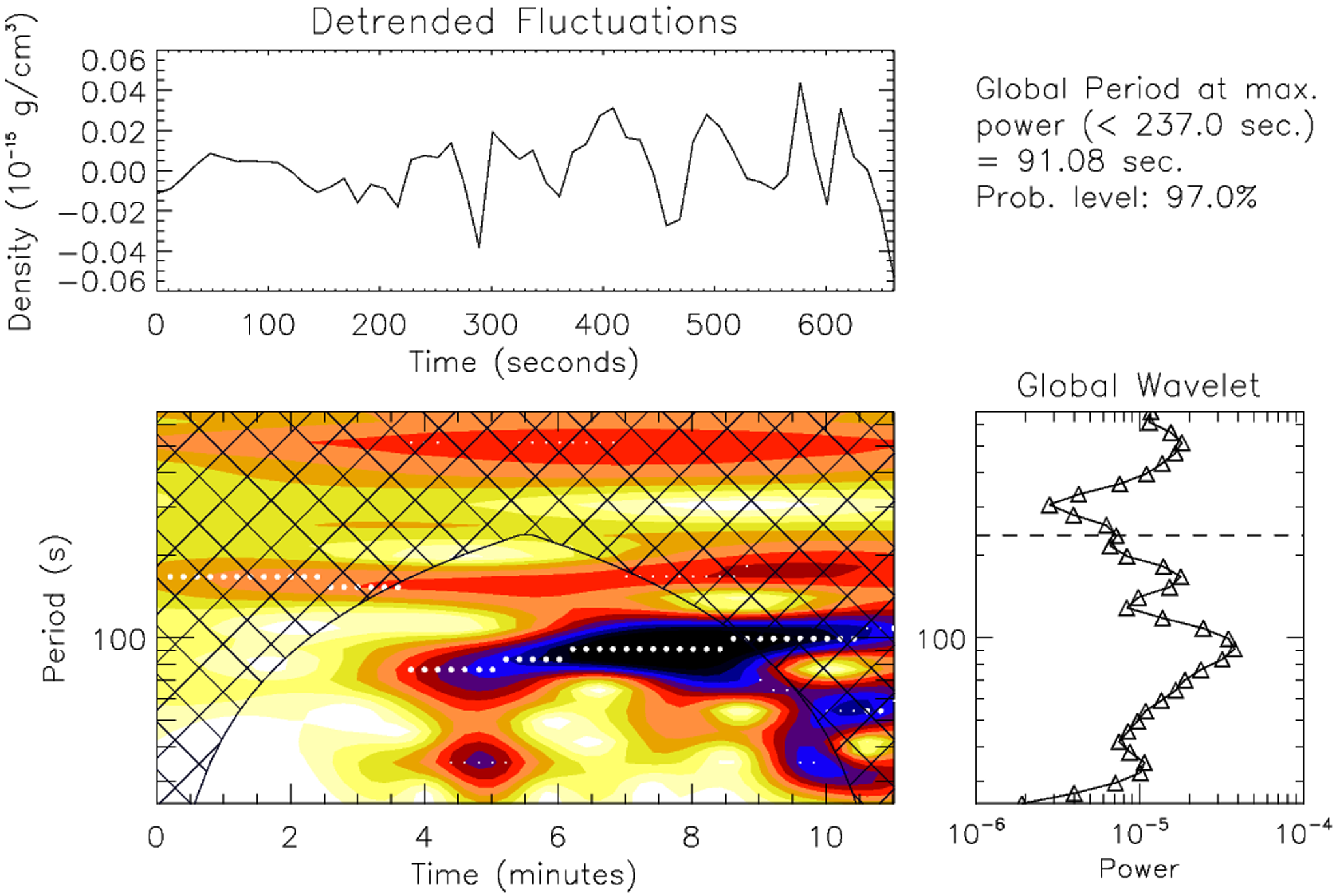}
         \hspace*{-0.01\textwidth}
         \includegraphics[height=7.2 cm,trim={0 0 0 0},clip]{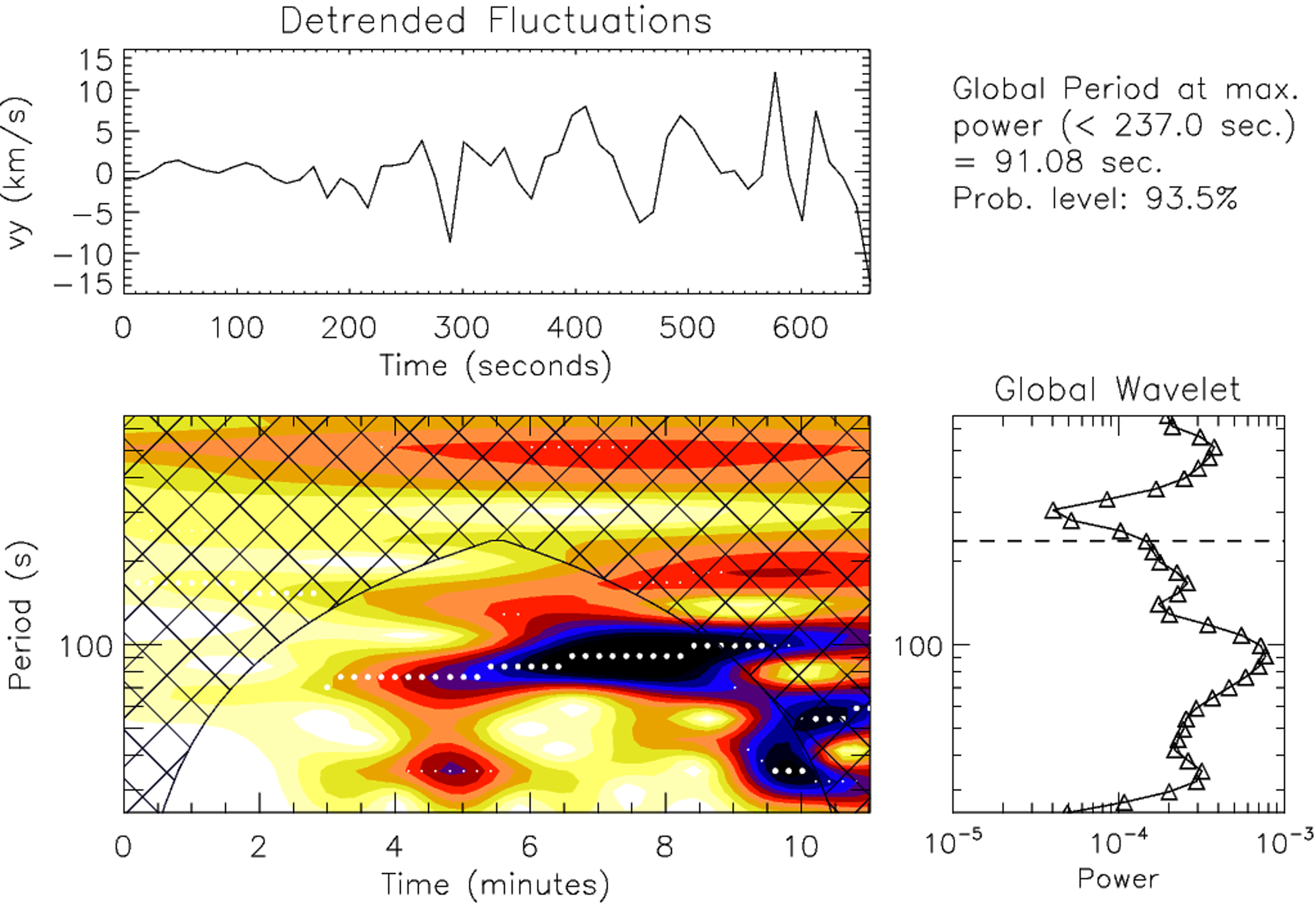}
        }
\vspace{-0.43\textwidth}
\centerline{ \large \bf      
\hspace{-0.1 \textwidth}  \color{black}{(e)}
\hspace{0.54\textwidth}  \color{black}{(f)}
   \hfill}
\vspace{0.42\textwidth}
\caption{The fluctuations in density and $v_y$ are obtained by subtracting the long-term background trends from the original profiles  at locations I, II and III in Figure~\ref{label 3}(a). Panels (a), (c) and (e) show the time series of fluctuations in density and their corresponding wavelet distributions, which exhibit dominant periods of 59.06 s, 70.23 s and 91.08 s at locations I, II and III, respectively. Panels (b), (d) and (f) show similar characteristics for $v_y$. In all of these figures, 0 s on the time axis stands for 1587 s, which is the same starting time as in the bottom panels of panel (b), (c) and (d) of Figure~\ref{label 3}. The unique randomization technique \citep{1985AJ.....90.2317L,2001A&A...368.1095O} estimates the significance levels of oscillations between 93.5 \% to 99.5\%. Similar characteristics for $B_x$ are shown in the Appendix.}
\label{label 5}
\end{figure*} 

\subsubsection{Periodicities of the Propagating Wave Trains}
At distances of 60 Mm along slits `s1', `s2' and `s3', i.e., at locations I, II and III, we extract the temporal variations of density, $x$- and $y$-components of velocity and  magnetic field. The long-term trend is subtracted from all of these profiles using a running average window in order to isolate the short-term fluctuations associated with the propagating waves (see Appendix C and Figure~\ref{label 10} and  the left panels of Figure~\ref{label 11} for more details). It is found that the fluctuations in density, $v_y$ (approximately parallel to the propagation direction of the wave) and $B_x$ (approximately perpendicular to the propagation direction) are in phase (see the upper panels of (a) and (b), (c) and (d), (e) and (f) in Figure~\ref{label 5} and the upper panels of (b), (d) and (f) in Figure~\ref{label 11}).  A wavelet analysis is carried out using the detrended fluctuations in each of the three cases. The periods of oscillation are around 59, 70 and 91 s (see the lower parts of panels (a) and (b), (c) and (d), (e) and (f) of Figure~\ref{label 5} for wavelet profiles of density and $v_y$  at locations I, II and III, respectively). All of these estimates have significance levels greater than 95 $\%$ except for the period of $v_y$ at location III whose significance level is 93.5 $\%$.

\subsubsection{Energy Associated with the Propagating Fast MHD Waves}
\begin{figure*}
\centerline{\hspace*{0.013\textwidth}
         \includegraphics[height=7 cm,trim={0 0 0 0},clip]{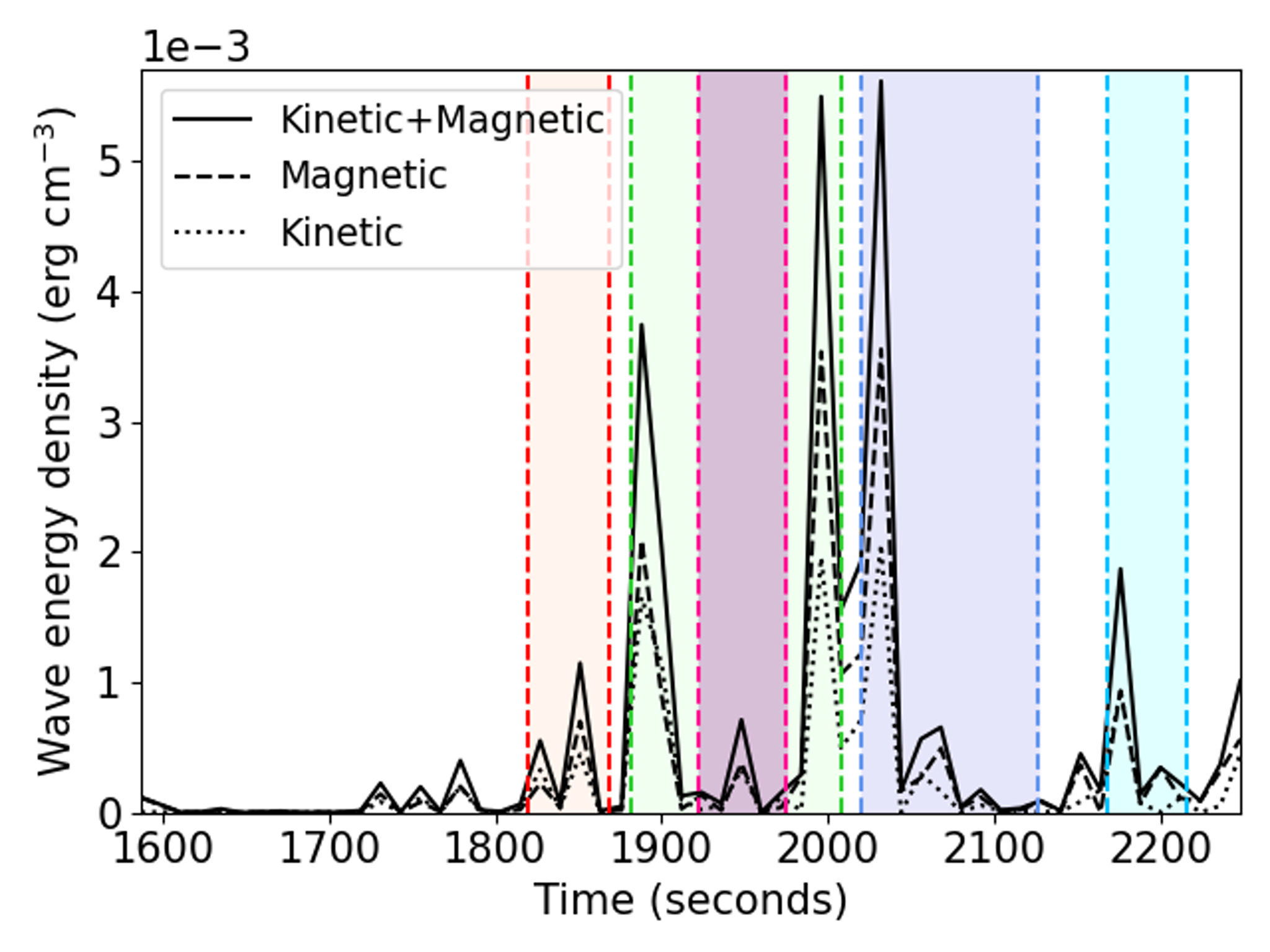}
         \hspace*{-0.01\textwidth}
         \includegraphics[height=7 cm,trim={0 0 0 0},clip]{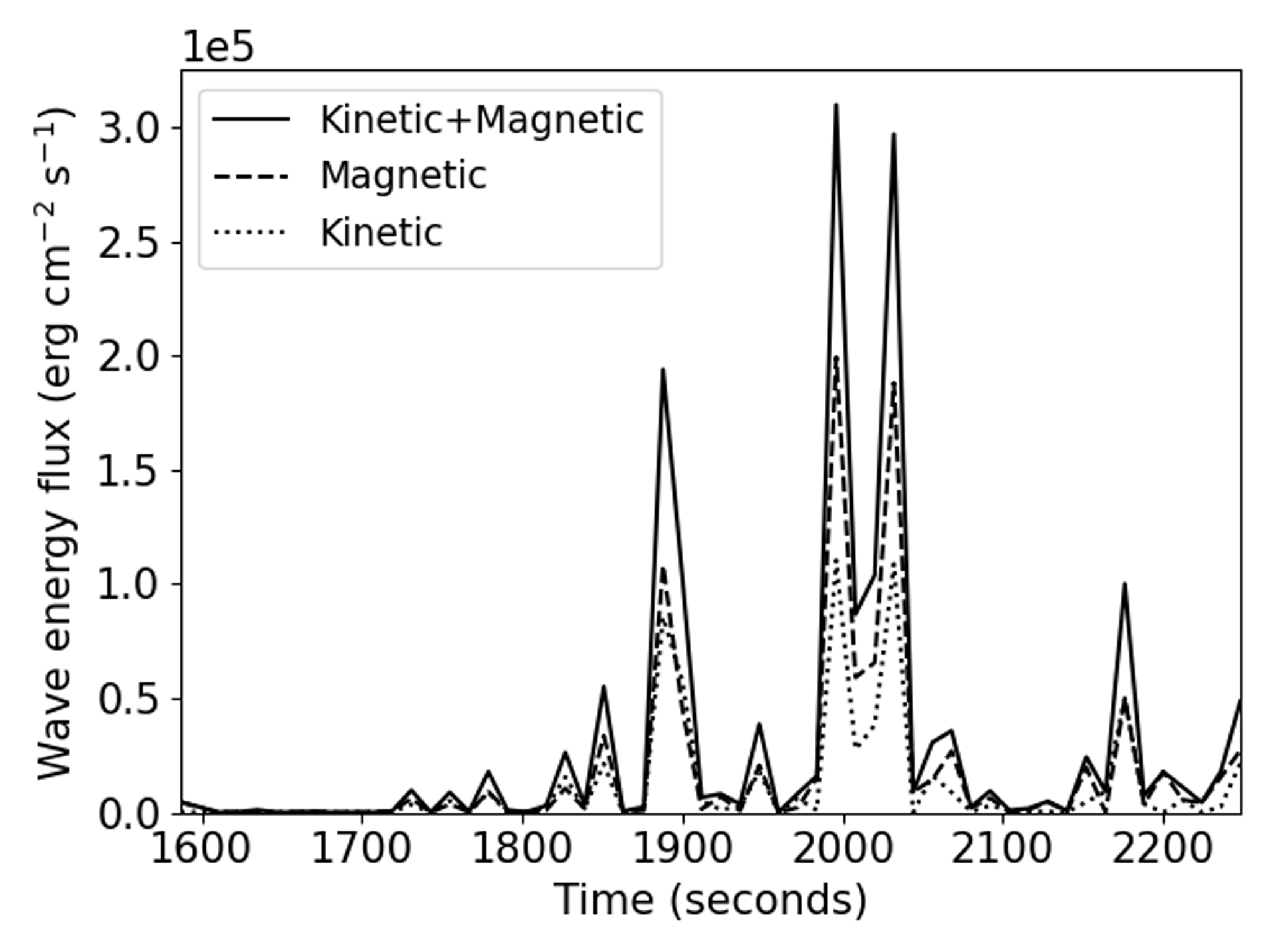}
       }
\vspace{-0.39\textwidth}
\centerline{ \large \bf      
\hspace{-0.03 \textwidth}  \color{black}{(a)}
\hspace{0.17 \textwidth}  \color{red}{Location I}
\hspace{0.16\textwidth}  \color{black}{(b)}
\hspace{0.17 \textwidth}  \color{red}{Location I}
   \hfill}
\vspace{0.38\textwidth}    
     
\centerline{\hspace*{0.013\textwidth}
         \includegraphics[height=7 cm,trim={0 0 0 0},clip]{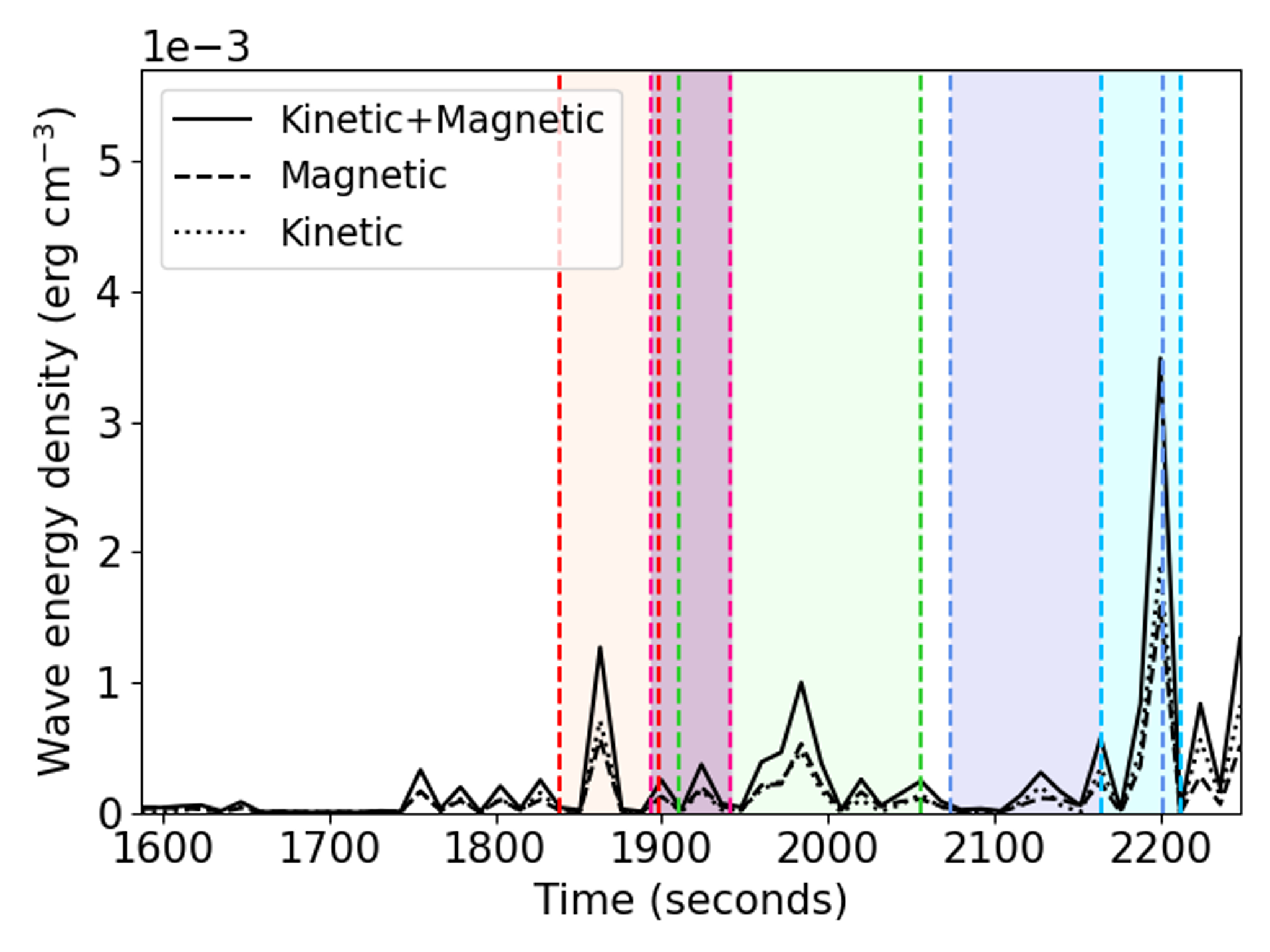}
         \hspace*{-0.01\textwidth}
         \includegraphics[height=7 cm,trim={0 0 0 0},clip]{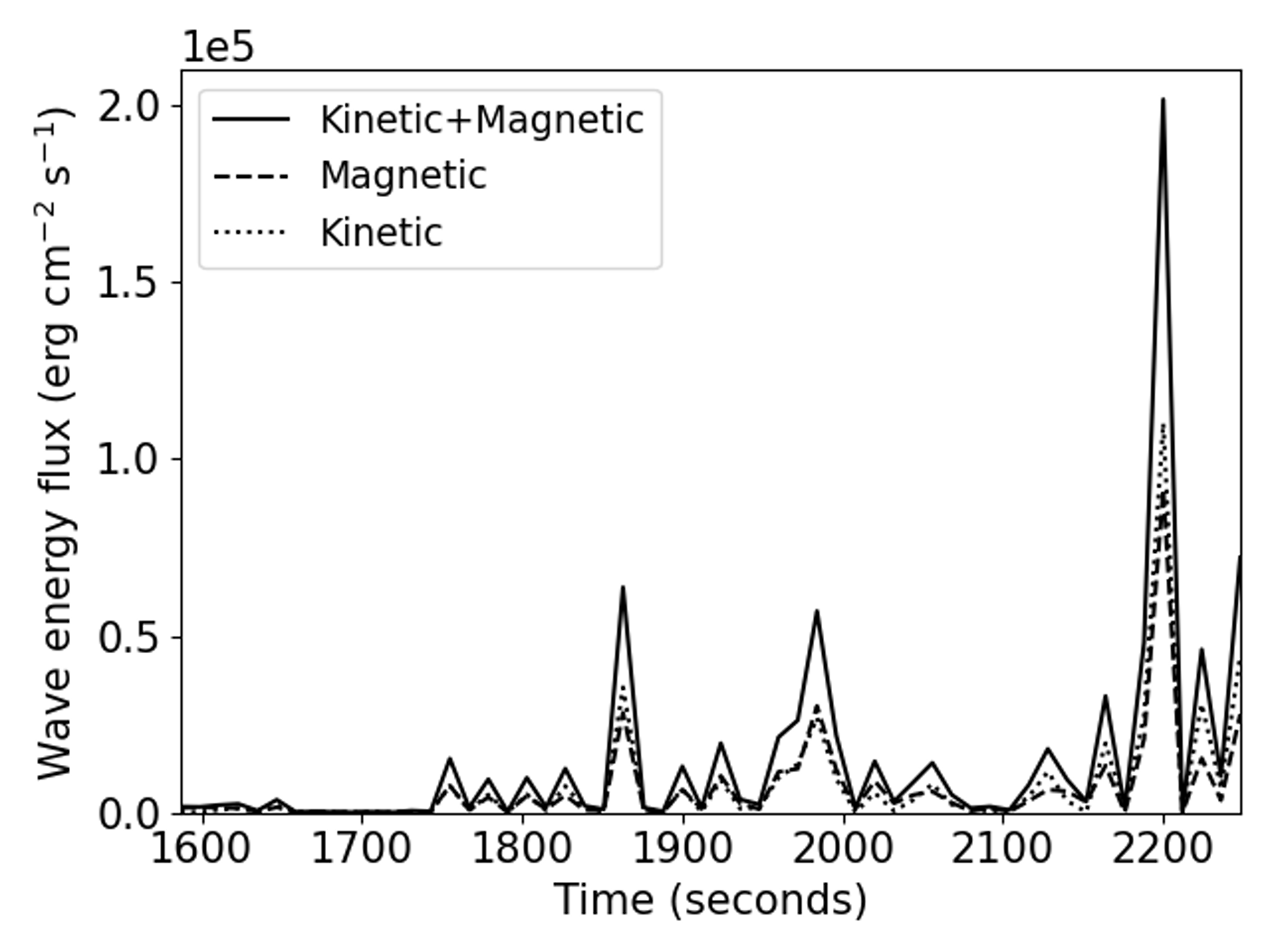}
        }
\vspace{-0.39\textwidth}
\centerline{ \large \bf      
\hspace{-0.03 \textwidth}  \color{black}{(c)}
\hspace{0.17 \textwidth}  \color{red}{Location II}
\hspace{0.16\textwidth}  \color{black}{(d)}
\hspace{0.17 \textwidth}  \color{red}{Location II}

   \hfill}
\vspace{0.38\textwidth}

\centerline{\hspace*{0.013\textwidth}
         \includegraphics[height=7 cm,trim={0 0 0 0},clip]{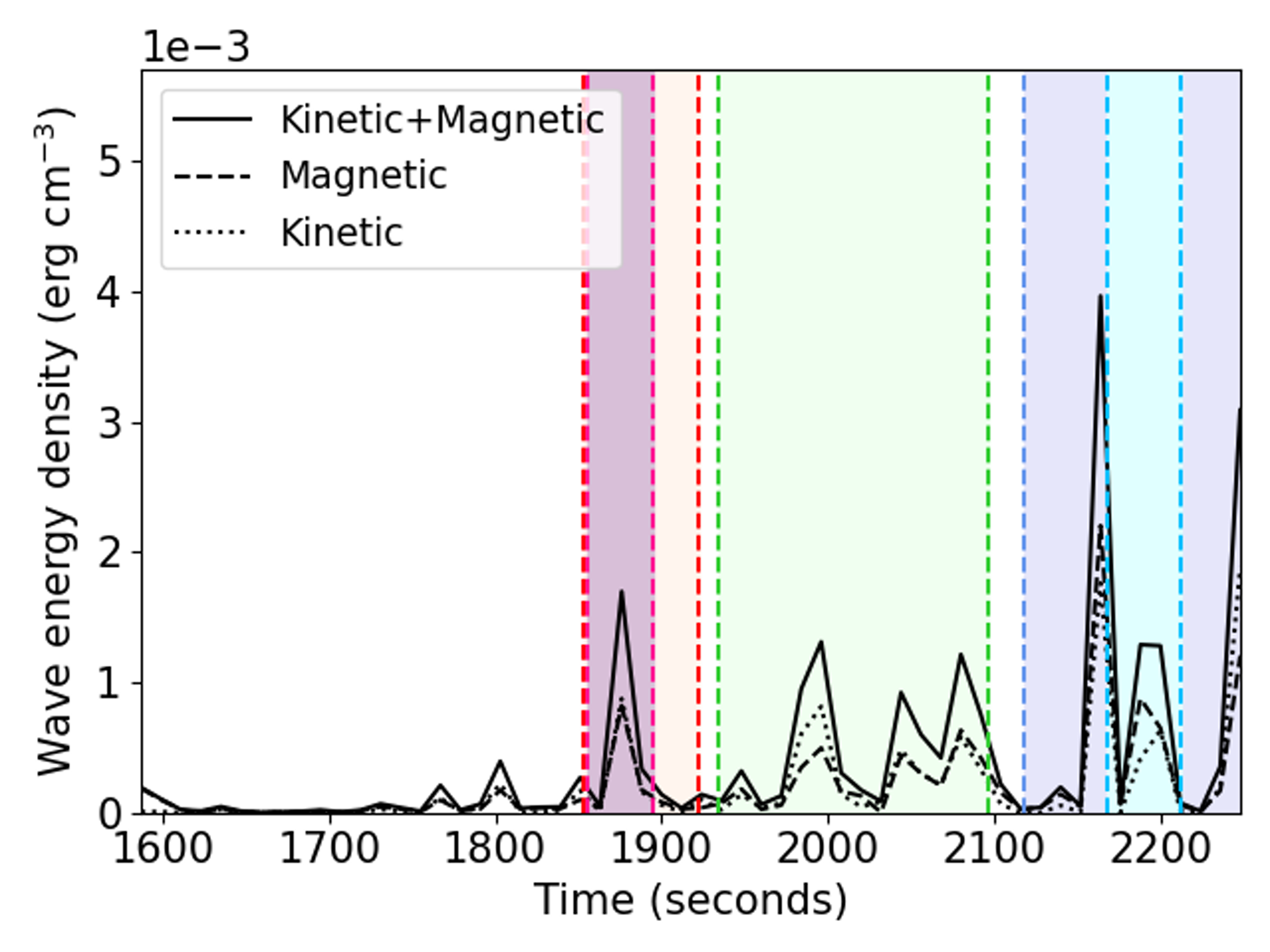}
         \hspace*{-0.01\textwidth}
         \includegraphics[height=7 cm,trim={0 0 0 0},clip]{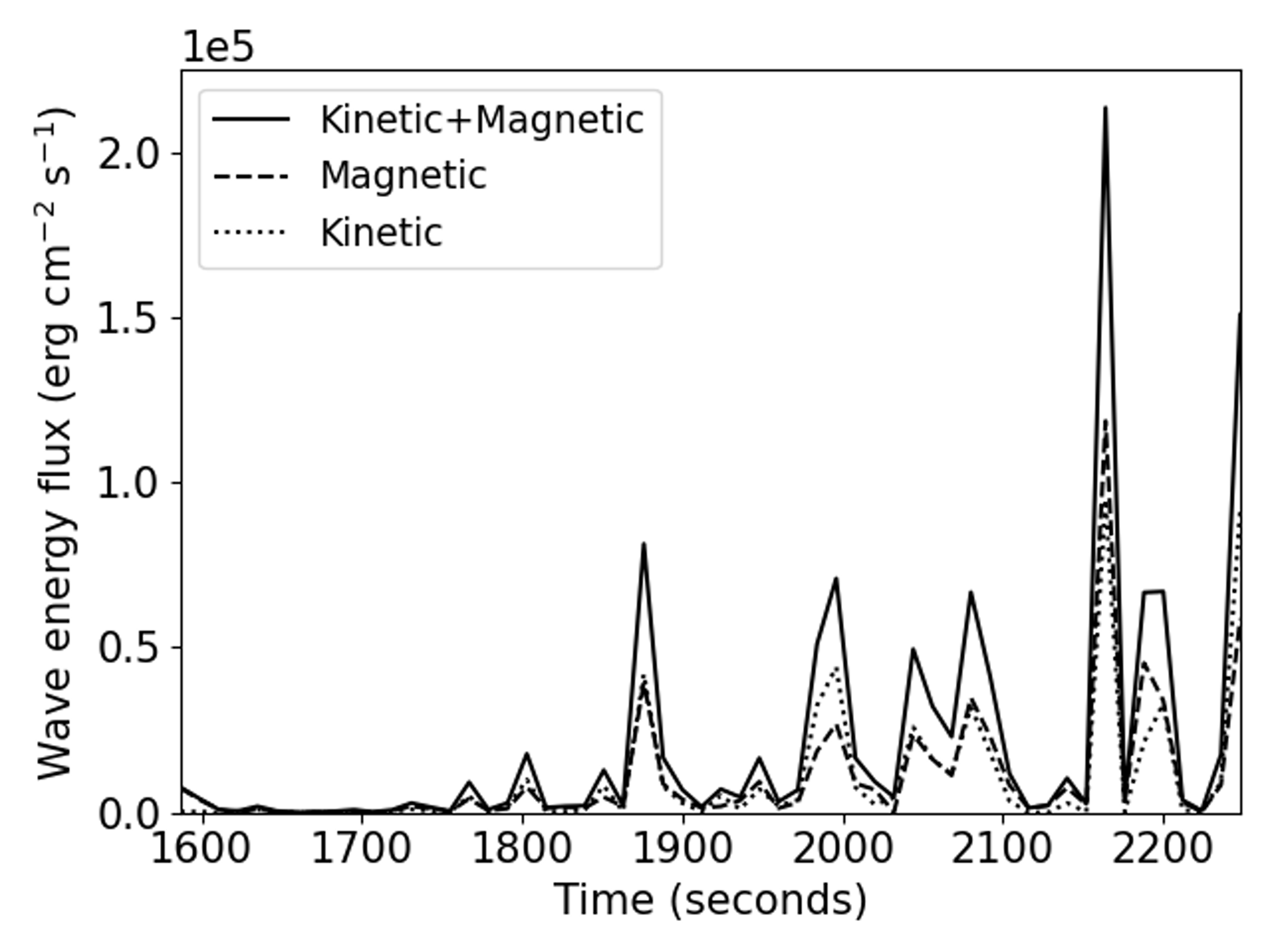}
        }
\vspace{-0.39\textwidth}
\centerline{ \large \bf      
\hspace{-0.03 \textwidth}  \color{black}{(e)}
\hspace{0.17 \textwidth}  \color{red}{Location III}
\hspace{0.16\textwidth}  \color{black}{(f)}
\hspace{0.17 \textwidth}  \color{red}{Location III}
   \hfill}
\vspace{0.38\textwidth}

\caption{Panels (a), (c) and (e) exhibit the temporal variation of wave energy densities estimated using Equation 13 at locations I, II and III, respectively, while panels (b), (d) and (f) give the corresponding wave energy fluxes  using Equation 14. The dashed curves show the profiles of magnetic energy, whereas the dotted curves show the profiles of kinetic energy. Solid curves denote the resultant wave energy quantities. The values of magnetic and kinetic energies are highly time-dependent. The shaded time-windows in panel (a), (c) and (e) basically correspond to different coalescence events, as discussed in Section 3.3.1.}
\label{label 6}
\end{figure*}

Since the fast MHD waves are propagating in the large-scale corona, it is interesting to find how much energy is transmitted to large distances. The wave energy density (WED) carried by the propagating fast magnetoacoustic wave is estimated at three locations (I), (II) and (III) (shown in Figure~\ref{label 3}(a)). We use the following formula to calculate the wave energy density (Equation (16) of \citet{2013A&A...558A..76R})
\begin{equation}
\mathrm{WED} = \frac{\rho}{2}(v_{x}^{2}+v_y^{2})+\frac{(b_{x}^{2}+b_{y}^{2})}{8\pi},
\end{equation}
where $v_{x}$, $v_{y}$, $b_{x}$ and $b_{y}$ are the perturbed parts of the corresponding variables. Basically, we again use the small-scale perturbations in velocity and magnetic field after subtraction of long-term trends due to the time-varying background. As shown in panels (a), (c) and (e) (corresponding to locations I, II and III) of Figure~\ref{label 6}, the estimated maximum wave energy densities are roughly $10^{-3}~\mathrm{erg~cm^{-3}}$ in all three positions. We further calculate the wave energy flux (WEF) at three different locations using
\begin{equation}
\mathrm{WEF} = \mathrm{WED} \times V_{f, \theta},
\end{equation}
where WED and $V_{f, \theta}$ at the above three different locations are estimated using Equations 13 and 10, respectively. The resulting wave energy fluxes are of order  $10^{5}~\mathrm{erg~cm^{-2}~s^{-1}}$ at each of the three locations which are roughly 60 Mm  from the source region (see panel (b), (d) and (f) of Figure~\ref{label 6}).
 
\section{DISCUSSION \& CONCLUSION} 
In this work, a region containing a magnetic null collapses to form a current sheet (CS) when it is subjected to a localised velocity perturbation.  As the sheet extends in length, it undergoes tearing instability and associated plasmoid formation. Multiple plasmoids of different sizes form and  move impulsively along the CS. Some of them cause inflow and thinning in their wakes followed by subsequent  plasmoid formation. Others rapidly merge with each other to form larger plasmoids.  Four important properties of the wavefronts generated by coalescence events identify them as compressive fast-mode waves, namely:

(i) their average propagation speeds are around 500 $\mathrm{km~s^{-1}}$ which is consistent with the values of the characteristic fast-mode speed; this ranges from 400 to 550 or 600 $\mathrm{km~s^{-1}}$ as estimated from Equation (10) (see Figure~\ref{label 3} and \ref{label 4});

(ii) the time-varying phase speeds of the propagating fluctuations at all three locations ((I), (II) and (III) in Figure~\ref{label 3}(a)) are always higher than the corresponding Alfv\'en speeds (Figure~\ref{label 4}b,d,f);

(iii) most importantly, the thermal and magnetic pressure perturbations are found to be in-phase with each other (bottom panels of Figure~\ref{label 3}b, c, d) \citep{2005psci.book.....A,2017ApJ...847...98J},

(iv) in addition, the small-scale fluctuations in density are in phase with $v_{y}$, i.e., the component of velocity approximately parallel to the direction of resultant propagation of waves and $b_{x}$, i.e., the component of magnetic field  approximately perpendicular to the direction of resultant propagation  \citep{1986A&A...164...77M,2015ApJ...800..111Y,2020ApJ...899...99O} (Figure~\ref{label 10} and left panels of Figure~\ref{label 11}). 

There are several observational studies in which quasi-periodic fast wave trains are reported to be associated with flares or corresponding radio bursts on the basis of their periodicities \citep{2012ApJ...753...53S,2013SoPh..288..585S,2016A&A...594A..96G}. In addition, there are a few studies in which the fast waves are launched from the loop top region, which indicates that magnetic reconnection may be a leading source of fast waves in the corona \citep{2013A&A...554A.144Y,2019ApJ...886L..25Y}. Since, in our case, there are multiple sources  of waves and the sources are time-dependent and dynamic, we have extracted the time-evolution of the wave energy flux at three locations on the left, middle and right in the large-scale model corona. Therefore, we have established a possible temporal and spatial correlation between the peaks in the estimated wave flux and the different events of plasmoid coalescence. To do so, we have compared the wave arrival times with a predication based on the average propagation speed of 500 $\mathrm{km~s^{-1}}$ and average distance of the coalesced plasmoids from the location under consideration. We consider the entire time window from the start of each coalescence to the time when the internal dynamics ceases. Several other articles have shown that merging of plasmoids or flux ropes may generate fast waves \citep{2016ApJ...823..150T,2017ApJ...847...98J,2022MNRAS.513.5224S}, but we have studied the temporal and spatial correlation between the wave energy flux at large distances from its source and the sources themselves, i.e., coalesced plasmoids. 

\citet{2011ApJ...736L..13L} reported a wave energy flux  of $(0.1-2.6) \times 10^{7}~\mathrm{erg~cm^{-2}~s^{-1}}$ at the coronal base of a funnel of coronal loops which they suggested to be sufficient for the steady-state heating of loops in an AR. \citet{2015ApJ...800..111Y} estimated the wave energy flux at a point 15 Mm away from the source of the waves and found it to be around $\mathrm{7 \times 10^{6}~\mathrm{erg~cm^{-2}~s^{-1}}}$. \citet{2018ApJ...860...54O} estimated a lower limit of $\mathrm{1.8 \times 10^{5}}~\mathrm{erg~cm^{-2}~s^{-1}}$ for the energy flux carried by counter-propagating quasi-periodic fast waves to the apex of a trans-equatorial loop connecting two epicentres of solar flares rooted in the AR corona, but mentioned that this may not be sufficient for  coronal heating due to their low rate of occurrence. \citet{1977ARA&A..15..363W} reported that the total energy loss in the quiet corona is roughly $\mathrm{3 \times 10^{5}~\mathrm{erg~cm^{-2}~s^{-1}}}$ when losses due to radiation, thermal conduction and solar wind are considered. In a coronal hole, the energy loss due to radiation and thermal conduction is of  order  $\mathrm{7 \times 10^{4}~\mathrm{erg~cm^{-2}~s^{-1}}}$. By comparison, the total energy loss in the AR corona is of order $10^{7}~\mathrm{erg~cm^{-2}~s^{-1}}$. However, the energy needed to drive solar wind in ARs can be less than $10^{5}~\mathrm{erg~cm^{-2}~s^{-1}}$ \citep{1977ARA&A..15..363W}. In our case, we have adopted a magnetic field strength appropriate for the quiet Sun and have estimated a wave energy flux at 60 Mm distances from the source region to be of order  $\mathrm{10^{5}~\mathrm{erg~cm^{-2}~s^{-1}}}$, which is therefore sufficient for heating coronal holes and the quiet Sun and driving solar winds in both quiet Sun and ARs. 

According to \citet{1994ApJ...435..482P}, the wave energy decay rate is independent of the background magnetic field ($B_{0}$) in case of viscous damping, whereas it is inversely proportional to $B_{0}^{2}$ for damping associated with thermal conduction. They have reported that the damping rate decreases by 1.5-4 times when the background field increases 10 times. In the present simulation, the background magnetic field in which waves are propagating varies roughly from 7 to 9 Gauss only. So, this change will hardly have any impact on wave damping. Moreover, they reported that the damping length for fast waves of period  10 s for a propagation angle 45\degree is around 130 Mm for a  background density $= 10^{9}~\mathrm{cm^{-3}}$, temperature $= 2 \times 10^{6}~\mathrm{K}$ and magnetic field  10 Gauss. In our case, waves of 90 s periodicity are propagating roughly at an angle of 45\degree along `s3'. Now, the damping rate is proportional to $T_{0}^{\alpha}\tau^{-2}$ where $2.5 \le \alpha \le 3.5$,  $T_{0}$ is the background temperature and $\tau$ is the wave period \citep{1994ApJ...435..482P}. Since the average background temperature is around $10^{6}~\mathrm{K}$ only and the period is 9 times higher in the presence of a similar background number density and magnetic field strength,  the decay rate will be much smaller and the damping length will be much longer than 130 Mm along `s3'. Similarly, for the other directions, the estimated periods are around 60 s and 70 s, and so they will also be hardly damped at all  within the distance covered by slits `s1' and `s2'. This explains why no wave damping has been detected in the field of view  in the present simulation. Nevertheless,  the waves are carrying substantial energy fluxes along all three directions to a distance of 60 Mm  from the source region, which will be potentially significant for coronal heating if it can be dissipated.

We conclude that, even though MHD waves and magnetic reconnection have usually been studied as separate independent processes, there can be a close relation between them. MHD waves can interact with a magnetic null to make it collapse to form a CS and to lead to the onset of reconnection. If the reconnection becomes impulsive and bursty, the repeated formation and coalescence of  plasmoids  can in turn generate fast-mode  waves. So, waves can act as a catalyst to initiate magnetic reconnection and \emph{vice versa}. The generated waves can carry energy fluxes to larger distances, which may contribute to coronal heating in the quiet Sun, coronal holes and active regions. This is one of the first numerical experiments to study both processes, i.e., reconnection initiated by a velocity perturbation and the resulting generation of fast-mode MHD waves by impulsive bursty reconnection. Thus it serves as an  example of SWAR (a Symbiosis of WAves and Reconnection) at work in the solar corona \citep{Sri24}. Since the locations of reconnection, namely, magnetic null points, separators and quasi-separators, are ubiquitous in the solar atmosphere together with the presence of MHD waves, these processes may be common. Hopefully, in future more focused studies with high-resolution space and ground-based observations  might provide more direct  signatures of this inter-relation. Also, since both waves and reconnection are basic plasma processes in many other plasma environments, it will be interesting to see the applicability of this idea in other plasma systems at astrophysical, space and laboratory scales.

Finally, we note that reconnection is a 3D process, and in the three-dimensional case tearing leads to the formation of flux ropes with a non-zero guide field component instead of the plasmoids produced in our 2D simulation. Some aspects of the dynamics of merging flux ropes will be qualitatively similar to what we find for plasmoids, but there will be extra, richer effects that arise from three dimensionality such as the launching of torsional waves along the flux ropes, i.e., in the direction of guide field \citep{2014PhPl...21h2114W}. Hence, it will be interesting to have more studies in the line of the Symbiosis of Waves and Reconnection (SWAR) using three-dimensional numerical simulations in future.

\section*{Acknowledgments}
The authors thank the anonymous referee for constructive suggestions which are helpful to improve the paper. We are thankful to open source MPI-AMRVAC 3.0 for providing a user-friendly flexible framework which enable us to write new routines required to simulate our scientific idea. S.M. would like to acknowledge the financial support provided by the Prime Minister's Research Fellowship (PMRF) of India. A.K.S acknowledges the ISRO grant
(DS/2B-13012(2)/26/2022-Sec.2) for the support of his scientific research. D.I.P. gratefully acknowledges support through an Australian Research Council Discovery
Project (DP210100709). R.K. acknowledges support by basic research funding from the Korea Astronomy and Space Science Institute (KASI2023185007). D.Y. is supported by the National Natural Science Foundation of China (NSFC; grant numbers 12173012, 12111530078 and 11803005), the Guangdong Natural Science Funds for Distinguished Young Scholar (grant number 023B1515020049),
the Shenzhen Technology Project (grant number GXWD20201230155427003-20200804151658001) and the Shenzhen Key Laboratory Launching Project (grant number ZDSYS20210702140800001).  

\vspace{5mm}
\software{MPI-AMRVAC, Paraview, Python}  
\appendix

\begin{figure*}
\centerline{\hspace*{0.013\textwidth}
         \includegraphics[height=5 cm,trim={0 0 0 0},clip]{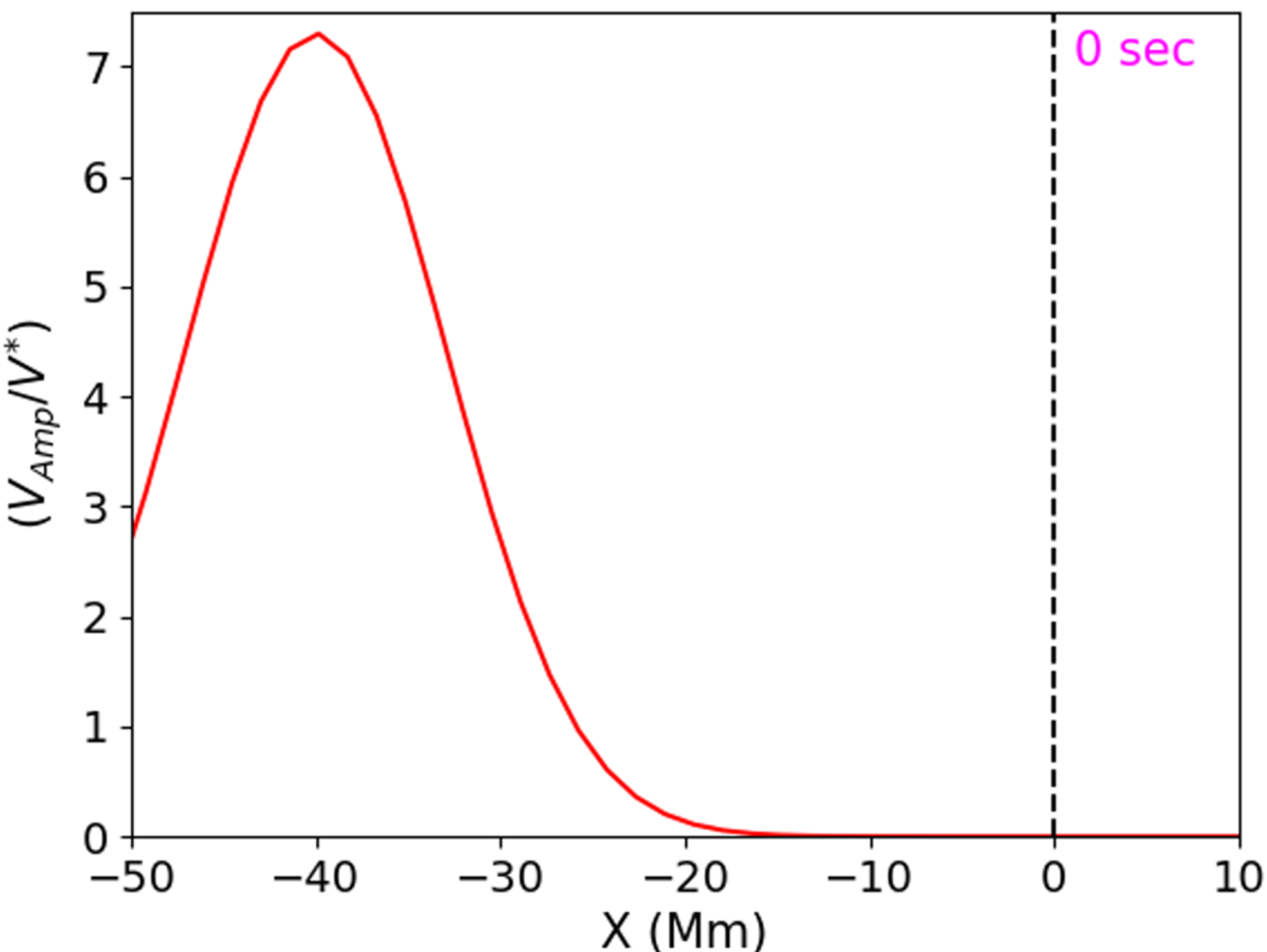}
         \hspace*{-0.01\textwidth}
         \includegraphics[height=5 cm,trim={0 0 0 0},clip]{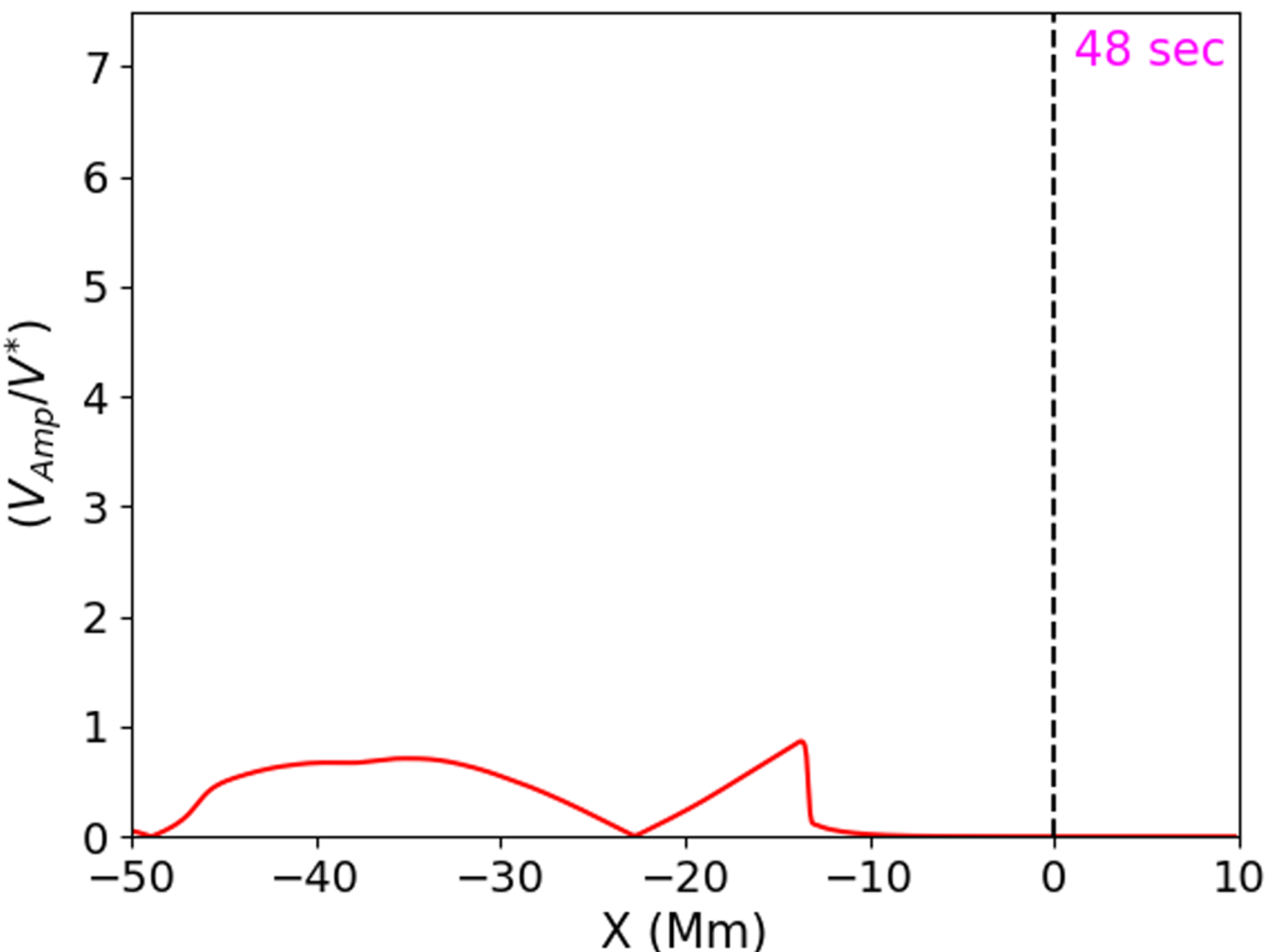}
         \hspace*{-0.01\textwidth}
         \includegraphics[height=5 cm,trim={0 0 0 0},clip]{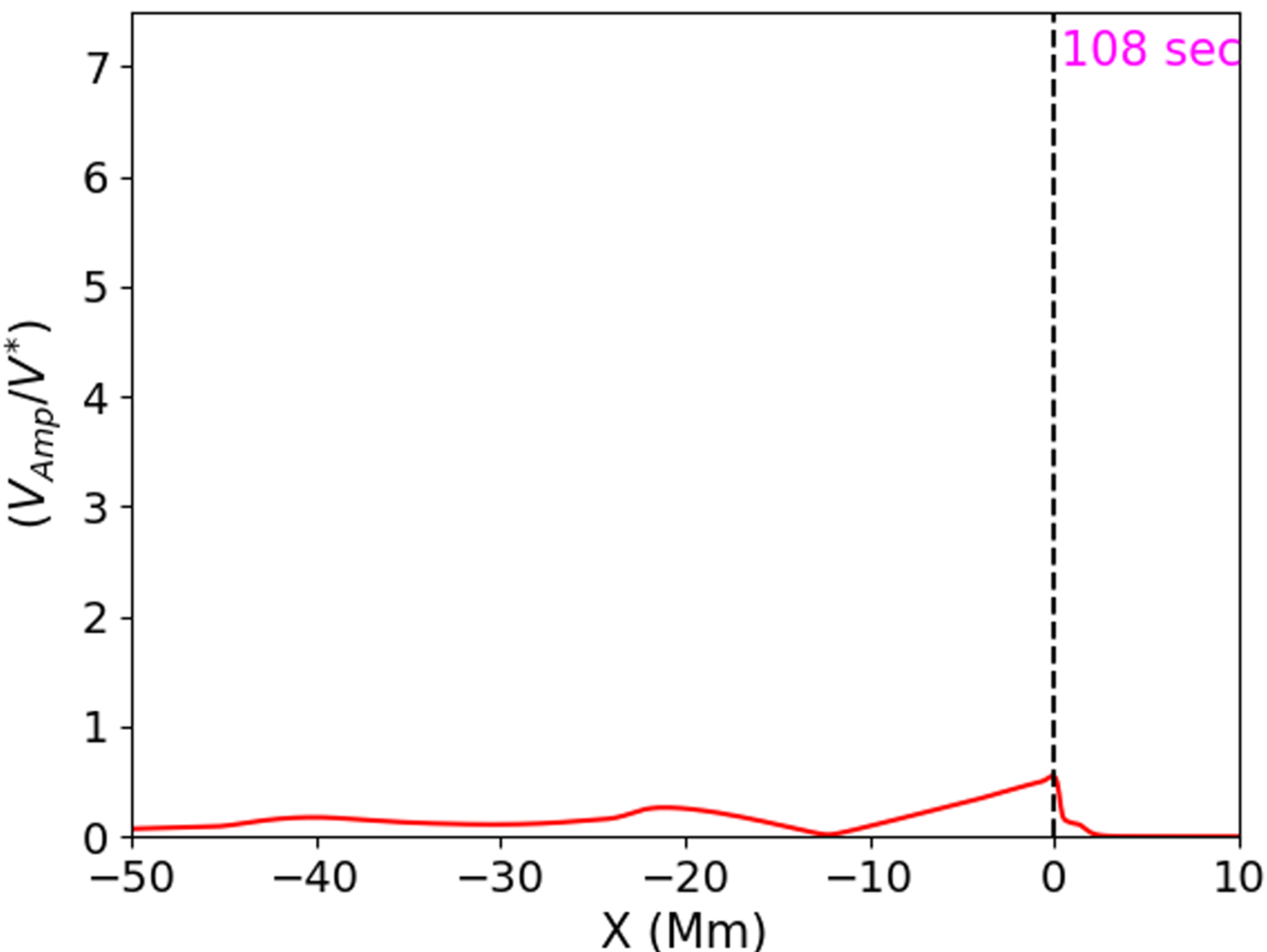}
         }
\vspace{-0.40\textwidth}
\centerline{ \large \bf      
   \hfill}
\vspace{0.37\textwidth}    
\caption{Temporal evolution of the initial velocity disturbance during its passage towards the magnetic null. At 0 s, the disturbance is properly Gaussian shaped centred at x = - 40 Mm. As it propagates, the leading edge undergoes steepening to form fast mode shock propagating across the field lines (48 s). Eventually at 108 s, the shock wave interacts with the magnetic null whose location is denoted as vertical dashed lines in all of three panels. An animation covering the entire passage of initial Gaussian pulse from its source region towards the magnetic null is available in the online version. The real time duration of the animation is 1 s.}
\label{label 7}
\end{figure*} 

\section{Variation of Resultant Lorentz force vector during Plasmoid Coalescence}
\begin{figure*}
\hspace{-2 cm}
\includegraphics[scale=0.9]{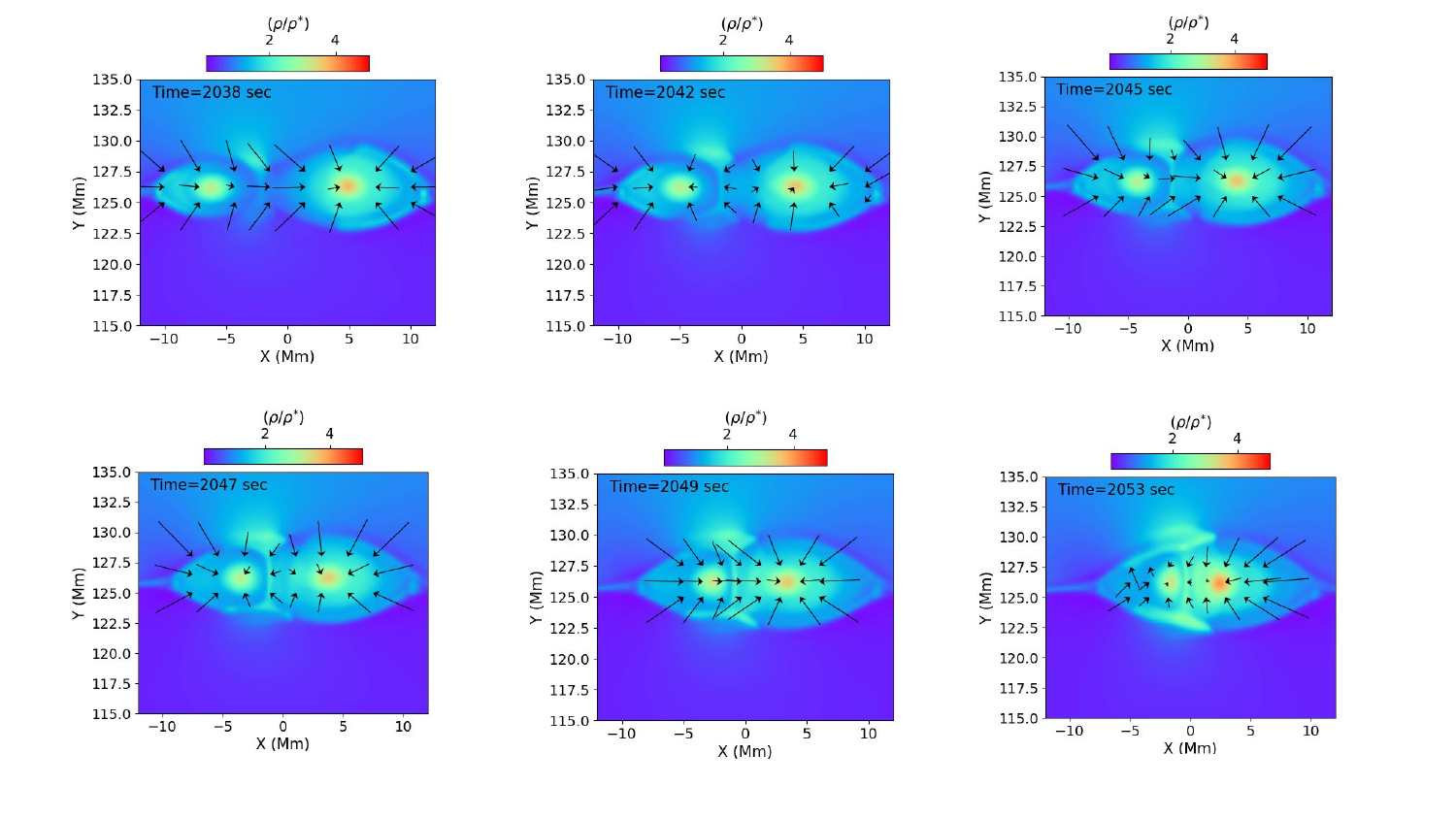}
\caption{The total resultant Lorentz force vector (black arrows) superposed on density maps during the  plasmoid coalescence V. An animation covering the variation  from 2024 s to 2065 s is available in the online version. The real-time duration of the animation is 2 s.}
\label{label 8}
\end{figure*}

We consider the coalescence  Event V as a representative case to show the variation of direction of resultant Lorentz force during the merging of two plasmoids (Figure ~\ref{label 8}). At 2038 s, the left plasmoid is being attracted towards the bigger right one. However, the oppositely directed Lorentz force vector in between two merging plasmoids suggest that an opposite force is opposing the merging process. Around 2045 s, again the attractive forces are dominating followed by a direction reversal again around 2047 s. At 2049 s, the vector signs suggest that attraction is higher than the repulsion. But then around 2053 s, the forces reverse direction. So, during merging of plasmoids, the resultant Lorentz force represents alternatively an attractive force promoting merging  and another force opposing it.
\section{Insignificant realization of wave propagation at the bottom part of the CS}
\begin{figure*}
\hspace{+0.2 cm}
\includegraphics[scale=0.22]{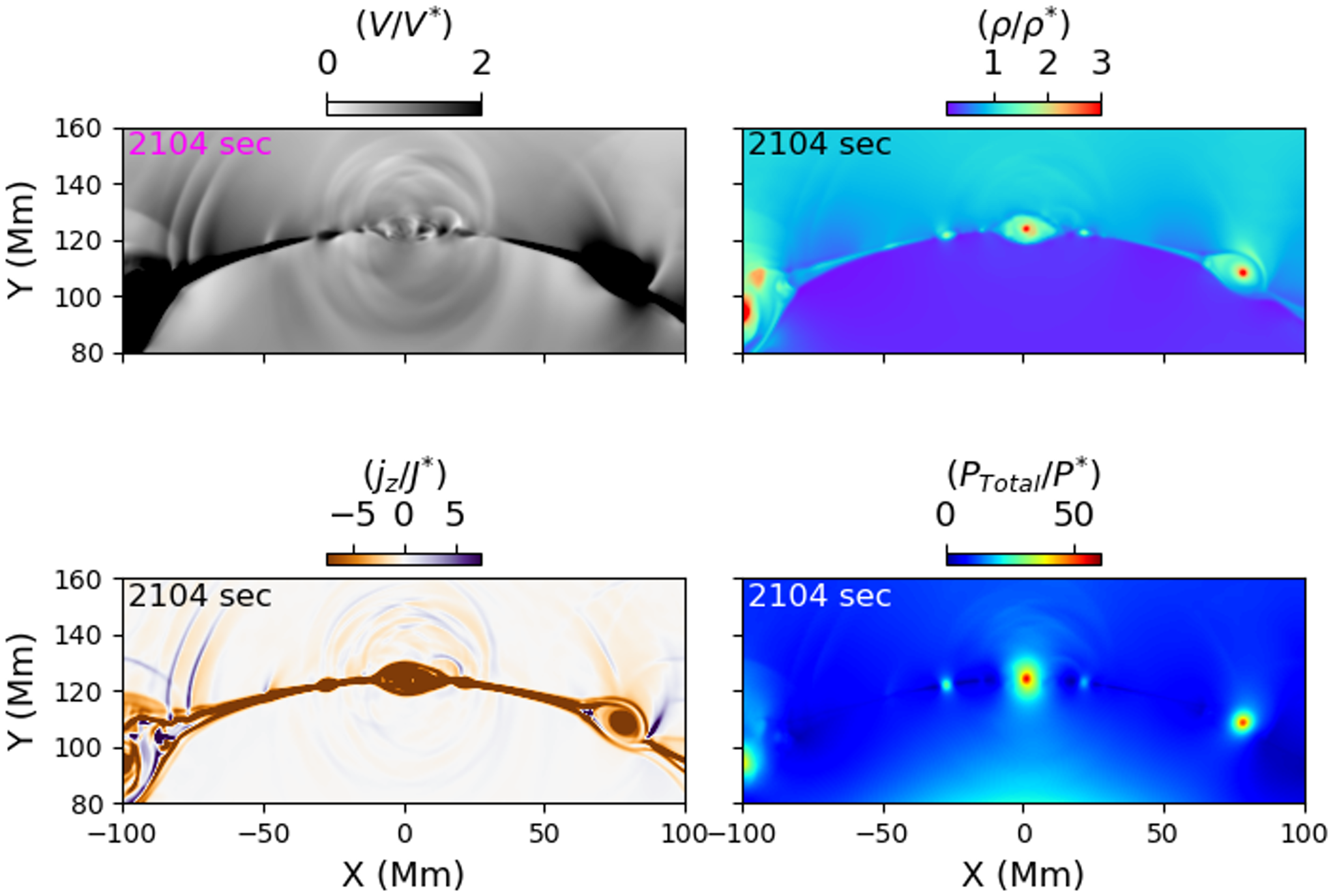}
\caption{Top left: Nearly arc-shaped velocity disturbances are evident in both upper and lower parts of the CS after  plasmoid coalescence. Top right: Density is considerably lower in the lower part of the CS than that in the upper part. Perturbations in density are significant in the upper part only. Bottom left: Capture of waves in terms of perturbations in magnetic field given by current density. It also shows that the perturbations are significant only in the upper part. Bottom right: Total pressure snapshot also exhibits perturbations in the form of arc-shaped features in upper part only. An animation covering the entire dynamics from 1695 s (start time of Event I) to 2140 s (a time after the end of Event V) is available in the online version. The real-time duration of the animation is 3 s.}
\label{label 9}
\end{figure*}

Arc-shaped velocity disturbances are found to propagate both in upper and lower part of the curved CS after each coalescence events or even with almost each moving plasmoid (See top left panel of Figure~\ref{label 9} and associated animation). However, we find that arc-shaped disturbances are not detected in the lower part as perturbations in density (See top right panel of Figure~\ref{label 9} and associated animation). Similarly, the perturbations in magnetic field (as depicted via current density) and total pressure in the lower part of the CS are much smaller than those in the upper part of the curved CS (See bottom left and right panel of Figure~\ref{label 9} and associated animation). The background physical conditions above and below the current sheet differ from each other as follows--[i] the density is considerably lower, and [ii] the magnetic field is much stronger below than above the CS. These give a higher Alfv\'en speed below the CS. Since the magnetic field is stronger in the bottom part, it is not  perturbed significantly, as shown in the current density snapshot. According to \citet{2011ApJ...736L..13L}, the maximum perturbation in density due to fast-mode waves can be estimated as $\delta \rho_{max}=(\rho\delta v)/V_{Ph}$, where $\delta v$, $\rho$ and $V_{Ph}$ are the velocity amplitude of the waves, background plasma density and phase speed of fast wave. So, even though velocity disturbances are propagating in both directions, due to low background density and high Alfv\'en speed, the wavefronts are less visible as density perturbations below the CS. Hence, wave propagation below the CS does not show up as significant perturbations in density, magnetic field and total pressure. Therefore, in the main paper we focus on the part of corona above the CS to understand the physics of modelled event. 

\section{Subtraction of Long-term Background Trends to Extract Fluctuations in Physical Variables during Wave Propagation}
\begin{figure*}
\centerline{\hspace*{0.013\textwidth}
         \includegraphics[height=6.7 cm,trim={0 0 0 0},clip]{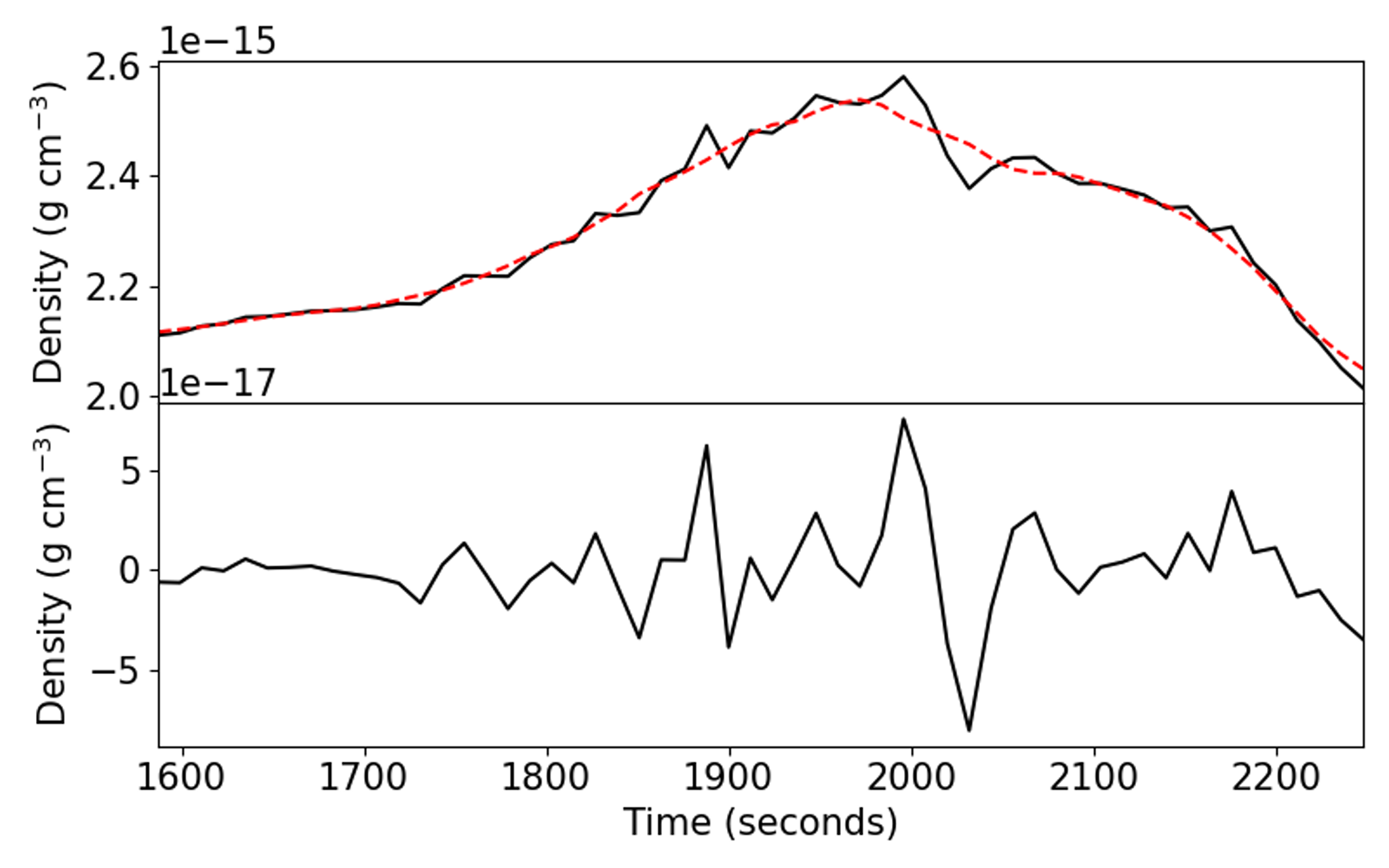}
         \hspace*{-0.01\textwidth}
         \includegraphics[height=6.5 cm,trim={0 0 0 0},clip]{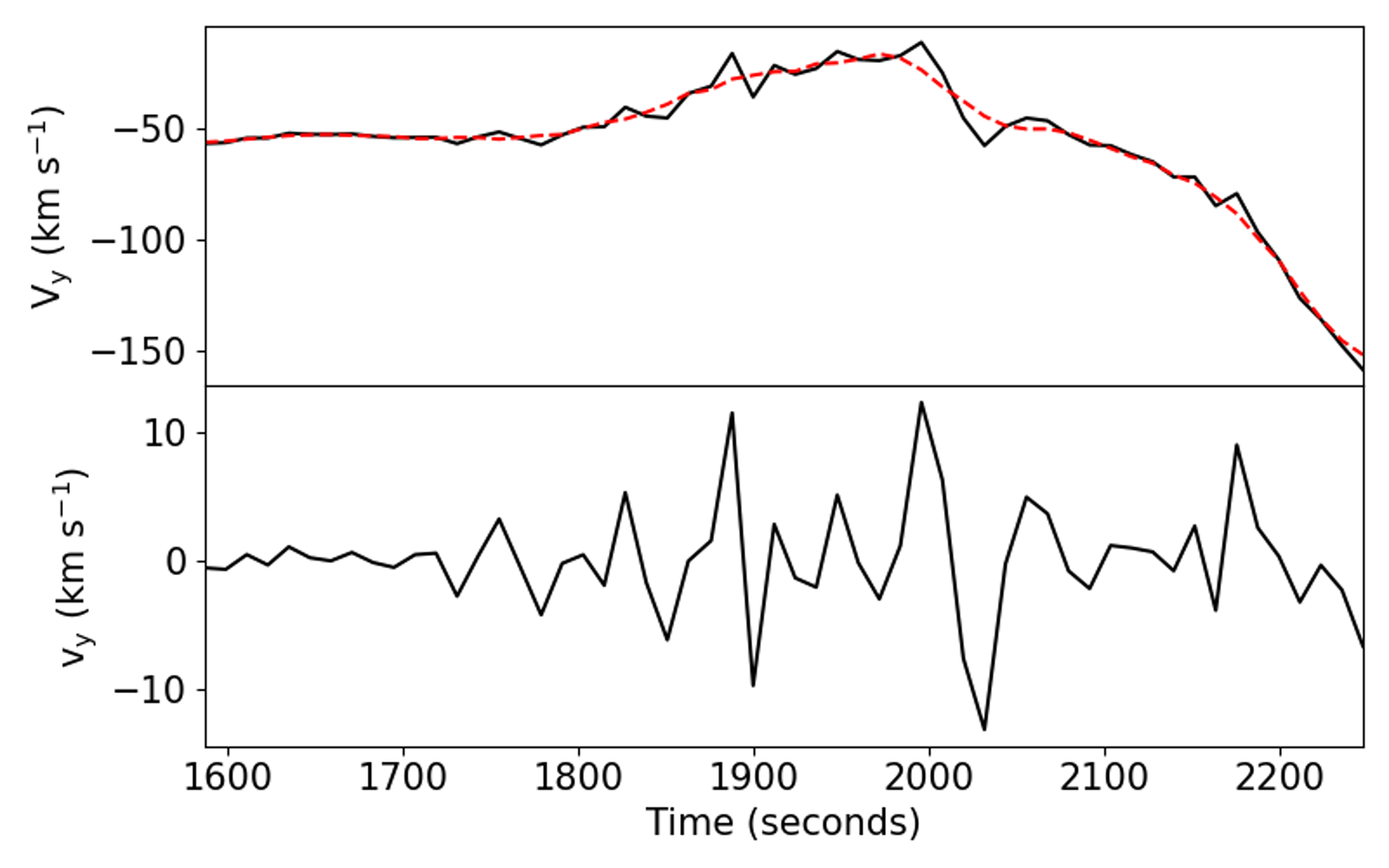}
       }
\vspace{-0.38\textwidth}
\centerline{ \large \bf      
\hspace{-0.09 \textwidth}  \color{black}{(a)}
\hspace{0.55\textwidth}  \color{black}{(b)}
   \hfill}
\vspace{0.38\textwidth}    
     
\centerline{\hspace*{0.013\textwidth}
         \includegraphics[height=6.7 cm,trim={0 0 0 0},clip]{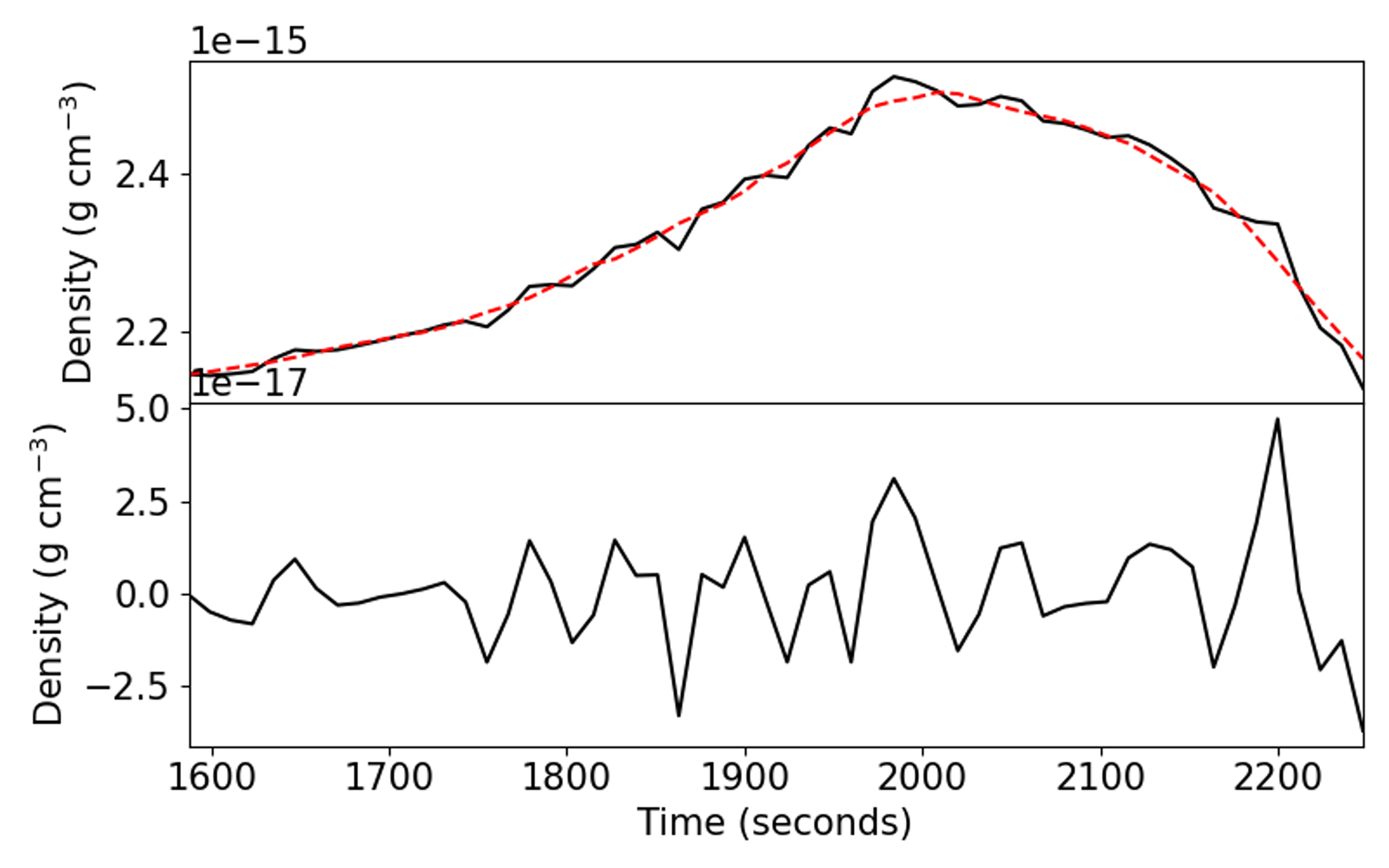}
         \hspace*{-0.01\textwidth}
         \includegraphics[height=6.5 cm,trim={0 0 0 0},clip]{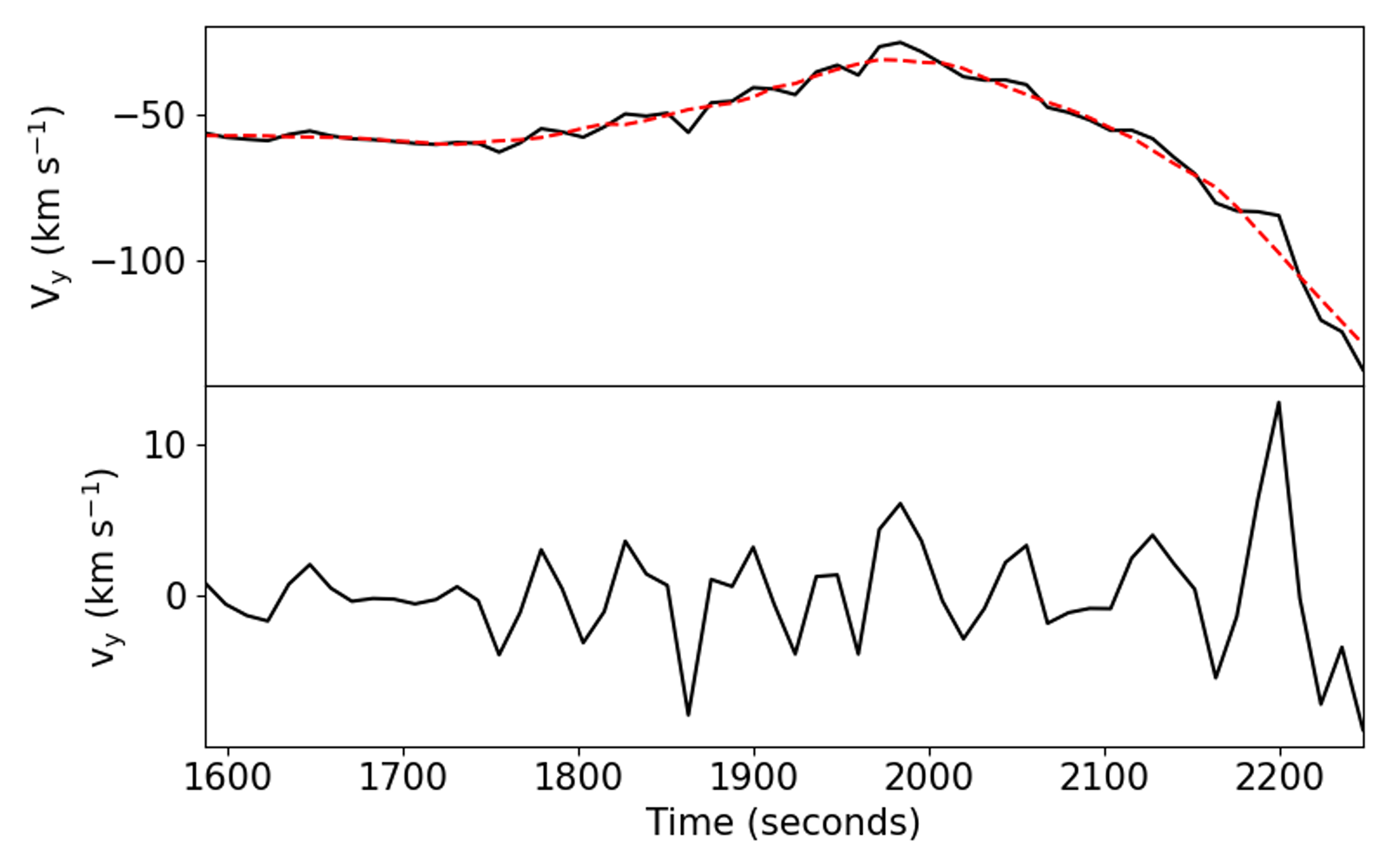}
        }
\vspace{-0.38\textwidth}
\centerline{ \large \bf      
\hspace{-0.09 \textwidth}  \color{black}{(c)}
\hspace{0.55\textwidth}  \color{black}{(d)}
   \hfill}
\vspace{0.38\textwidth}

\centerline{\hspace*{0.013\textwidth}
         \includegraphics[height=6.7 cm,trim={0 0 0 0},clip]{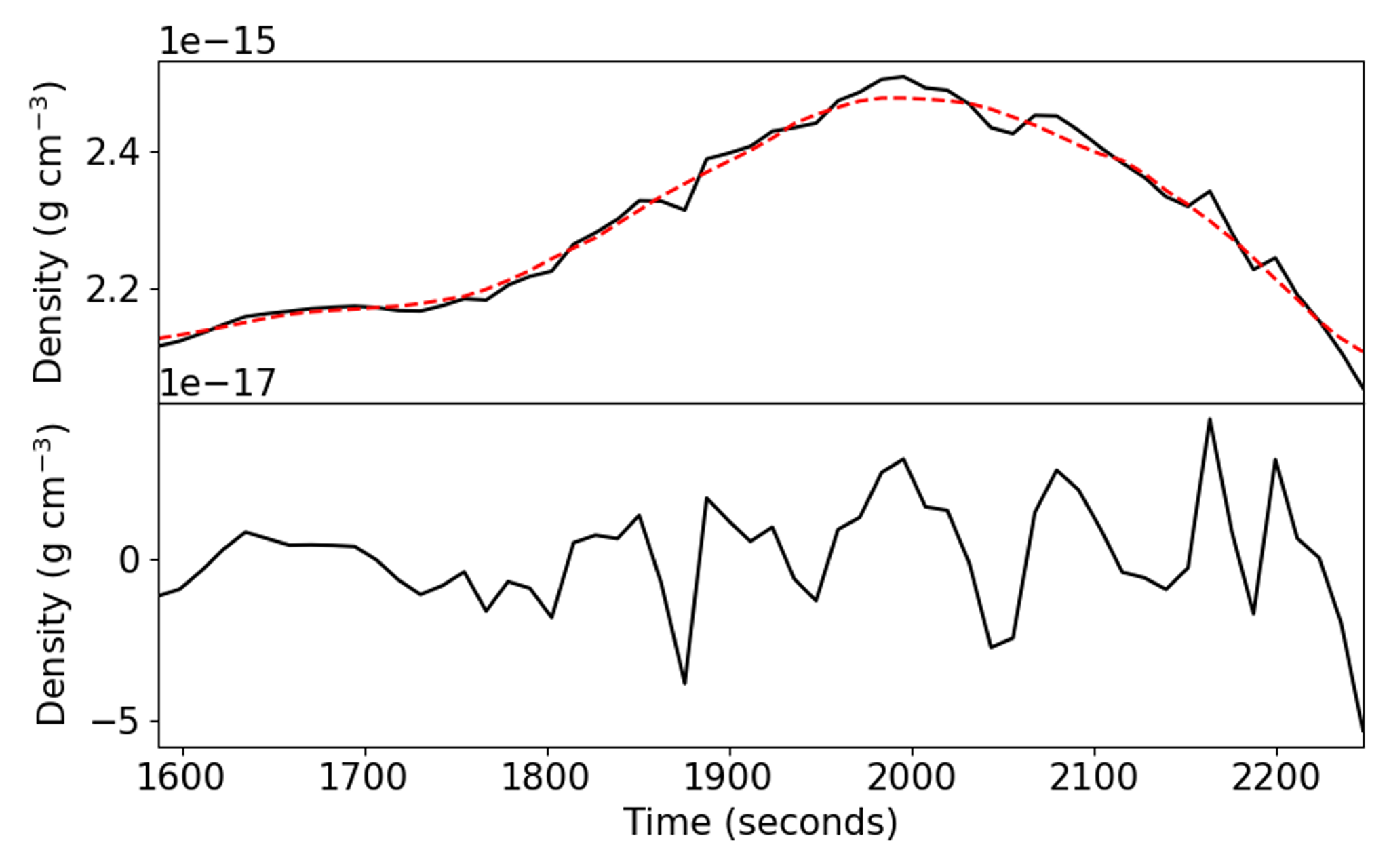}
         \hspace*{-0.01\textwidth}
         \includegraphics[height=6.5 cm,trim={0 0 0 0},clip]{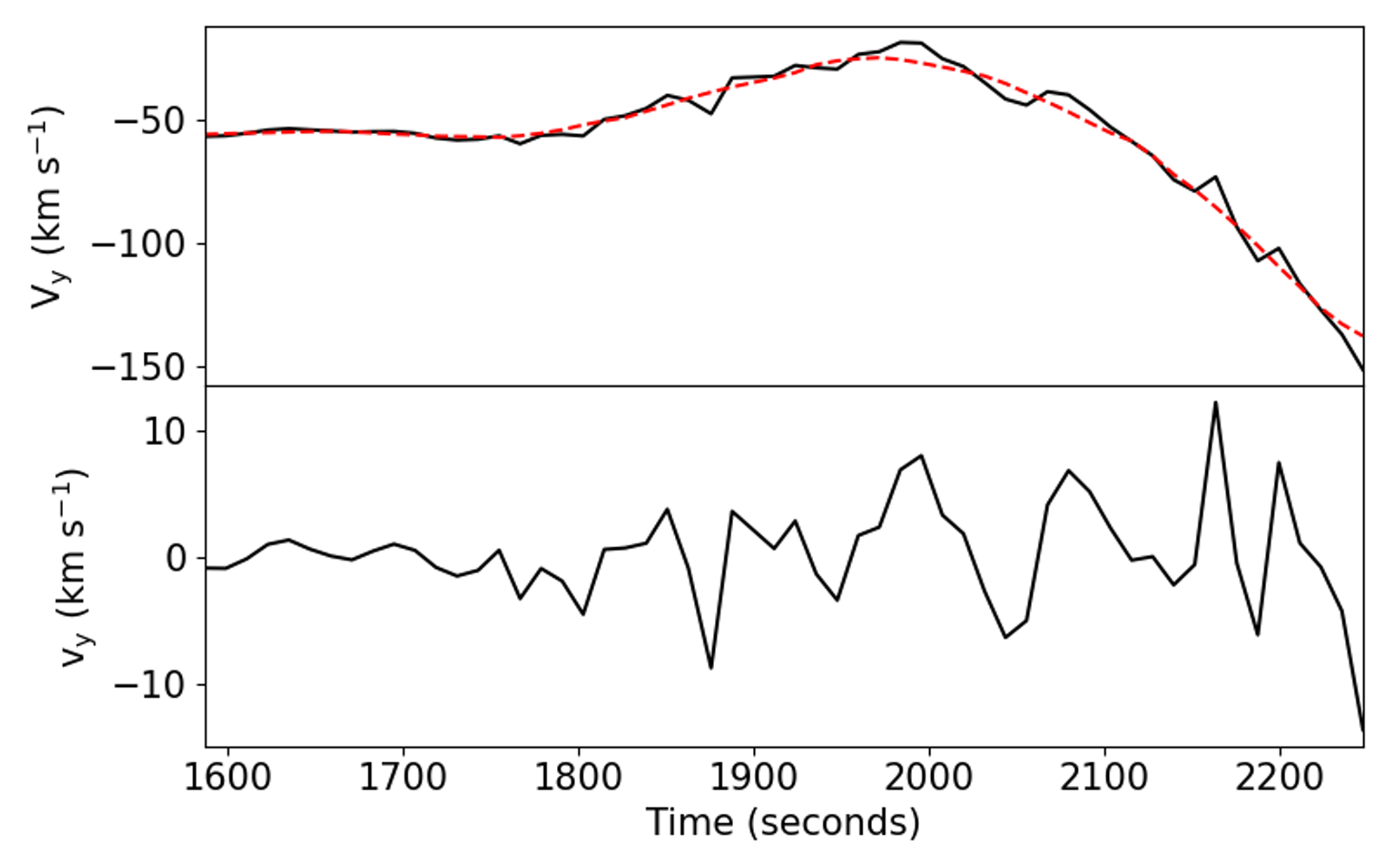}
        }
\vspace{-0.38\textwidth}
\centerline{ \large \bf      
\hspace{-0.09 \textwidth}  \color{black}{(e)}
\hspace{0.55\textwidth}  \color{black}{(f)}
   \hfill}
\vspace{0.38\textwidth}
\caption{The upper panels of  (a), (c) and (e) show the density variations (black solid curve) measured at locations I, II and III, respectively. The red dashed curves in those panels denote the long-term background trends, which are subtracted in order to produce the  detrended small-scale density fluctuations shown in the bottom panels of (a), (c) and (e),  which are a signature of wave propagation. The panels (b), (d) and (f) show the corresponding results for $v_y$.}
\label{label 10}
\end{figure*}

One important similarity between our simulation and the corona is the presence of a dynamically evolving background. As a result, subtraction of that relatively long-term trend due to contributions from the time-dependent background is necessary while analysing the perturbations due to propagating waves in this simulation. Basically, the extracted small-scale fluctuations  are associated with the propagation of multiple wavefronts -- propagating on top of a slowly evolving background. We have fitted the long-term trends of the derived time-series associated with density, $V_{y}$ and $B_{x}$ in terms of the best fit using a box-car average window width 7 for locations I and II, and 8 for location III. These choices are based on the fact that the long-term trend passes almost through the Gaussian width of each fluctuation peak or dip present in the time series. The subtracted long-term trends are depicted as red dashed curves in the top sub-panels of  Figure~\ref{label 10} and the left panels of Figure~\ref{label 11}. We further check the analysis by varying the window-width between 4 to 10. For location I, the dominant period remains 59.06 s for window width 4 to 7 with significance level varying between 96 and 99.5 $\%$ for density, $v_{y}$ and $b_{x}$. For widths higher than 7, the long-term curves do not properly pass through the Gaussian width and the period changes to 99.32 s with the significance level around 94-95 $\%$. For location II, the estimated period remains 70.23 s for window width 6 and 7 with significance level 96-97.5 $\%$. Beyond width 7, the period is estimated to be 76.59 s up to window width 9 with significance level being 93-95 $\%$. But the long term curve does not fit well for window width 8 onwards. For location III, the dominant period remains 91.08 s for window width 6 to 10 with significance level  92-97.5 $\%$ for all three aforementioned physical variables.  Under the above detailed analyses, the most reliable and significant periods at locations I, II, and III are 59.06 s, 70.23 s, and 91.08 s respectively.

\section{Perturbation in Magnetic Field due to Wave Propagation}
\begin{figure*}
\centerline{\hspace*{0.01\textwidth}
         \includegraphics[height=6.5 cm,trim={0 0 0 0},clip]{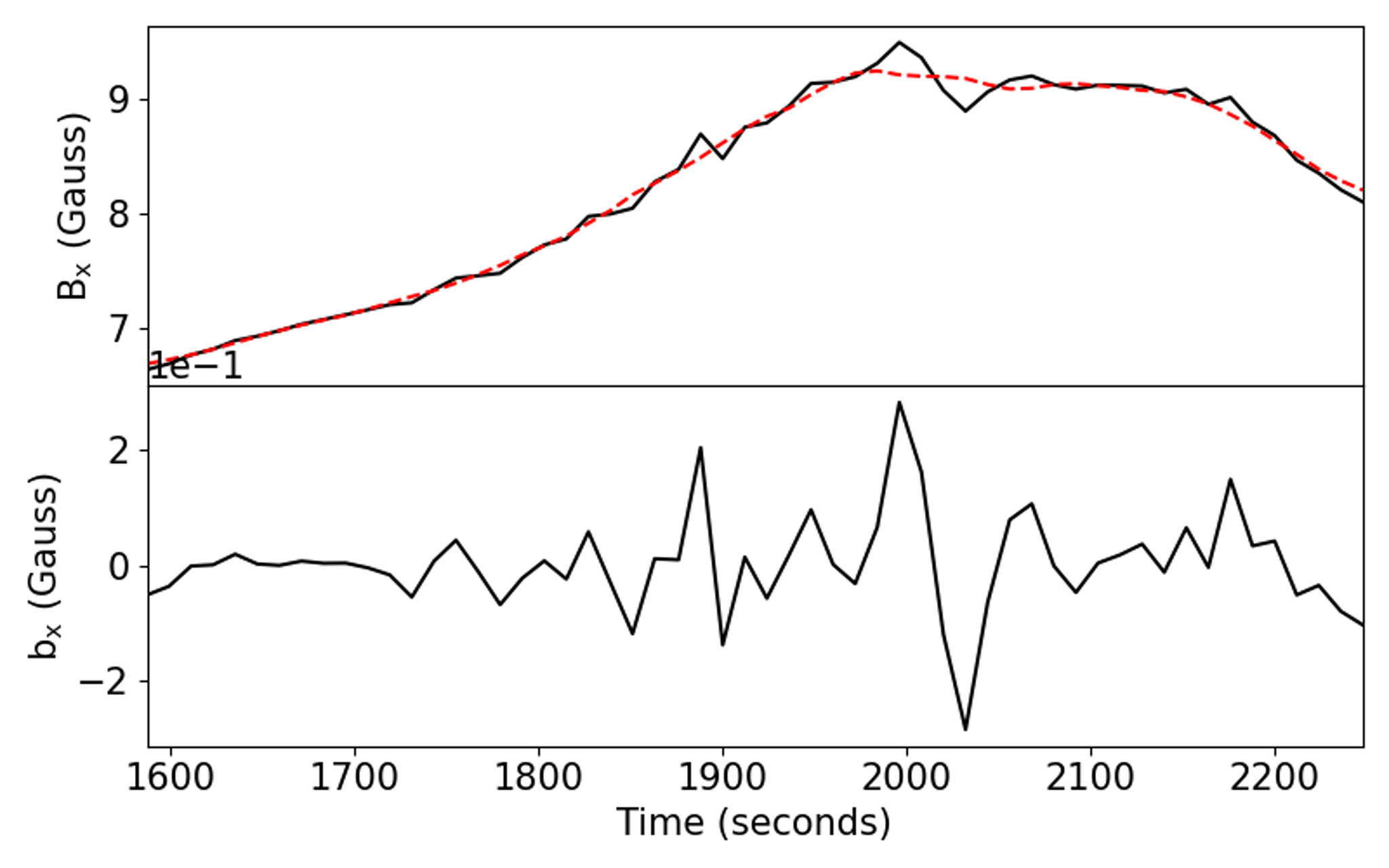}
         \hspace*{-0.01\textwidth}
         \includegraphics[height=7.2 cm,trim={0 0 0 0},clip]{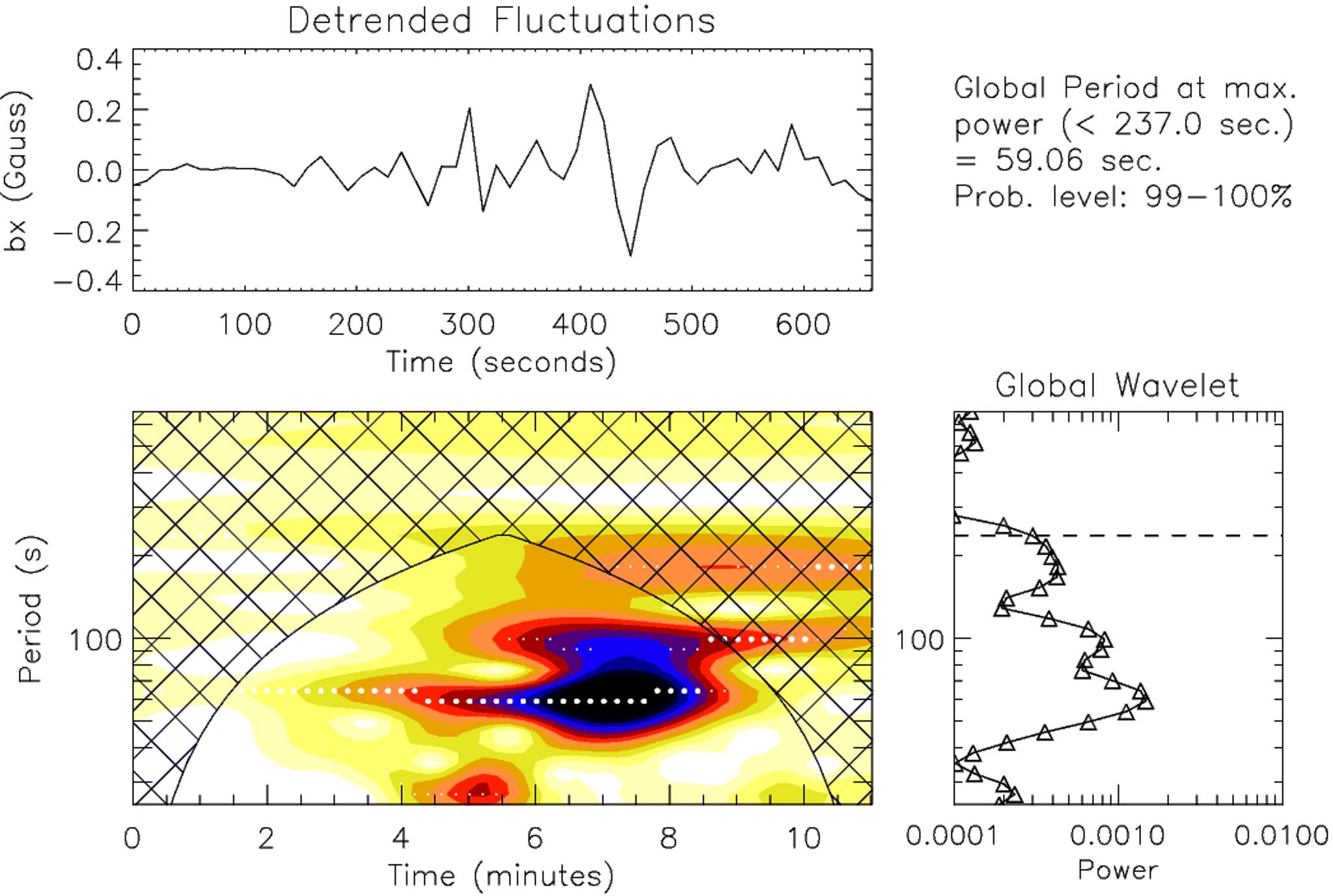}
       }
\vspace{-0.40\textwidth}
\centerline{ \large \bf      
\hspace{-0.09 \textwidth}  \color{black}{(a)}
\hspace{0.50\textwidth}  \color{black}{(b)}
   \hfill}
\vspace{0.39\textwidth}    

\centerline{\hspace*{0.01\textwidth}
         \includegraphics[height=6.5 cm,trim={0 0 0 0},clip]{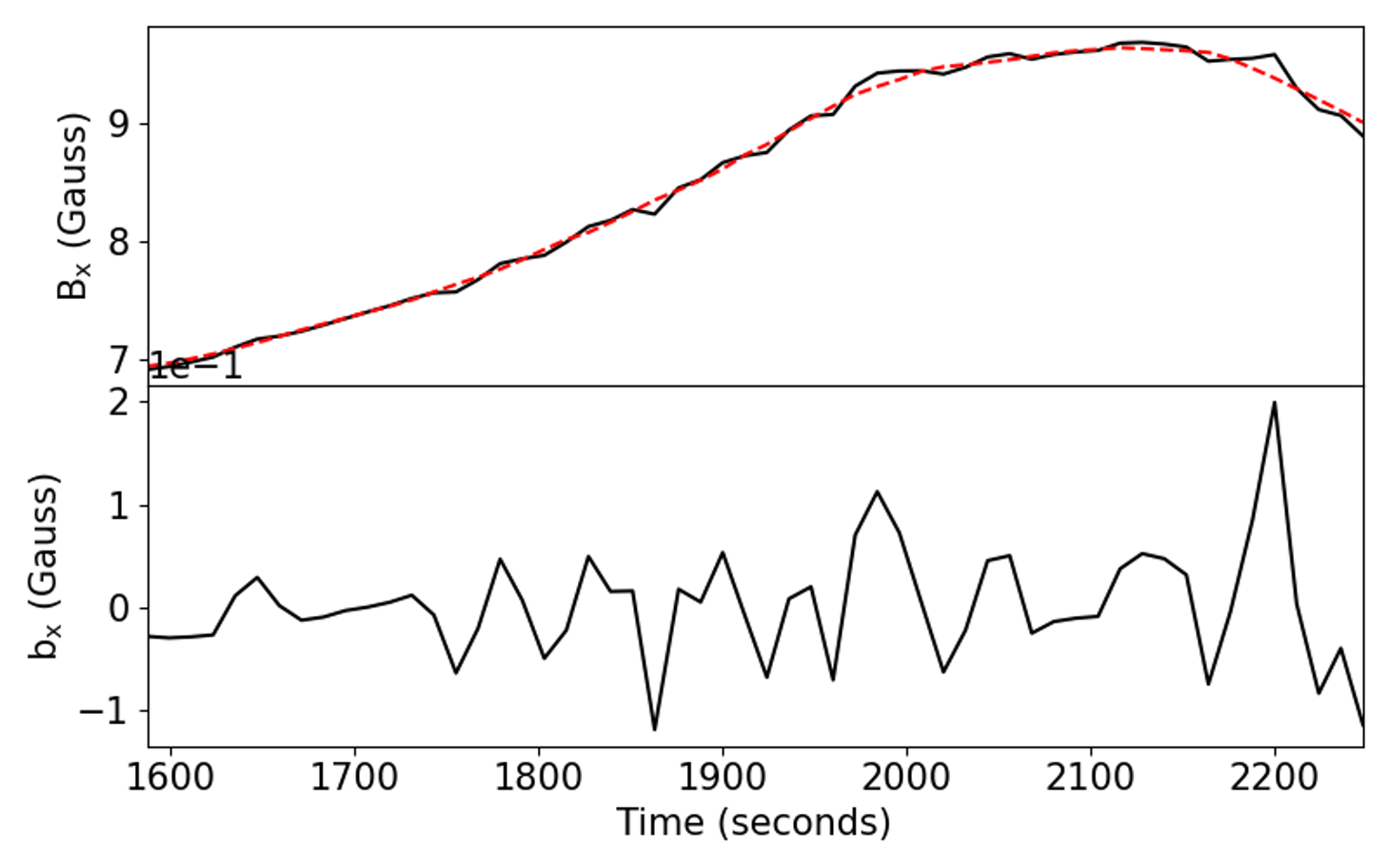}
         \hspace*{-0.01\textwidth}
         \includegraphics[height=7.2 cm,trim={0 0 0 0},clip]{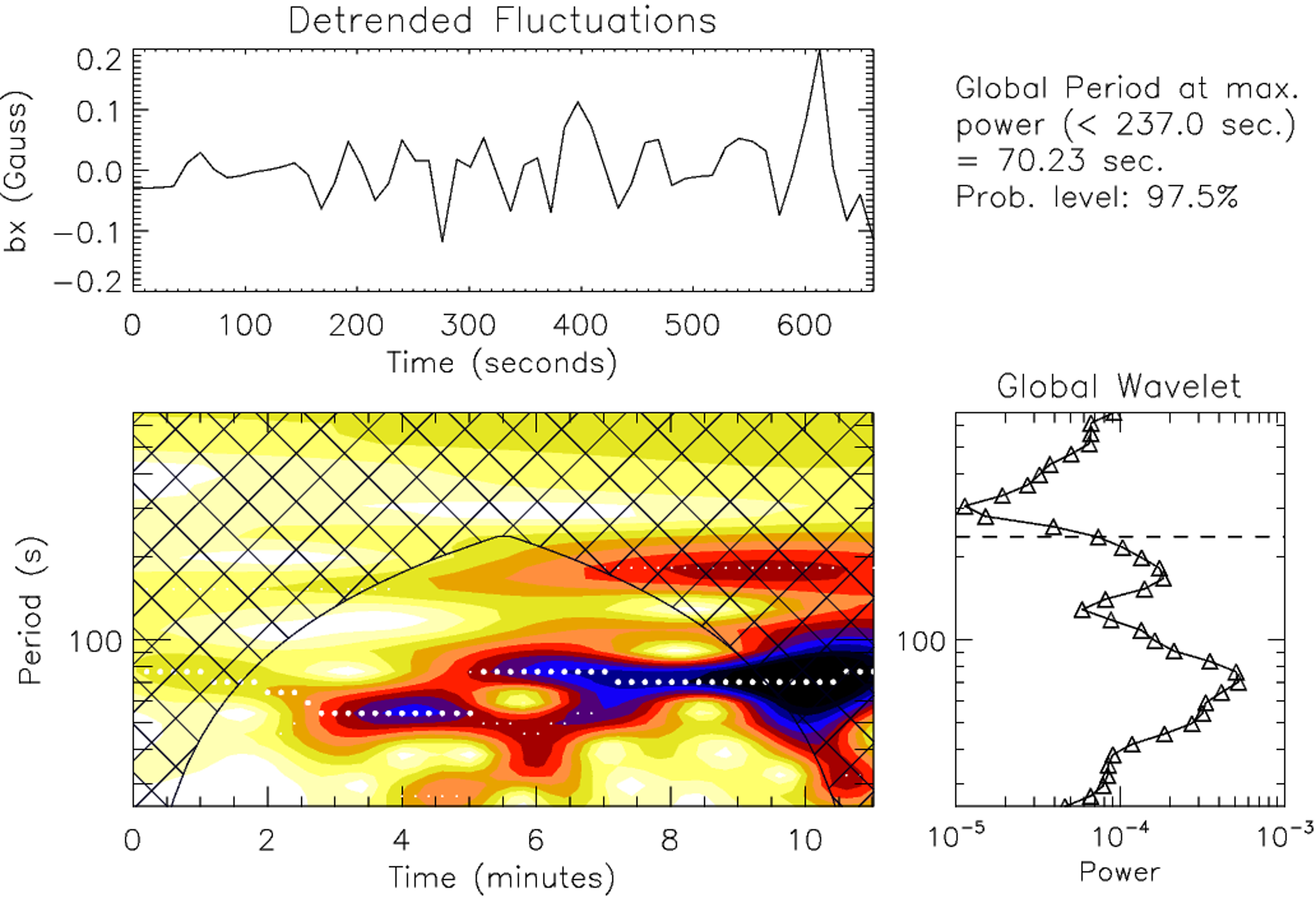}
       }
\vspace{-0.40\textwidth}
\centerline{ \large \bf      
\hspace{-0.09 \textwidth}  \color{black}{(c)}
\hspace{0.50\textwidth}  \color{black}{(d)}
   \hfill}
\vspace{0.39\textwidth}    

\centerline{\hspace*{0.01\textwidth}
         \includegraphics[height=6.5 cm,trim={0 0 0 0},clip]{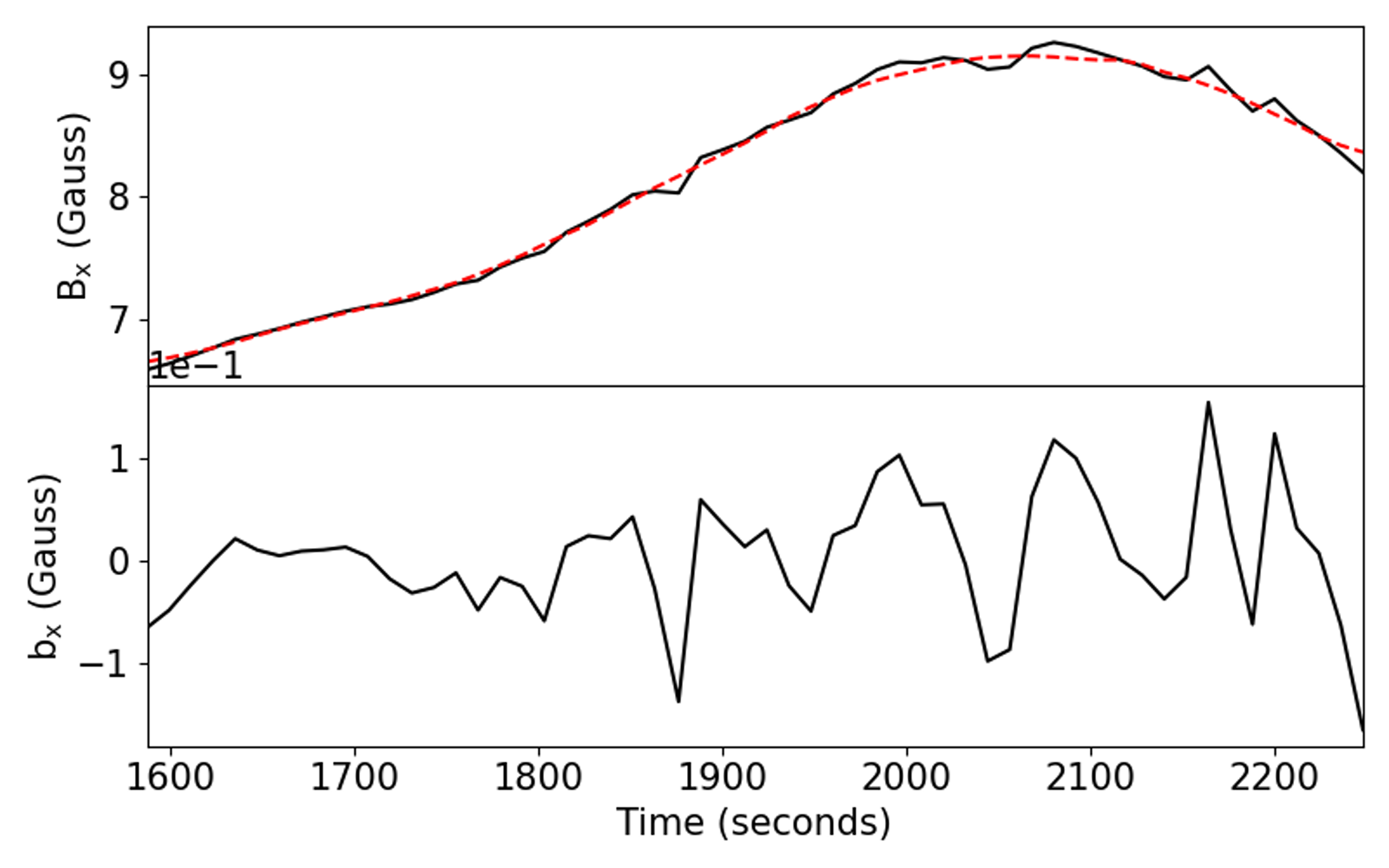}
         \hspace*{-0.01\textwidth}
         \includegraphics[height=7.2 cm,trim={0 0 0 0},clip]{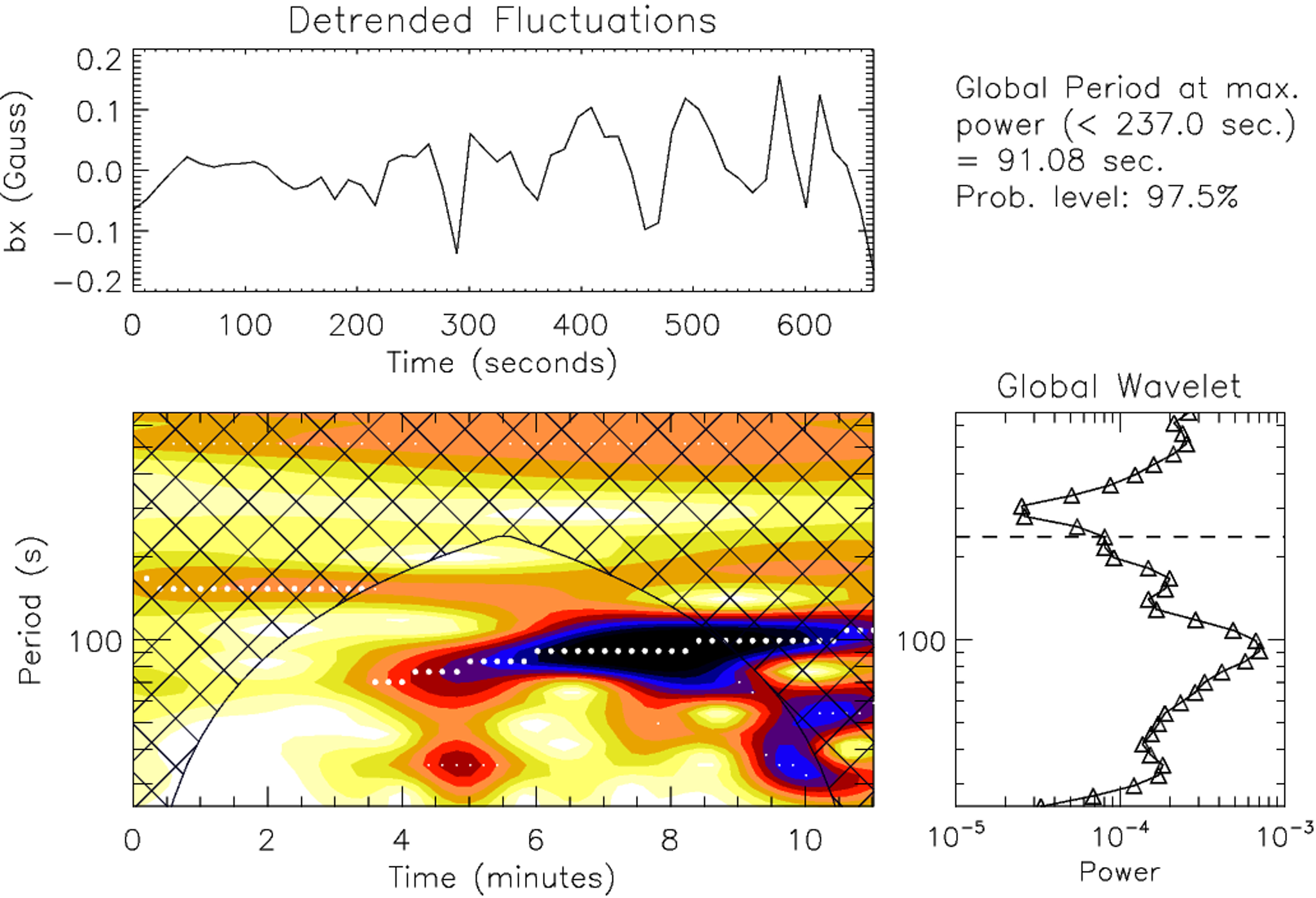}
       }
\vspace{-0.40\textwidth}
\centerline{ \large \bf      
\hspace{-0.09 \textwidth}  \color{black}{(e)}
\hspace{0.50\textwidth}  \color{black}{(f)}
   \hfill}
\vspace{0.39\textwidth}           

\caption{The fluctuations in $B_x$ are produced by subtracting the long-term background trends (red dashed curves) from the original profiles measured by taking an average of 3 pixels at the locations I, II and III denoted in Figure~\ref{label 3}(a) (see panels (a), (c) and (e)). The top panels of  (b), (d) and (e) repeat the bottom panels of  (a), (c) and (e) and are placed above the corresponding wavelet distributions for comparison. Note that 0 s in these figures corresponds to 1587 s, i.e., the starting time of the figures in the left panels. The wavelet distributions exhibit dominant periods of 59.06 s, 70.23 s and 91.08 s at locations I, II and III, respectively,  as in Figure~\ref{label 5}.}
\label{label 11}
\end{figure*}

As we discuss in Section 3.3.3 and Appendix C, we extracted fluctuations in density, $y$-component of velocity and $x$-component of magnetic field. We exhibit those fluctuations in density and $y$-component of velocity in Figure~\ref{label 5}. Here, we present the perturbations in the $x$-component of the magnetic field, i.e., the component (approximately) perpendicular to the direction of propagation of the waves. This is estimated in the same manner as described in Appendix C by subtracting the long-term background trend to leave only the `wave component', $b_x$ (see left panels of Figure~\ref{label 11}). We find similar profiles and periodicities in $b_{x}$ as we find for density and $v_{y}$. It is to be noted that the temporal variation of density perturbations, perturbations in $v_{y}$ and $b_{x}$ are all in phase with each other at  the three locations which is a fundamental characteristics of fast-mode waves.

\section{Spatio-temporal Evolution of Magnitude of Total Magnetic Field Before and During Wave Propagation}
\begin{figure*}
\centerline{\hspace*{0.013\textwidth}
         \includegraphics[height=7.2 cm,trim={0 0 0 0},clip]{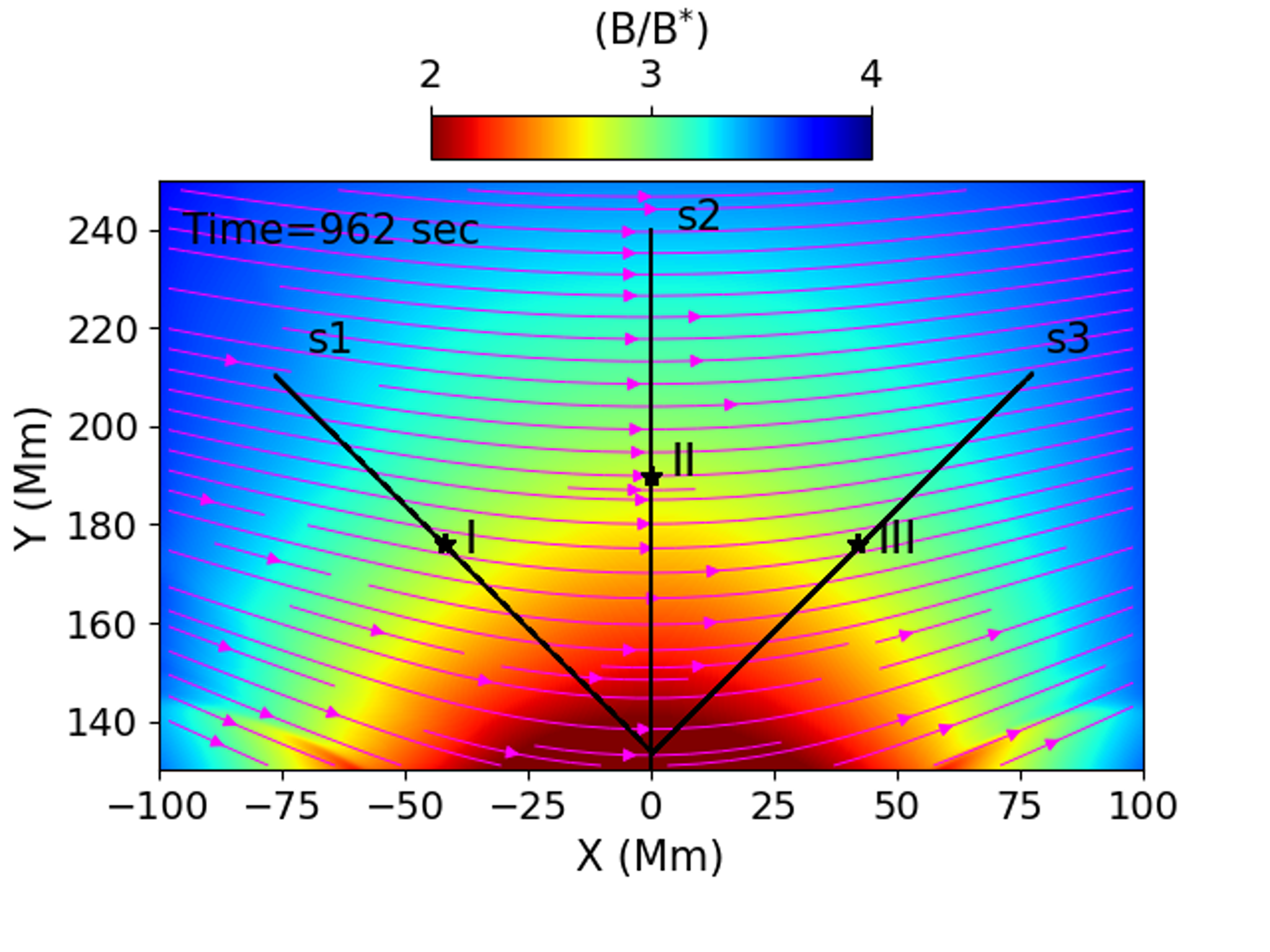}
         \hspace*{-0.01\textwidth}
         \includegraphics[height=7.2 cm,trim={0 0 0 0},clip]{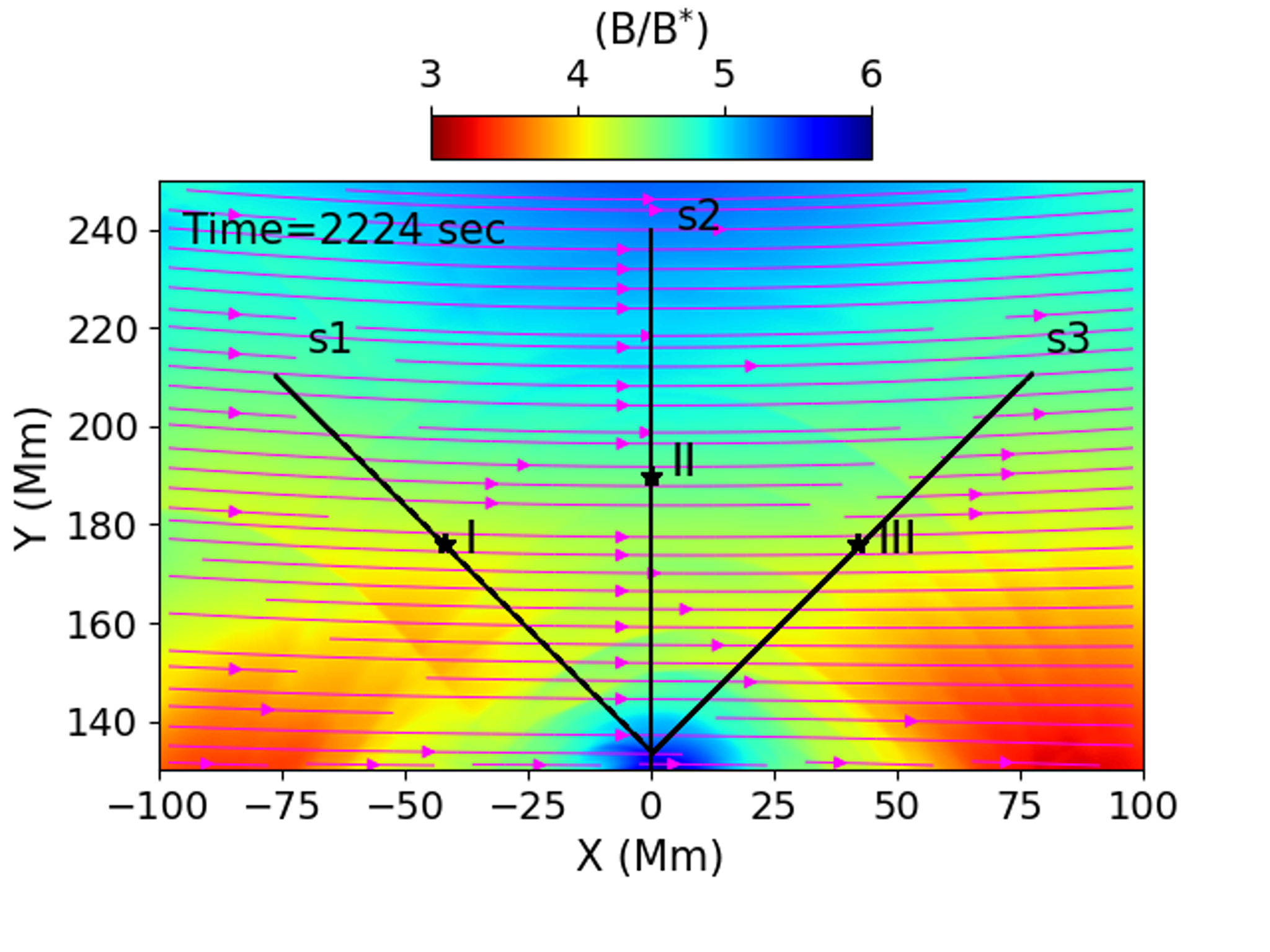}
         }
\vspace{-0.40\textwidth}
\centerline{ \large \bf      
\hspace{-0.05 \textwidth}  \color{black}{(a1)}
\hspace{0.48\textwidth}  \color{black}{(a2)}
   \hfill}
\vspace{0.37\textwidth}    

\centerline{\hspace*{0.013\textwidth}
         \includegraphics[height=5.6 cm,trim={0 0 0 0},clip]{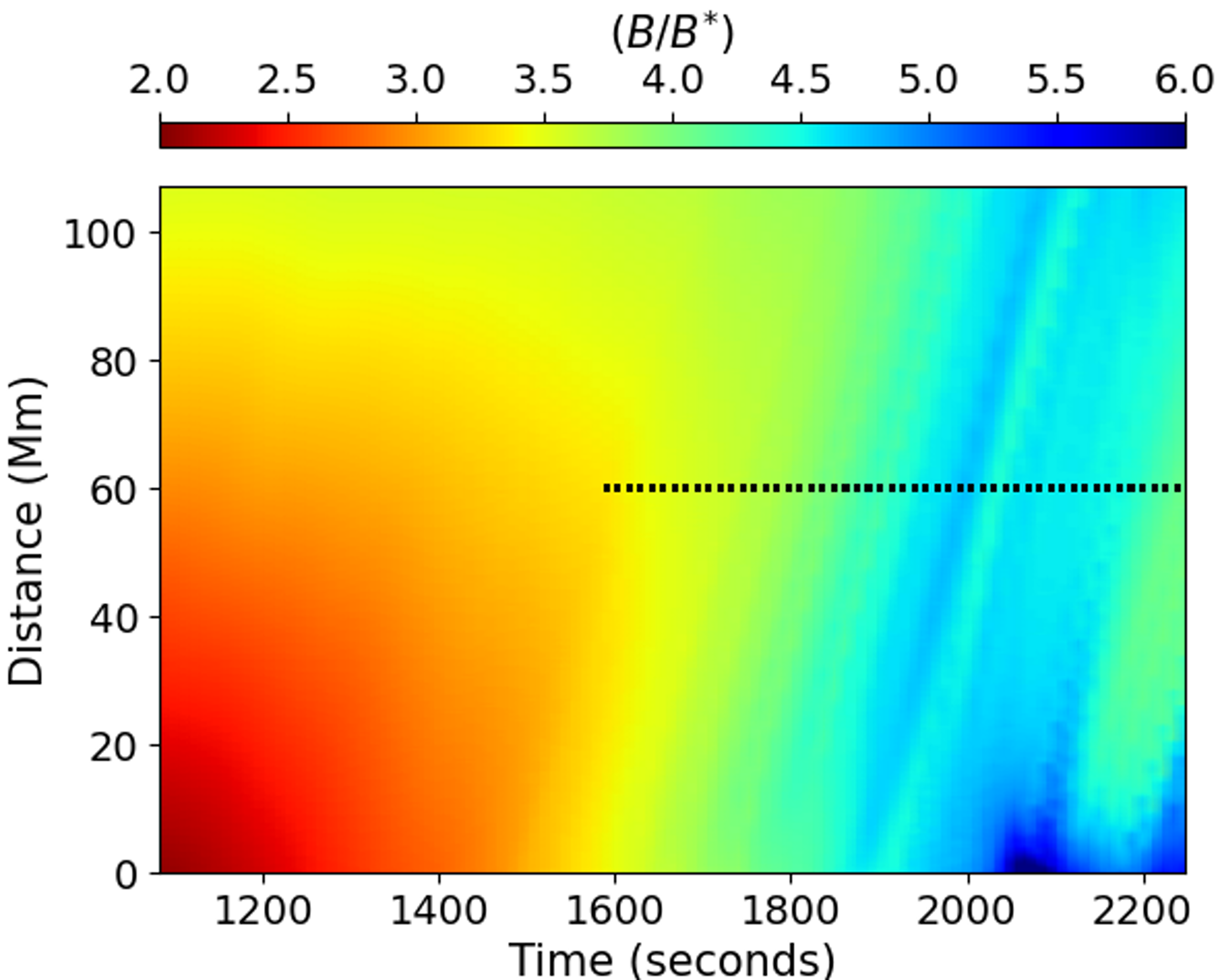}
         \hspace*{-0.01\textwidth}
         \includegraphics[height=5.6 cm,trim={0 0 0 0},clip]{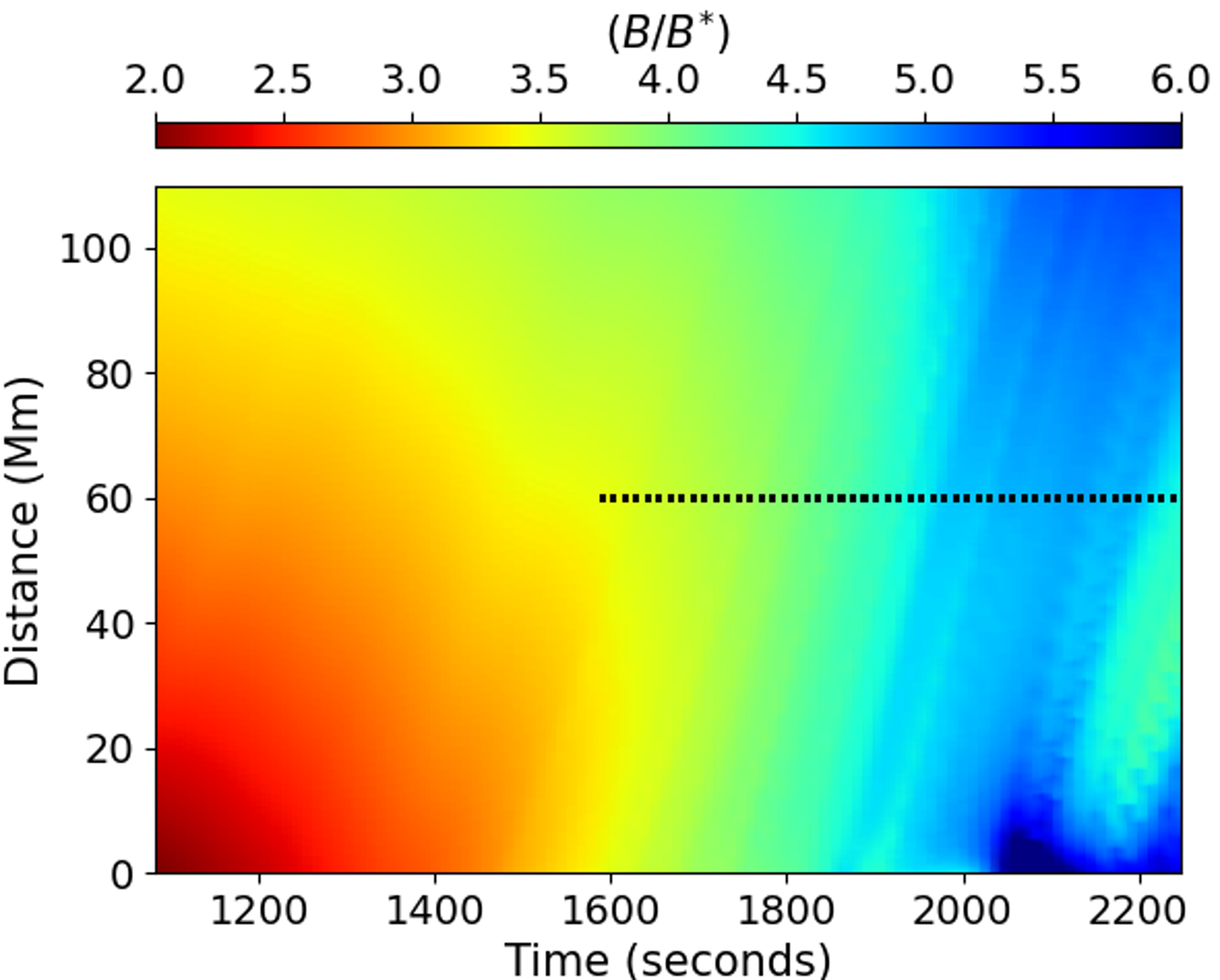}
         \hspace*{-0.01\textwidth}
         \includegraphics[height=5.6 cm,trim={0 0 0 0},clip]{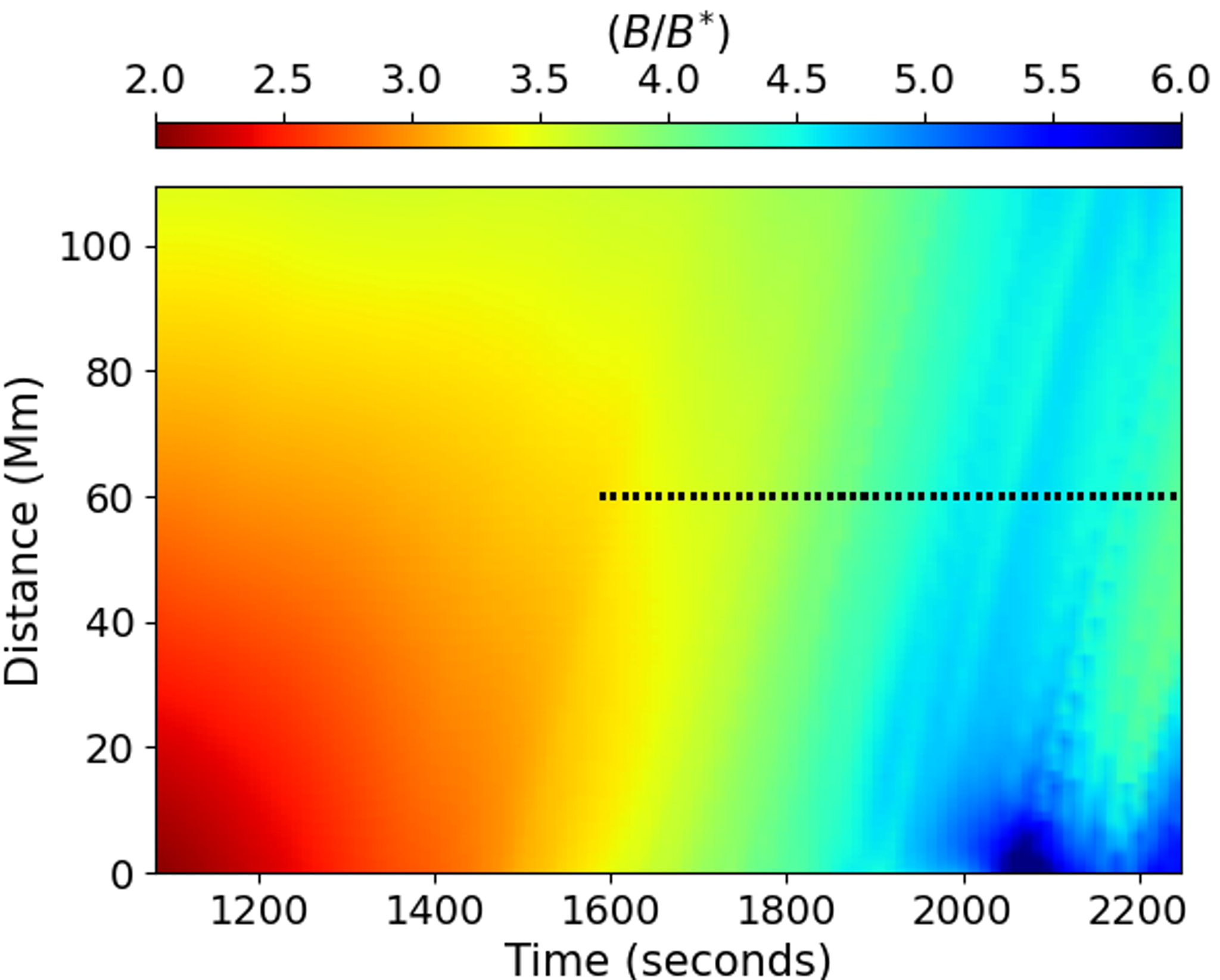}
        }
\vspace{-0.33\textwidth}
\centerline{ \large \bf      
\hspace{-0.09 \textwidth}  \color{black}{(b)}
\hspace{0.34\textwidth}  \color{black}{(c)}
\hspace{0.34\textwidth}  \color{black}{(d)}
   \hfill}
\vspace{0.31\textwidth}
\caption{The spatial distribution of total magnetic field at (a1) 962 s when the current sheet has been formed (see right panel of Figure~\ref{label 1}a) and at (a2) 2224 s during propagation of the waves (the same time as in Figure~\ref{label 3}a). The colormap scales for these plots are different due to the magnetic evolution. Magenta streamlines indicate the  magnetic fields at 962 and 2224 s in (a1) and (a2), respectively. Panels (b), (c) and (d) show distance-time diagrams of the magnetic field magnitude along the slits s1, s2 and s3, respectively.}
\label{label 12}
\end{figure*}

Since the background magnetic field is very dynamic as evident from long-term trends in $B_{x}$ at three different locations shown as red dashed curves in panels (a), (c) and (e) of Figure~\ref{label 11}, we determine the spatial distribution of the total magnetic field at two different times (see Figure~\ref{label 12}) (one after formation of the current sheet but before start of wave propagation and another during wave propagation) as follows:

[i] at 962 s when the current sheet has formed but has not yet become impulsive and bursty due to the onset of the plasmoid instability;

[ii] at 2224 s, when multiple wavefronts are propagating in the domain as a result of multiple coalescences in the tearing current sheet.

So panels (a1) and (a2) of Figure~\ref{label 12} clearly provide evidence of spatial non-uniformity of the total magnetic field before and during wave propagation. Magenta curves indicate the  magnetic field at 962 and 2224 s. Even though at 962 s the magnetic field is initially curved, it becomes almost horizontal by the time wave propagation starts at around 1587 s (see also Figure~\ref{label 4}a, c, e). Therefore, we further provide a distance-time diagram of the total magnetic field to visualise the spatial as well as temporal evolution of the magnetic field along slits `s1', `s2' and `s3' oriented in three different directions as shown in panel (a1) and (a2) of Figure~\ref{label 12}. These slits are the same ones used in producing the distance-time maps of density in Figure~\ref{label 3}. Hence, from panels (b), (c) and (d) of Figure~\ref{label 12}, it is clear that the spatial distribution of the magnitude of the magnetic field is not uniform along the slits before the start of wave propagation and varies from 4 to 6 Gauss roughly from 0 to 60 Mm. On contrary, from 1587 s, i.e., the starting time of wave propagation, the spatial variation of magnetic field diminishes and it becomes almost uniform along all of the three slits. Moreover, the total magnetic field is almost horizontally oriented with only a 2-3\degree inclination to the positive $x$-direction, while the magnitude of the total magnetic field varies only from 7 to 9 Gauss from 1587 s to the end of the simulation at the three locations of measurements. In addition, the temporal evolutions of the background magnetic field at the three locations I, II and III are very similar, as depicted by the long-term trends (red dashed curves) in Figure~\ref{label 11}.

\begin{figure*}
\hspace{-0.8 cm}
\includegraphics[scale=0.8]{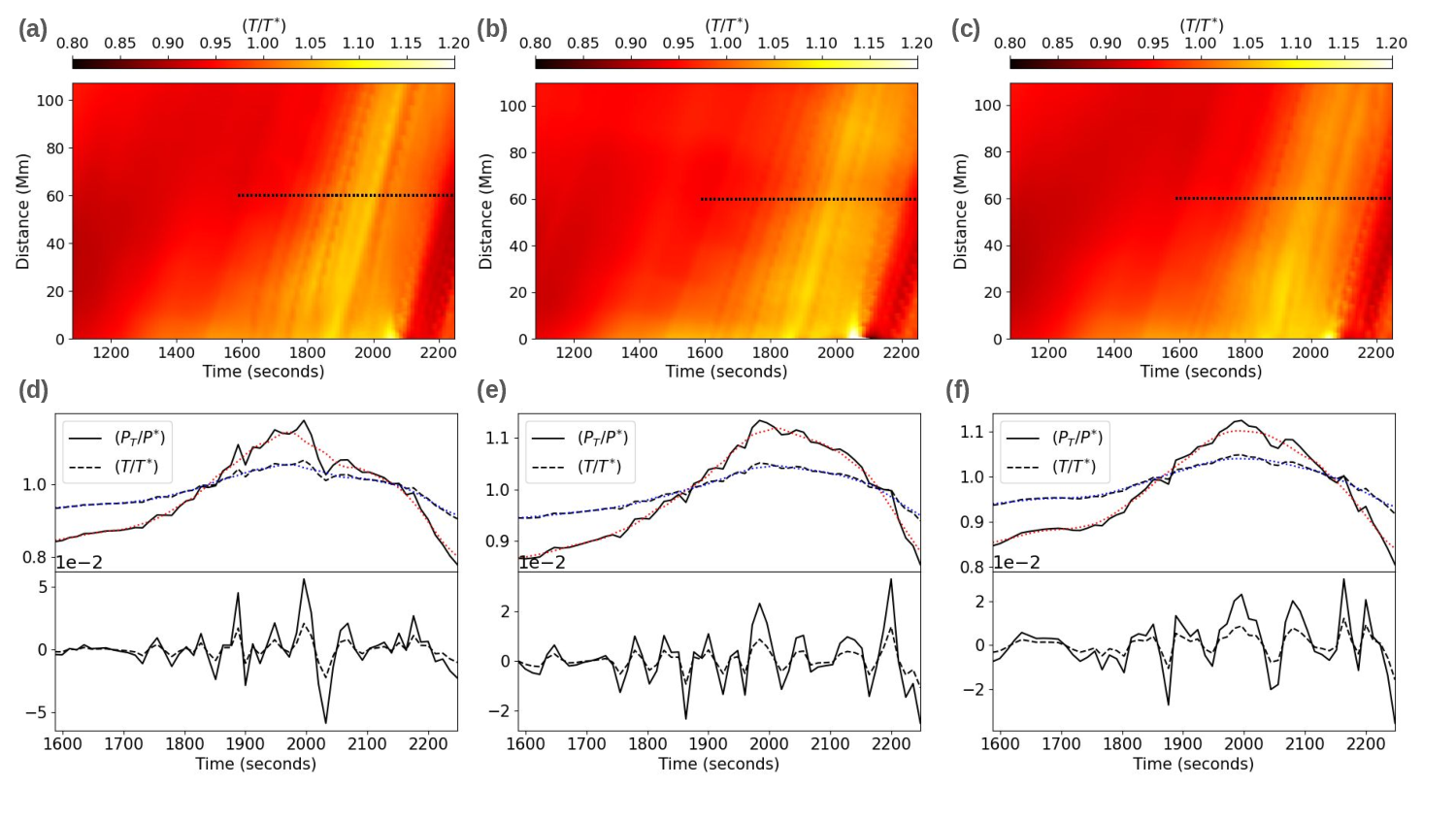}
\caption{Panels (a), (b) and (c) give the spatio-temporal variations of temperature along slits `s1', `s2' and `s3', respectively. The black dotted horizontal lines are same as those in Figure~\ref{label 3} and Figure~\ref{label 9}. Panels (d), (e) and (f) represents the in-phase relation between perturbations in thermal pressure and temperature. Since these two quantities have different dimensions physically, we plot them in dimensionless form for consistency. Red and blue dotted curves are the background trends which are subtracted to extract the perturbed quantities using same time windows as described in Appendix C.}
\label{label 13}
\end{figure*}


\begin{thebibliography}{}
\expandafter\ifx\csname natexlab\endcsname\relax\def\natexlab#1{#1}\fi
\providecommand{\url}[1]{\href{#1}{#1}}
\providecommand{\dodoi}[1]{doi:~\href{http://doi.org/#1}{\nolinkurl{#1}}}
\providecommand{\doeprint}[1]{\href{http://ascl.net/#1}{\nolinkurl{http://ascl.net/#1}}}
\providecommand{\doarXiv}[1]{\href{https://arxiv.org/abs/#1}{\nolinkurl{https://arxiv.org/abs/#1}}}

\end{thebibliography}


\begin{thebibliography}{}
\bibitem[Antolin \& Shibata(2010)]{2010ApJ...712..494A} Antolin, P. \& Shibata, K.\ 2010, \apj, 712, 494. doi:10.1088/0004-637X/712/1/494
\bibitem[Aschwanden(2005)]{2005psci.book.....A} Aschwanden, M.~J.\ 2005, Physics of the Solar Corona. An Introduction with Problems and Solutions (2nd edition), by M.J. Aschwanden.  892 pages.  ISBN 3-540-30765-6, Library of Congress Control Number: 2005937065.  Praxis Publishing Ltd., Chichester, UK; Springer, New York, Berlin, 2005.
\bibitem[Bhattacharjee et al.(2009)]{2009PhPl...16k2102B} Bhattacharjee, A., Huang, Y.-M., Yang, H., et al.\ 2009, Physics of Plasmas, 16, 112102. doi:10.1063/1.3264103
\bibitem[Biskamp \& Welter(1980)]{1980PhRvL..44.1069B} Biskamp, D. \& Welter, H.\ 1980, \prl, 44, 1069. doi:10.1103/PhysRevLett.44.1069
\bibitem[Biskamp(1982)]{1982PhLA...87..357B} Biskamp, D.\ 1982, Physics Letters A, 87, 357. doi:10.1016/0375-9601(82)90844-1
\bibitem[Craig \& McClymont(1991)]{1991ApJ...371L..41C} Craig, I.~J.~D. \& McClymont, A.~N.\ 1991, \apjl, 371, L41. doi:10.1086/185997

\bibitem[Davila(1987)]{1987ApJ...317..514D} Davila, J.~M.\ 1987, \apj, 317, 514. doi:10.1086/165295

\bibitem[Delaboudini{\`e}re et al.(1995)]{1995SoPh..162..291D} Delaboudini{\`e}re, J.-P., Artzner, G.~E., Brunaud, J., et al.\ 1995, \solphys, 162, 291. doi:10.1007/BF00733432
\bibitem[Edwin \& Zheliazkov(1992)]{1992SoPh..140....7E} Edwin, P.~M. \& Zheliazkov, I.\ 1992, \solphys, 140, 7. doi:10.1007/BF00148426

\bibitem[Finn \& Kaw(1977)]{1977PhFl...20...72F} Finn, J.~M. \& Kaw, P.~K.\ 1977, Physics of Fluids, 20, 72. doi:10.1063/1.861709
\bibitem[Forbes \& Priest(1982)]{1982SoPh...81..303F} Forbes, T.~G. \& Priest, E.~R.\ 1982, \solphys, 81, 303. doi:10.1007/BF00151304
\bibitem[Forbes \& Priest(1983)]{1983SoPh...84..169F} Forbes, T.~G. \& Priest, E.~R.\ 1983, \solphys, 84, 169. doi:10.1007/BF00157455
\bibitem[Forbes \& Priest(1987)]{1987RvGeo..25.1583F} Forbes, T.~G. \& Priest, E.~R.\ 1987, Reviews of Geophysics, 25, 1583. doi:10.1029/RG025i008p01583
\bibitem[Gallagher \& Long(2011)]{2011SSRv..158..365G} Gallagher, P.~T. \& Long, D.~M.\ 2011, \ssr, 158, 365. doi:10.1007/s11214-010-9710-7
\bibitem[Goddard et al.(2016)]{2016A&A...594A..96G} Goddard, C.~R., Nistic{\`o}, G., Nakariakov, V.~M., et al.\ 2016, \aap, 594, A96. doi:10.1051/0004-6361/201628478
\bibitem[Guo et al.(2019)]{2019ApJ...870L..21G} Guo, Y., Xia, C., Keppens, R., et al.\ 2019, \apjl, 870, L21. doi:10.3847/2041-8213/aafabf
\bibitem[Harten(1983)]{1983JCoPh..49..357H} Harten, A.\ 1983, Journal of Computational Physics, 49, 357. doi:10.1016/0021-9991(83)90136-5
\bibitem[Heyvaerts \& Priest(1983)]{1983A&A...117..220H} Heyvaerts, J. \& Priest, E.~R.\ 1983, \aap, 117, 220

\bibitem[Hollweg(1975)]{1975RvGSP..13..263H} Hollweg, J.~V.\ 1975, Reviews of Geophysics and Space Physics, 13, 263. doi:10.1029/RG013i001p00263
\bibitem[Hollweg(2007)]{Holl07} Hollweg, J.~V.\ 2007, J. Geophys. Res., 112, A8

\bibitem[Jel{\'\i}nek et al.(2017)]{2017ApJ...847...98J} Jel{\'\i}nek, P., Karlick{\'y}, M., Van Doorsselaere, T., et al.\ 2017, \apj, 847, 98. doi:10.3847/1538-4357/aa88a6



\bibitem[Keppens et al.(2023)]{2023A&A...673A..66K} Keppens, R., Popescu Braileanu, B., Zhou, Y., et al.\ 2023, \aap, 673, A66. doi:10.1051/0004-6361/202245359
\bibitem[Kumar et al.(2017)]{2017ApJ...844..149K} Kumar, P., Nakariakov, V.~M., \& Cho, K.-S.\ 2017, \apj, 844, 149. doi:10.3847/1538-4357/aa7d53
\bibitem[Leake et al.(2020)]{2020ApJ...891...62L} Leake, James E., Daldorff, Lars K. S., \& Klimchuk, James A.\ 2020, \apj, 891, 62. doi:10.3847/1538-4357/ab7193
\bibitem[Leake et al.(2024)]{2024ApJ...973...21L} Leake, James E., Daldorff, Lars K. S., \& Klimchuk, James A.\ 2024, \apj, 973, 21. doi:10.3847/1538-4357/ad5e71

\bibitem[Lemen et al.(2012)]{2012SoPh..275...17L} Lemen, J.~R., Title, A.~M., Akin, D.~J., et al.\ 2012, \solphys, 275, 17. doi:10.1007/s11207-011-9776-8
\bibitem[Liberatore et al.(2023)]{2023ApJ...957..110L} Liberatore, A., Liewer, P.~C., Vourlidas, A., et al.\ 2023, \apj, 957, 110. doi:10.3847/1538-4357/acf8bf
\bibitem[Li et al.(2018)]{2018ApJ...868L..33L} Li, L., Zhang, J., Peter, H., et al.\ 2018, \apjl, 868, L33. doi:10.3847/2041-8213/aaf167
\bibitem[Linnell Nemec \& Nemec(1985)]{1985AJ.....90.2317L} Linnell Nemec, A.~F. \& Nemec, J.~M.\ 1985, \aj, 90, 2317. doi:10.1086/113936

\bibitem[Liu et al.(2011)]{2011ApJ...736L..13L} Liu, W., Title, A.~M., Zhao, J., et al.\ 2011, \apjl, 736, L13. doi:10.1088/2041-8205/736/1/L13
\bibitem[Liu et al.(2012)]{2012ApJ...753...52L} Liu, W., Ofman, L., Nitta, N.~V., et al.\ 2012, \apj, 753, 52. doi:10.1088/0004-637X/753/1/52
\bibitem[Liu \& Ofman(2014)]{2014SoPh..289.3233L} Liu, W. \& Ofman, L.\ 2014, \solphys, 289, 3233. doi:10.1007/s11207-014-0528-4
\bibitem[Longcope \& Priest(2007)]{2007PhPl...14l2905L} Longcope, D.~W. \& Priest, E.~R.\ 2007, Physics of Plasmas, 14, 122905. doi:10.1063/1.2823023

\bibitem[Loureiro et al.(2007)]{2007PhPl...14j0703L} Loureiro, N.~F., Schekochihin, A.~A., \& Cowley, S.~C.\ 2007, Physics of Plasmas, 14, 100703. doi:10.1063/1.2783986
\bibitem[Marsch(1986)]{1986A&A...164...77M} Marsch, E.\ 1986, \aap, 164, 77

\bibitem[Mondal et al.(2024)]{2024ApJ...963..139M} Mondal, S., Srivastava, A.~K., Pontin, D.~I., Ding Yuan \& Priest, E.R.\ 2024, \apj, 963, 139. doi:10.3847/1538-4357/ad2079
\bibitem[Nakariakov \& Verwichte(2005)]{2005LRSP....2....3N} Nakariakov, V.~M. \& Verwichte, E.\ 2005, Living Reviews in Solar Physics, 2, 3. doi:10.12942/lrsp-2005-3
\bibitem[Nistic{\`o} et al.(2014)]{2014A&A...569A..12N} Nistic{\`o}, G., Pascoe, D.~J., \& Nakariakov, V.~M.\ 2014, \aap, 569, A12. doi:10.1051/0004-6361/201423763
\bibitem[Ofman et al.(1998)]{1998ApJ...493..474O} Ofman, L., Klimchuk, J.~A., \& Davila, J.~M.\ 1998, \apj, 493, 474. doi:10.1086/305109

\bibitem[Ofman \& Liu(2018)]{2018ApJ...860...54O} Ofman, L. \& Liu, W.\ 2018, \apj, 860, 54. doi:10.3847/1538-4357/aac2e8
\bibitem[Ofman \& Kucera(2020)]{2020ApJ...899...99O} Ofman, L. \& Kucera, T.~A.\ 2020, \apj, 899, 99. doi:10.3847/1538-4357/aba2eb
\bibitem[O'Shea et al.(2001)]{2001A&A...368.1095O} O'Shea, E., Banerjee, D., Doyle, J.~G., et al.\ 2001, \aap, 368, 1095. doi:10.1051/0004-6361:20010073
\bibitem[Patsourakos \& Vourlidas(2012)]{2012SoPh..281..187P} Patsourakos, S. \& Vourlidas, A.\ 2012, \solphys, 281, 187. doi:10.1007/s11207-012-9988-6
\bibitem[Pek{\"u}nl{\"u} et al.(2001)]{2001MNRAS.326..675P} Pek{\"u}nl{\"u}, E.~R., {\c{C}}ak{\i}rl{\i}, {\"O}., \& {\"O}zetken, E.\ 2001, \mnras, 326, 675. doi:10.1046/j.1365-8711.2001.04639.x

\bibitem[Pontin et al.(2024)]{2024ApJ...960...51P} Pontin, D.~I., Priest, E.~R., Chitta, L.~P., et al.\ 2024, \apj, 960, 51. doi:10.3847/1538-4357/ad03eb
\bibitem[Porter et al.(1994)]{1994ApJ...435..482P} Porter, L.~J., Klimchuk, J.~A., \& Sturrock, P.~A.\ 1994, \apj, 435, 482. doi:10.1086/174830
\bibitem[Powell et al. (1999)]{1999JCoPh.154..284P} Powell, Kenneth G., Roe, Philip L., Linde, Timur J., et al.\ 1999, Journal of Computational Physics, 154, 284. doi:10.1006/jcph.1999.6299
\bibitem[Priest(1986)]{1986MitAG..65...41P} Priest, E.~R.\ 1986, Mitteilungen der Astronomischen Gesellschaft Hamburg, 65, 41
\bibitem[Priest(2014)]{2014masu.book.....P} Priest, E.\ 2014, Magnetohydrodynamics of the Sun, by Eric Priest, Cambridge, UK: Cambridge University Press, 2014. doi:10.1017/CBO9781139020732
\bibitem[Russell \& Stackhouse(2013)]{2013A&A...558A..76R} Russell, A.~J.~B. \& Stackhouse, D.~J.\ 2013, \aap, 558, A76. doi:10.1051/0004-6361/201321916
\bibitem[Sakai(1983)]{1983JPlPh..30..109S} Sakai, J.-I.\ 1983, Journal of Plasma Physics, 30, 109. doi:10.1017/S0022377800001033

\bibitem[Sen \& Keppens(2022)]{2022A&A...666A..28S} Sen, S. \& Keppens, R.\ 2022, \aap, 666, A28. doi:10.1051/0004-6361/202244152
\bibitem[Sen et al.(2023)]{2023A&A...678A.132S} Sen, S., Jenkins, J., \& Keppens, R.\ 2023, \aap, 678, A132. doi:10.1051/0004-6361/202347038

\bibitem[Shen \& Liu(2012)]{2012ApJ...753...53S} Shen, Y. \& Liu, Y.\ 2012, \apj, 753, 53. doi:10.1088/0004-637X/753/1/53
\bibitem[Shen et al.(2013)]{2013SoPh..288..585S} Shen, Y.-D., Liu, Y., Su, J.-T., et al.\ 2013, \solphys, 288, 585. doi:10.1007/s11207-013-0395-4
\bibitem[Srivastava et al.(2021)]{2021JGRA..12629097S} Srivastava, A.~K., Ballester, J.~L., Cally, P.~S., et al.\ 2021, Journal of Geophysical Research (Space Physics), 126, e029097. doi:10.1029/2020JA029097
\bibitem[Srivastava et al.(2024)]{Sri24} Srivastava, A.~K., Priest, E.~R., Ofman, L., Mondal, Sripan, Kwon, R.-Y., Pontin, D., Murawski, K., Mishra, S.~K., Yuan, Ding, Asai, A., 2024, COSPAR 44th Scientific Assembly, Busan, S. Korea, E2.7-0002-24.
\bibitem[Stewart et al.(2022)]{2022MNRAS.513.5224S} Stewart, J., Browning, P.~K., \& Gordovskyy, M.\ 2022, \mnras, 513, 5224. doi:10.1093/mnras/stac1286

\bibitem[Takasao \& Shibata(2016)]{2016ApJ...823..150T} Takasao, S. \& Shibata, K.\ 2016, \apj, 823, 150. doi:10.3847/0004-637X/823/2/150
\bibitem[Van Doorsselaere et al.(2020)]{2020SSRv..216..140V} Van Doorsselaere, T., Srivastava, A.~K., Antolin, P., et al.\ 2020, \ssr, 216, 140. doi:10.1007/s11214-020-00770-y
\bibitem[van Leer(1979)]{1979JCoPh..32..101V} van Leer, B.\ 1979, Journal of Computational Physics, 32, 101. doi:10.1016/0021-9991(79)90145-1
\bibitem[Warmuth(2015)]{2015LRSP...12....3W} Warmuth, A.\ 2015, Living Reviews in Solar Physics, 12, 3. doi:10.1007/lrsp-2015-3
\bibitem[Withbroe \& Noyes(1977)]{1977ARA&A..15..363W} Withbroe, G.~L. \& Noyes, R.~W.\ 1977, \araa, 15, 363. doi:10.1146/annurev.aa.15.090177.002051
\bibitem[Wuelser et al.(2004)]{2004SPIE.5171..111W} Wuelser, J.-P., Lemen, J.~R., Tarbell, T.~D., et al.\ 2004, \procspie, 5171, 111. doi:10.1117/12.506877
\bibitem[Wyper \& Pontin(2014)]{2014PhPl...21h2114W} Wyper, P.~F. \& Pontin, D.~I.\ 2014, Physics of Plasmas, 21, 082114. doi:10.1063/1.4893149

\bibitem[Xia et al.(2012)]{2012ApJ...748L..26X} Xia, C., Chen, P.~F., \& Keppens, R.\ 2012, \apjl, 748, L26. doi:10.1088/2041-8205/748/2/L26

\bibitem[Yang et al.(2015)]{2015ApJ...800..111Y} Yang, L., Zhang, L., He, J., et al.\ 2015, \apj, 800, 111. doi:10.1088/0004-637X/800/2/111
\bibitem[Yuan et al.(2013)]{2013A&A...554A.144Y} Yuan, D., Shen, Y., Liu, Y., et al.\ 2013, \aap, 554, A144. doi:10.1051/0004-6361/201321435
\bibitem[Yuan et al.(2019)]{2019ApJ...886L..25Y} Yuan, D., Feng, S., Li, D., et al.\ 2019, \apjl, 886, L25. doi:10.3847/2041-8213/ab5648
\bibitem[Zhao et al.(2017)]{2017ApJ...841..106Z} Zhao, X., Xia, C., Keppens, R., et al.\ 2017, \apj, 841, 106. doi:10.3847/1538-4357/aa7142

\bibitem[Zheng et al.(2018)]{2018ApJ...858L...1Z} Zheng, R., Chen, Y., Feng, S., et al.\ 2018, \apjl, 858, L1. doi:10.3847/2041-8213/aabe87

\end{thebibliography}


\end{document}